\documentclass[twocolumn,floatfix,prx,aps,amsmath,amssymb,longbibliography,eqsecnum, footinbib]{revtex4-1}
\usepackage{float}
\usepackage{graphicx}
\usepackage{amsmath}
\usepackage{amssymb}
\usepackage{color}
\usepackage[pdftex,breaklinks,bookmarks=false,colorlinks,linkcolor=blue,citecolor=blue,urlcolor=blue]{hyperref}

\usepackage{mathtools} 

\DeclarePairedDelimiter\floor{\lfloor}{\rfloor}

\newcommand{\abs}[1]{\left\lvert#1\right\rvert} 
\DeclareMathOperator{\Tr}{Tr}

\begin{document}
\title{Topological Multipartite Entanglement in a Fermi Liquid}

\author{Pok Man Tam, Martin Claassen and Charles L. Kane}
\affiliation{Department of Physics and Astronomy, University of Pennsylvania, Philadelphia, PA 19104}

\begin{abstract}
We show that the topology of the Fermi sea of a $D$-dimensional Fermi gas is reflected in the multipartite entanglement characterizing $D+1$ regions that meet at a point.   For odd $D$ we introduce the multipartite mutual information, and show that it exhibits a $\log^D L$ divergence as a function of system size $L$ with a universal coefficient that is proportional to the Euler characteristic $\chi_F$ of the Fermi sea.  This provides a generalization, for a Fermi gas, of the well known result for $D=1$ that expresses the $\log L$ divergence of the bipartite entanglement entropy in terms of the central charge $c$ characterizing a conformal field theory.    For even $D$ we introduce a charge-weighted entanglement entropy that is manifestly odd under a particle-hole transformation.   We show that the corresponding charge-weighted mutual information exhibits a similar $\log^D L$ divergence proportional to $\chi_F$.   Our analysis relates the universal behavior of the multipartite mutual information in the absence of interactions to the $D+1$'th order equal-time density correlation function, which we show exhibits a universal behavior in the long wavelength limit proportional to $\chi_F$.   Our analytic results are based on the replica method.  In addition we perform a numerical study of the charge-weighted mutual information for $D=2$ that confirms several aspects of the analytic theory.    Finally, we consider the effect of interactions perturbatively within the replica theory.   We show that for $D=3$ the $\log^3 L$ divergence of the topological mutual information is not perturbed by weak short-ranged interactions, though for $D=2$ the charge-weighted mutual information is perturbed.   Thus, for $D=3$ the multipartite mutual information provides a robust classification that distinguishes distinct topological Fermi liquid phases.
\end{abstract}

\maketitle
\tableofcontents

\section{Introduction}
\label{sec I}

A powerful method for characterizing the phases of quantum many particle systems is to identify the patterns of long-range entanglement present in the ground state wavefunction.   A hallmark for this type of analysis is the theory of the topological entanglement entropy of a gapped $2+1$ dimensional topological phase \cite{Kitaev2006, Levin2006}.   In that case, the total quantum dimension of the quasiparticle excitations, which is a topological quantity characterizing the phase, is related to a measure of the entanglement in the ground state wavefunction given by the mutual information between three subregions in the plane. This analysis has subsequently been generalized for higher-dimensional gapped topological phases \cite{Castelnovo2008, Grover2011}.

Long-range entanglement also occurs in gapless systems.   It is well known that in $1+1$ dimensional conformal field theory (CFT) the bipartite entanglement entropy exhibits a logarithmic divergence as a function of system size $L$ with a coefficient proportional to the central charge $c$ \cite{Holzhey1994, Vidal2003, Calabrese2004, *Calabrese2009}.  $c$ is a topologically quantized quantity that characterizes the low energy degrees of freedom responsible for long-ranged entanglement.   There has been considerable interest in generalizing this type of analysis to higher dimensions \cite{Cardy1988ctheorem, Ryu2006, Solodukhin2008, Casini2009, Myers2011, Casini2011, Liu2013, Casini2015}.

Fermi gasses provide a tractable setting to characterize entanglement, and have a broad application to electronic materials.    A 1D Fermi gas is a simple CFT, in which $c$  counts the number of (right moving) Fermi points, or equivalently the number of components of the 1D Fermi sea.   In $D$ dimensions, the bipartite entanglement entropy of a Fermi gas exhibits an area law with a logarithmically divergent coefficient that probes the projected area of the Fermi surface \cite{Wolf2006, Klich2006, Swingle2010, Ding2012, Calabrese2012a,Yang2016}.   This can be understood simply by considering a quasi 1D geometry with periodic boundary conditions for the remaining $D-1$ dimensions.   Then, since the transverse momentum eigenstates decouple into independent 1D modes, the coefficient of the log in the entanglement entropy simply counts the number of 1D modes below the Fermi level, given by $A A_F/(2\pi)^D$, where $A \sim L^{D-1}$ is the real space area of the boundary and $A_F \sim k_F^{D-1}$ is the projected area of the Fermi sea.   For a more general bipartition, this result can be expressed as an integral over both the real space boundary and the Fermi surface in a form analogous to the Widom formula from the theory of signal processing \cite{Klich2006}.

Unlike the 1D case, the coefficient of the logarithmic divergence of the bipartite entanglement entropy for a $D>1$ dimensional Fermi gas is \textit{not} a topological quantity.   Since it depends on both the dimensions of the Fermi surface and the dimensions of the partition boundary, it will vary continuously as non-universal parameters are adjusted.   Moreover, this coefficient does not distinguish qualitatively distinct patterns of entanglement.   For example, a 3D system composed of independent 1D wires clearly exhibits long-ranged entanglement along the wires, but there is no entanglement between the wires.   The bipartite entanglement entropy is not sensitive to this distinction.

The quasi 1D and 3D Fermi surfaces described above are distinguished by the topology of the filled Fermi sea, which can be characterized by its {\it Euler characteristic}, $\chi_F$\cite{Kane2022}.   $\chi_F$ is an integer topological invariant defined as
\begin{equation}
\chi_F = \sum_{l=0}^D (-1)^l b_l,
\label{chif betti}
\end{equation}
where $b_l$ is the $l$'th Betti number, given by the rank of the $l$'th homology group, which counts the topologically distinct $l$-cycles \cite{Nakahara1990}.  A 3D spherical Fermi sea has $\chi_F = 1-0+0-0 = 1$, while a quasi 1D Fermi sea, which spans the 3D Brillouin zone in two directions, has  $\chi_F = 1-2+1-0 = 0$.  
According to the Morse theory, $\chi_F$ can also be expressed in terms of the critical points in the electronic dispersion $E_{\bf k}$ \cite{Milnor1963, Nash1988}.  $\chi_F$ is related to the critical points ${\bf k}_c$ in $E_{\bf k}$, where ${\bf v}_{\bf k} = \nabla_{\bf k} E_{\bf k} = 0$.   
\begin{equation}
\chi_F = \sum_c f_{{\bf k}_c} \eta_c,
\label{chif morse}
\end{equation}
where $f_{\bf k} = \theta(E_F - E_{\bf k})$ specifies the Fermi sea, and the signature of each critical point is given by $\eta_c = {\rm sgn}\det {\mathbb H}_c$, where the Hessian ${\mathbb H}_c$ is the matrix of second derivatives of $E_{\bf k}$.   It follows that $\chi_F$ changes at a Lifshitz transition, where a minimum, maximum or saddle point in $E_{\bf k}$ passes through $E_F$ \cite{Lifshitz1960, Volovik2017}.

\begin{figure}
\includegraphics[width=0.8\columnwidth]{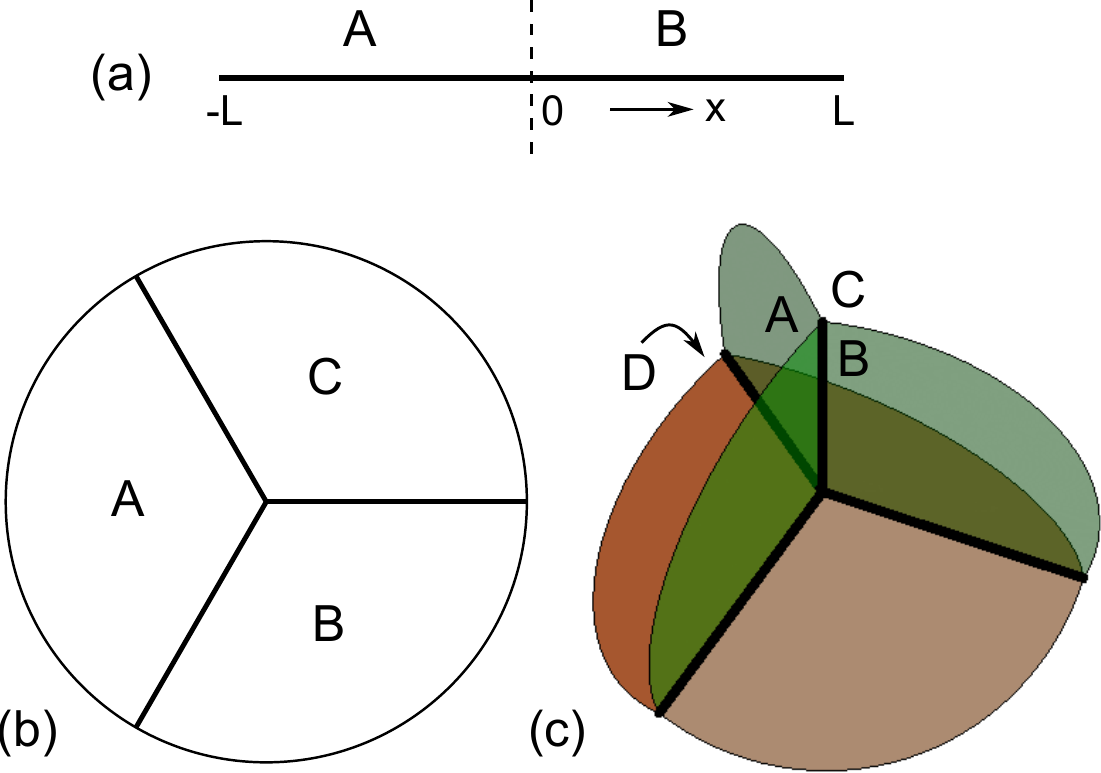}
\caption{The $D$-dimensional Fermi gas for (a) $D=1$, (b) $D=2$ and (c) $D=3$ is partitioned into $D+1$ regions that meet at a single point, with any $k$ regions sharing a flat boundary of dimension $D+1-k$.  Here we study an entanglement measure, known as the mutual information, that captures the intrinsic correlations among all $D+1$ regions. The mutual information is \textit{topological} in that it exhibits a leading logarithmic divergence proportional to the Euler characteristic $\chi_F$ of the Fermi sea. }
\label{Fig1}
\end{figure}

In this paper we introduce an entanglement measure characterizing the ground state wavefunction that is sensitive to the Fermi sea topology.    The key difference between $D=1$ and higher dimensions from the point of view of entanglement is that in $D=1$ two regions generically meet at a point, while for $D>1$ they meet on a $D-1$ dimensional plane.   Thus the bipartite entanglement entropy scales as $(k_F L)^{D-1}$, which for $D>1$ depends on the non-universal parameter $k_F$.    However, as indicated in Fig. \ref{Fig1}, for $D=2$ {\it three} regions generically meet at a point, and for $D=3$ {\it four} regions meet at a point.   This motivates us to consider the multipartite entanglement between $D+1$ regions that meet at a point.   Multipartite entanglement measures have been the subject of increasing current interest \cite{Walter2016, Rota2016, Bayat2017, Pezze2017, Wang2018, Shirley2019, Zou2021,Liu2022}.   Here we introduce the {\it topological mutual information} characterizing  $D+1$ regions that meet at a point and show that it exhibits a logarithmic divergence that is proportional to $\chi_F$.

The mutual information ${\mathcal I}_{D+1}$ characterizing $D+1$ regions is designed so that the pairwise (or higher) correlations between different regions are subtracted off, leaving only the intrinsic correlations between all $D+1$ regions.   For $D=1$, the mutual information characterizing two regions $A$ and $B$ is
\begin{equation}
{\mathcal I}_2 = S_A + S_B - S_{AB},
\label{i2 def}
\end{equation}
where $S_A$ is the bipartite von Neumann entanglement entropy associated with subregion $A$ \cite{Casini2009_MI, Swingle2010MI, Swingle2012MI, Casini2015}.   Since in this paper we consider ground state properties, and the entire system $AB$ is in a pure state, we have $S_{AB}=0$ and $S_A = S_B$.   Thus ${\mathcal I}_2$ is the same (up to a factor of two definition) as the bipartite entanglement entropy.   Using the fact that for free fermions $c= \chi_F$, the Calabrese-Cardy formula for the bipartite entanglement entropy can then be expressed as \cite{Calabrese2004,*Calabrese2009}
\begin{equation}
{\mathcal I}_2 
= \frac{\chi_F}{3} \log \Lambda,
\label{i2 cardy}
\end{equation}
where $\Lambda \sim k_F L$.   

To generalize this to higher dimensions we first consider $D=3$, where for four regions the mutual information is defined as
\begin{align}
{\mathcal I}_4 = &S_A + S_B + S_C + S_D - S_{AB} - S_{AC} - S_{AD} \nonumber\\
&- S_{BC} - S_{BD} - S_{CD} + S_{ABC} + S_{ABD} \nonumber\\
&+ S_{ACD} + S_{BCD} - S_{ABCD}.
\label{i4 def}
\end{align}
We will show that the topological mutual information ${\mathcal I}_4$ 
characterizing four regions that meet at a point
exhibits a universal logarithmic divergence of the form
\begin{equation}
{\mathcal I}_4 = \frac{\chi_F}{5\pi^2} \log^3 \Lambda.
\end{equation}

The situation for $D=2$ is different.   The natural extension of (\ref{i2 def}) and (\ref{i4 def}) for three regions $A$, $B$ and $C$ that meet at a point in 2D does not work.  For a pure state $S_{ABC}=0$ and $S_{AB} = S_C$, so it follows that 
\begin{equation}
 S_A + S_B + S_C - S_{AB} - S_{AC} - S_{BC} + S_{ABC} = 0.
\label{i3 def}
\end{equation}
Perhaps this should not come as a surprise because for $D=1$ and $D=3$ $\chi_F$ can equally well be regarded as a property of the {\it Fermi surface}.   For odd $D$ the Euler characteristic $\chi_{\partial F}$ of the $D-1$ dimensional Fermi {\it surface} is simply related: $\chi_{\partial F} = 2\chi_F$.    For even $D$, however, the Euler characteristic of any closed $D-1$ dimensional surface is zero.   For instance, in 2D, all Fermi surfaces are topologically the same as circles, with $\chi_{\partial F}= 1-1 = 0$.   Thus, in even dimensions we have {\it neither} a topological entanglement measure analogous to (\ref{i2 def},\ref{i4 def}) nor a topological invariant characterizing the Fermi surface.

Nonetheless, in even dimensions the Euler characteristic of the {\it Fermi sea} can be non-zero, and it distinguishes different topological classes.   For example in two dimensions, the Euler characteristic counts the difference between the number of electron-like and hole-like Fermi surfaces, with open Fermi surfaces contributing zero.   It is able to distinguish a 2D circular Fermi sea ($\chi_F=1-0+0=1$) from a quasi 1D Fermi sea that would arise for a 2D array of decoupled 1D wires ($\chi_F=1-1+0=0$), which has only 1D entanglement.  We therefore seek an entanglement measure that probes $\chi_F$ in two dimensions.

An important property of $\chi_F$ in even dimensions is that it is odd under a particle-hole transformation, which exchanges the inside and outside of the Fermi surface.  Clearly, the bipartite entanglement entropy is even under the exchange of particles and holes.  This motivates us to consider a new entanglement measure that is odd under particle-hole symmetry.   We will introduce a ``charge-weighted" bipartite entanglement entropy, $S^Q_A$, which weights $S_A$ according to the charge $Q_A - \langle Q_A \rangle$.   By construction, this quantity is odd under a particle-hole transformation.   This allows us to define a
{\it charge-weighted topological mutual information} in even dimensions.   Specifically, for $D=2$ we define
\begin{equation}
{\mathcal I}^Q_3 = S^Q_A + S^Q_B + S^Q_C - S^Q_{AB} - S^Q_{AC} - S^Q_{BC} + S^Q_{ABC}.
\label{i3q def}
\end{equation}
We will see that $S^Q_{AB} = - S^Q_C$, so that unlike in (\ref{i3 def}), the terms do not cancel.  
We will show that ${\mathcal I}^Q_2$ exhibits a universal logarithmic divergence of the form
\begin{equation}
{\mathcal I}^Q_3 = \frac{3 \chi_F}{4\pi^2} \log^2 \Lambda.
\end{equation}

To compute the topological mutual information, we will employ the replica method used by Calabrese and Cardy \cite{Calabrese2004,*Calabrese2009}.   This leads naturally to a formulation in terms of the equal-time correlations between the numbers of particles in the different regions:  ${\mathcal I}_2 \propto \langle Q_A Q_B \rangle_c$, ${\mathcal I}_3^Q \propto \langle Q_A Q_B Q_C \rangle_c$, and ${\mathcal I}_4 \propto \langle Q_A Q_B Q_C Q_D\rangle_c$.  (Here the subscript $c$ indicates the connected correlation function).   The connection between the bipartite entanglement entropy for free fermions and number correlations (including higher-order cumulants) has been noted and extensively studied in Refs. \cite{Klich2009, Song2011, Song2012}.   Our analysis shows that these correlation functions exhibit a universal logarithmic divergence as a function of system size, with a coefficient proportional to $\chi_F$.   We will show that this is related to a universal (and to our knowledge unexplored) feature of the long wavelength density correlations of an infinite free Fermi gas.    Specifically, we will show that in $D$ dimensions, the $D+1$'th order density correlation function in momentum space exhibits a universal behavior for small ${\bf q}$,
\begin{align}
s_{D+1}({\bf q}_1,...,{\bf q}_D) &\equiv \int \frac{d^D{\bf q}_{D+1}}{(2\pi)^D}
\langle \rho_{{\bf q}_1} \rho_{{\bf q}_2}....\rho_{{\bf q}_{D+1}}\rangle_c \nonumber\\
&= \frac{\chi_F}{(2\pi)^D} |\det {\mathbb Q}|,
\label{sd formula0}
\end{align}
where ${\mathbb Q}$ is the $D\times D$ matrix built out of the column vectors ${\bf q}_1$, ..., ${\bf q}_D$.   Note that due to translation symmetry the expectation value is proportional to $\delta({\bf q}_1 + ... + {\bf q}_{D+1})$, and (\ref{sd formula0}) is the same when regarded as a function of any $D$ of the vectors ${\bf q}_1$, ..., ${\bf q}_{D+1}$.   This result is insensitive to continuous changes in the shape of the Fermi sea.  The long wavelength density correlations contain {\it topological} information about the Fermi sea.

In order to confirm the predictions of our analytic replica theory analysis it is desirable to develop an independent numerical method for computing the topological mutual information. While numerics is computationally challenging for $D=3$ we find that the efficient numerical methods for computing the bipartite entanglement entropy for free fermions developed in Refs. \cite{Peschel2001,Peschel2003,Henley2004,Peschel2009}  can be adapted to the computation of the charge-weighted mutual information in two dimensions.   This will be discussed in detail in Section \ref{sec IV}, where we demonstrate consistency between the numerics and several aspects of the analytic predictions.    

Finally, it is important to address the stability of our results in the presence of electron-electron interactions.   In one dimension, it is known that short-ranged electron-electron interactions do {\it not} affect the bipartite entanglement entropy.  The central charge $c$ retains its integer quantization in a Luttinger liquid, despite the fact that the log divergence of the number correlations $\langle Q_A Q_B \rangle_c$ loses its quantization \cite{Giamarchi2004, Song2012}.   In Section \ref{sec V} we will study the effects of electron-electron interactions perturbatively within the replica theory.   We will explain how the replica theory resolves that discrepancy, and we will show that in $D=3$ the situation is similar.   We predict that in a three dimensional Fermi liquid the $\log^3 \Lambda$ divergence of ${\mathcal I}_4$ remains quantized to $\chi_F$ in the presence of short-ranged interactions, despite the fact that the Fermi liquid parameters modify the divergence of $\langle Q_A Q_B Q_C Q_D \rangle_c$.   Thus, the divergence of the topological mutual information provides a sharp characterization that distinguishes distinct {\it topological Fermi liquid} phases, even in the presence of interactions.   In contrast, we will show that the charge-weighted mutual information defined for $D=2$ is {\it not} robust in the presence of interactions. Short-ranged interactions will modify the coefficient of the $\log^2 \Lambda$ divergence of ${\mathcal I}^Q_2$.

The paper is organized as follows.   In Section \ref{sec II} we review the replica method and apply it to the topological mutual information in $D=1$ and $D=3$.   We also introduce the charge-weighted entanglement entropy, along with the charge-weighted mutual information and show how they are computed in the replica theory.   These calculations relate ${\mathcal I}_{2,4}$ and ${\mathcal I}^Q_3$ to the number correlations, which are in turn related to the universal long wavelength density correlator, $s_{D+1}$, defined in (\ref{sd formula0}).   In Section \ref{sec III}, we discuss $s_3$ and $s_4$ in detail and establish (\ref{sd formula0}).   Alternative derivations of these results are presented in Appendix \ref{sec A} and {\ref{geoproof}}.   In addition, some lengthy parts of the calculation, including the Fourier transform of (\ref{sd formula0}), as well as a subsequent real space integration are described in Appendices \ref{sec B} and \ref{sec C}.   In Section \ref{sec IV} we describe the numerical computation of ${\mathcal I}_3^Q$.   We begin with a discussion of the computational method, and then demonstrate consistency with the predicted quantization of $\log^2 \Lambda$ divergence of  ${\mathcal I}_3^Q$, as well as its dependence on the topology of the Fermi sea and the structure of the real space partitions.   In Section \ref{sec V} we address the effect of electronic interactions, and show that within the replica theory the topological mutual information remains quantized for a three dimensional Fermi liquid.
Finally, in Section \ref{sec VI}, we close with a discussion of further questions.

\section{Entanglement and Number Correlations}
\label{sec II}

\subsection{Replica Theory for Bipartite Entanglement Entropy}
\label{sec IIA}

In this section we review the replica method for computing the entanglement entropy and show how for free fermions it relates to the number correlations.
Consider a system of free fermions that is partitioned into two regions $A$ and $B$.  The reduced density matrix is
\begin{equation}
\rho_A = {\rm Tr}_B [ \rho_{AB} ],
\end{equation}
where $\rho_{AB} = |\psi_{AB}\rangle\langle\psi_{AB}|$ is the density operator for the pure state $|\psi_{AB}\rangle$.   The von Neumann entanglement entropy for subsystem A is then
\begin{equation}
S_A = - {\rm Tr}_A[\rho_A \log \rho_A].
\label{sa}
\end{equation}
Following Calabrese and Cardy\cite{Calabrese2004}, we evaluate $S_A$ using the replica trick.   We introduce
\begin{equation}
S_{A,n} = {\rm Tr}_A[\rho_A^n].
\end{equation}
Noting that ${\rm Tr}_A[\rho_A]=1$, the von Neumann entropy is recovered by analytically continuing as a function of $n$:
\begin{equation}
S_A = - \left. \frac{\partial S_{A,n}}{\partial n} \right|_{n=1}.
\label{replica limit}
\end{equation}
$S_{A,n}$ is related to, but defined slightly differently from the R\'enyi entropy, which is given by 
$(\log S_{A,n})/(1-n)$.

\begin{figure}
\includegraphics[width=0.6\columnwidth]{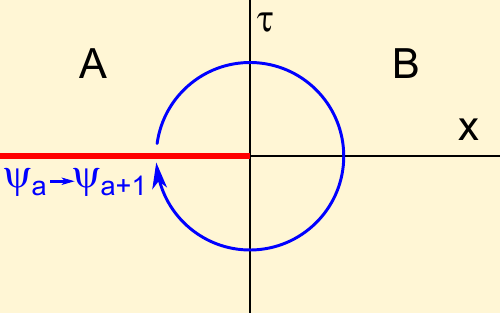}
\caption{Replica method for calculating the bipartite entanglement entropy, illustrated for spatial dimension $D=1$. To calculate $\Tr_A[\rho^n_A]$ in the Euclidean path integral formalism, $n$ replicas of the original system are introduced. This quantity can then be calculated as the partition function defined on an $n$-sheeted Riemann surface, constructed as follows: for region $A$, each replica (with label $a$) at time $\tau=-\epsilon$ is glued to the next replica (with label $a+1$ mod $n$) at time $\tau=+\epsilon$; for region $B$, each replica is glued back to itself.}
\label{Fig2}
\end{figure}

The key insight of Calabrese and Cardy is that for integer $n$ we can consider the partition function describing $n$ replicas of the original system as a Euclidean space-time path integral.   Then, the  trace ${\rm Tr}_A[({\rm Tr}_B \rho_{AB})^n]$ can be interpreted as describing a system in which the replicas are joined together such that inside region A replica $a$ at time $\tau = -\varepsilon$ is connected to replica $a+1 \mod n$ at time $\tau = +\varepsilon$.  In $D=1$ spatial dimension the point-like boundary between $A$ and $B$ then resembles a screw dislocation in a three dimensional space-time-replica index space as indicated in Fig. \ref{Fig2}.  For $D>1$ the $D-1$ dimensional boundary between $A$ and $B$ behaves similarly.   The effect of the partial trace is then to introduce a {\it twist operator} $T_{A,n}$ into the partition function at time $\tau=0$, whose action on the fermion operators $\psi_a({\bf r})$ is given by
\begin{equation}
T_{A,n}^\dagger \psi_a({\bf r}) T_{A,n} = \left\{\begin{array} {ll}
\psi_a({\bf r}) & {\bf r} \in B \\
\psi_{a+1}({\bf r}) & {\bf r}\in A, \  a < n \\
(-1)^{n+1}\psi_1({\bf r}) & {\bf r}\in A, \ a=n.
 \end{array}\right.
\end{equation}
Here the sign $(-1)^{n+1}$ reflects the anticommutation of fermion operators \cite{Casini2005, Larsen1995}, which in a path integral requires an antiperiodic temporal boundary condition.   This is accounted for by taking $\psi_a \rightarrow (-1)^a \psi_a$ in the product ${\rm Tr}[ \prod_{a=1}^n \rho_A]$, allowing an interpretation in terms of an $n$ component fermion field $\psi_a$ that satisfy $\psi_a(\tau=\beta) = -\psi_a(\tau=0)$.
We then have
\begin{equation}
S_{A,n} =  \langle T_{A,n} \rangle.
\label{renyi twist}
\end{equation}

This can be simplified by doing a Fourier transform of the fermion operators in replica space, which diagonalizes the twist operator.   We  introduce
\begin{equation}
\tilde\psi_p({\bf r}) = \frac{1}{\sqrt{n}}\sum_{a=1}^{n} e^{-2\pi i p a/n} \psi_a({\bf r}),
\end{equation}
where the ``replica momentum" $p$ is an integer (half-integer) modulo $n$ when $n$ is odd (even).   The action of the twist operator is then, 
\begin{equation}
T_{A,n}^\dagger \tilde\psi_p({\bf r}) T_{A,n} = \tilde\psi_p({\bf r})  
\left\{\begin{array}{ll} e^{\frac{2\pi i p}{n}} & {\bf r}\in A \\
1 & {\bf r}\in B\end{array}\right..
\end{equation}
We thus conclude that the twist operator has the form
\begin{equation}
T_{A,n} = \prod_{p=-(n-1)/2}^{(n-1)/2} e^{ \frac{2\pi i p}{n} Q_{A,p}}
\label{twist}
\end{equation}
where
\begin{equation}
Q_{A,p} = \int_{{\bf r}\in A} d^D{\bf r} \tilde\psi^\dagger_p({\bf r})\tilde\psi_p({\bf r})
\end{equation}
is the total charge in region $A$ in the $p$'th replica momentum channel.

\begin{figure}
\includegraphics[width=2 in]{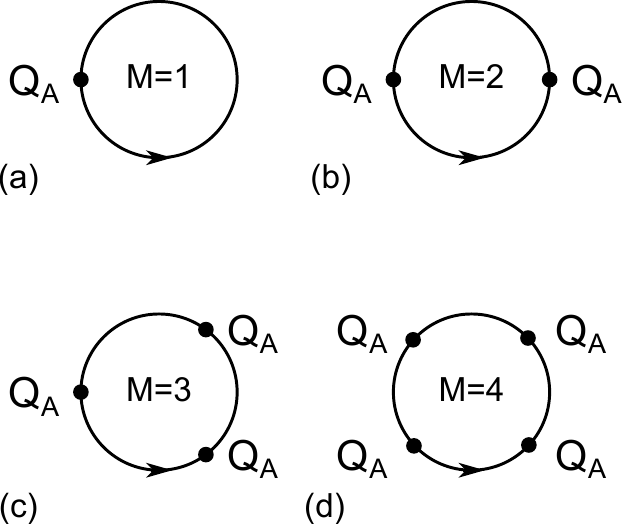}
\caption{Feynman diagrams for the connected correlation functions $\langle Q_A^M \rangle_c$ generated by the cumulant expansion (\ref{cumulant sum}).  Solid lines are free fermion propagators, and small circles are $Q_A$-vertices.}
\label{Fig3}
\end{figure}

Now, for free fermions, two key simplifications occur.   The first is that the Hamiltonian decouples into $n$ independent and identical copies in each replica-momentum channel.   It follows from (\ref{renyi twist}) and (\ref{twist}) that
\begin{equation}
S_{A,n} = \prod_p \langle e^{\frac{2\pi i p}{n} Q_{A,p}}\rangle,
\label{sanqp}
\end{equation}
where each of the expectation values in the product is evaluated with respect to the same Hamiltonian, and can be computed in the unreplicated theory.    The second simplification is that the expectation value of the exponent of $Q_{A,p}$ can be expanded in a cumulant expansion,
\begin{equation}
\langle T_{A,n} \rangle = \exp \left[\sum_p \sum_{M=1}^\infty \frac{1}{M!}\left(\frac{2\pi i p}{n}\right)^M \langle Q_{A,p}^M \rangle_c \right],
\label{cumulant sum}
\end{equation}
where the expectation value in the exponent includes only the connected terms.  Since $Q_{A,p}$ is quadratic in the fermion operators, the connected terms involve evaluating a Feynman diagram that consists of a single fermion loop with $M$ $Q_{A,p}$-vertices, see Fig. \ref{Fig3}.

Since $\langle Q_{A,p}^M \rangle_c$ is independent of $p$ we can drop the subscript $p$ and evaluate the sum on $p$ over integers (half-integers) modulo $n$ when $n$ is odd (even).  To do this it is necessary to choose a range for $p$.   Since $Q_{A,p}$ has integer eigenvalues it follows that $e^{2\pi i Q_{A,p}} = 1$. Thus, it is clear that changing the range (for example from $-(n-1)/2\le p \le (n-1)/2$ to $-(n-3)/2 \le p \le (n+1)/2$) does not affect (\ref{sanqp}).   However, changing the range {\it does} affect the coefficients of the $\langle Q_{A,p}^M \rangle_c$ terms in the cumulant expansion (\ref{cumulant sum}).   This discrepancy can be resolved by noting that the cumulant expansion of $\langle e^{2\pi i Q_{A,p}}\rangle = 1$ implies a non-trivial identity obeyed by the terms in the cumulant expansion.   In the following, we will choose the range of $p$ to respect the symmetry under $p \rightarrow -p$.  

The sum over replica-momentum channels can now be evaluated by noting that 
\begin{equation}
C_{n,M} \equiv \sum_{p=-\frac{n-1}{2}}^{\frac{n-1}{2}} p^M = (1 + (-1)^M) H_{\frac{n-1}{2},-M},
\label{harmonic number}
\end{equation}
where the generalized harmonic number is $H_{m,l} = \sum_{p=1}^m p^{-l}$.   Thus, $C_{n,M}=0$ for odd $M$, and the first few even terms are,
\begin{align}
C_{n,2} &= \frac{n(n^2-1)}{12},\\
C_{n,4} &= \frac{n(n^2-1)(3n^2-7)}{240}.
\end{align}
$C_{n,M}$ can be analytically continued as a function of $n$, which allows us to evaluate the $n \rightarrow 1$ limit required for (\ref{replica limit}).   For $n \rightarrow 1$ it is found that for arbitrary positive even integers $M$,
\begin{equation}
C_{n,M} = - (n-1) M \zeta_{1-M} + O(n-1)^2,
\label{h to zeta}
\end{equation}
where $\zeta_{1-M}$ is the Riemann zeta function.  Using the identity $\zeta_{1-M} = 2 (2\pi)^{-M} \cos (\pi M/2) (M-1)! \zeta_M$ we conclude that to order $(n-1)$,
\begin{equation}
\sum_{p=-\frac{n-1}{2}}^{\frac{n-1}{2}} \frac{1}{M!} \left(\frac{2\pi i p}{n}\right)^M = \left\{\begin{array}{ll}
0 & M \ {\rm odd} \\
-2 (n-1) \zeta_M & M \ {\rm even},
\end{array}\right.
\label{harmonic identity}
\end{equation}
with $\zeta_2 = \pi^2/6$ and $\zeta_4 = \pi^4/90$.

Combining Eqs. \ref{replica limit}, \ref{renyi twist}, \ref{cumulant sum} and \ref{harmonic identity}, we can now express the von Neumann entropy as a sum over the cumulants of the number correlations.  Writing $M=2l$ we obtain
\begin{equation}
S_A = \sum_{l=1}^\infty 2 \zeta_{2l} \langle Q_A^{2l} \rangle_c.
\label{s cumulant}
\end{equation}
This result has been derived previously using a different method in connection with the theory of full counting statistics and its relation to the entanglement entropy \cite{Klich2009, Song2011, Song2012}.   The derivation presented here, based on the replica method, has the advantage that it is straightforward to examine the effects of electron-electron interactions, which will be considered in Section \ref{sec V}.

We will now argue that terms in the cumulant expansion (\ref{s cumulant}) are arranged in decreasing order of divergence in the system size $L$.   Therefore, to extract the leading divergence, it is only necessary to consider the first non-zero term in the expansion.   This can be seen using a simple power counting argument.

Consider the $M=2l$'th term in the expansion.   This can be determined by integrating the $M$'th order correlation function of the density $\rho({\bf r})$ over the $D M$ coordinates ${\bf r}_{i=1,...,M}$.   The order of divergence in the system size $L$ will depend on how fast this correlator goes to zero when $|{\bf r}_i - {\bf r}_j|$ is large.   In Section \ref{sec III} and Appendix \ref{sec A}, we study the density correlation function in detail, and show that the momentum space correlator
\begin{equation}
s_{M}({\bf q}_1,...,{\bf q}_{M-1}) = \int \frac{d^D{\bf q}_M}{(2\pi)^D} 
\langle \rho({\bf q}_1) ... \rho({\bf q}_M) \rangle_c 
\end{equation}
vanishes in the small $q$ limit like
\begin{equation}
s_{M}({\bf q}_1, ...{\bf q}_{M-1})    \sim k_F^{D+1-M} q^{M-1},
\end{equation}
where $k_F^{-1}$ is a short distance length scale set by the dimensions of the Fermi surface, and
 we note that due to translation symmetry ${\bf q}_M = -{\bf q}_1 - ... - {\bf q}_{M-1}$. A similar scaling for 3-point functions of the density has been discussed in Ref. \onlinecite{Son2022}.  Fourier transforming over the $M-1$ independent variables leads to a real space correlator that depends only on differences ${\bf r}_i - {\bf r}_j$, and for large $r$ scales as
\begin{align}
\langle \rho({\bf r}_1) ... \rho({\bf r}_M) \rangle_c \sim &k_F^{D+1-M} \int  q ^{M-1} e^{i q r} d^{(M-1)D}q  \\ & \sim 
\frac{k_F^{D+1-M}}{r^{(D+1)(M-1)}}.
\end{align}
Integrating this over ${\bf r}_i$ then gives
\begin{equation}
\langle Q^M \rangle_c \sim k_F^{D+1-M} \int \frac{d^{D M} r }{ r^{(D+1)(M-1)}} \sim (k_F L)^{D+1-M}.
\end{equation}
This shows that for $M > D+1$ the $M$'th term in the cumulant expansion converges for large $L$, while for $M < D+1$ it diverges as $(k_F L)^{D+1-M}$.   For example, for $M=2$, we expect an area-law (for $D=3$) contribution $(k_F L)^{D-1}$, which is proportional to the area of the Fermi surface \cite{Klich2006}.

The term with $M = D+1$ is marginal, and we shall see that it gives rise to universal logarithmic divergences.   In the following we will show that for $D=1$ this recovers the Calabrese-Cardy formula.  For $D=3$ we will introduce an entanglement measure that subtracts off the leading divergent $M=2$ term, leaving the universal $M=4$ term.   For $D=2$, the above power counting argument suggests that there is a universal divergence for $M=3$.   However, the $M=3$ term is not present in (\ref{s cumulant}).  In Section \ref{sec IID} we will explain why this is, and we will introduce a modified entanglement measure that probes the $M=3$ term.

\subsection{Topological mutual information in one dimension}
\label{sec IIB}

As a warmup, in this section we will use (\ref{s cumulant}) to recover the result of Calabrese and Cardy \cite{Calabrese2004}.   While this result is well known, our derivation will set the stage for our later results.  We consider a one dimensional system of free fermions defined on a line segment, and we partition the system into two subregions $A$ and $B$ that meet at a single point.   The topological mutual information, defined in (\ref{i2 def}) is simply related to the bipartite entanglement entropy, ${\mathcal I}_2 = 2 S_A$.
To evaluate this, we consider the first term in the cumulant expansion, with $M=2$.   We will argue that this term captures the leading logarithmic divergence of ${\mathcal I}_2$.  Using the facts that the total charge $Q = Q_A + Q_B$ is conserved and therefore drops out of the connected correlation function and that $[Q_A,Q_B]=0$, we can write
\begin{equation}
{\mathcal I}_2 = -4 \zeta_2 \langle Q_A Q_B \rangle_c.
\label{S1AB}
\end{equation}
This can be determined by first evaluating the equal-time density-density correlation function in momentum space.  Defining
\begin{equation}
\rho_q = \int dx e^{-i q x} \rho(x) = \int \frac{dk}{2\pi} c^\dagger_k c_{k+q},
\end{equation}
where $\rho(x)=\psi^\dagger(x)\psi(x)$ and $c_k$ is the momentum space fermion operator we may write this as
\begin{align}
s_2(q) &= \int \frac{dq'}{2\pi} \langle \rho(q) \rho(q') \rangle_c = \int \frac{dk}{2\pi} (1-f_{k+q}) f_{k},
\end{align}
where $f_k = \langle c_k^\dagger c_k\rangle = \theta(E_F-E_k)$ is the Fermi occupation factor.
This is simply the length in momentum space that is inside the Fermi sea, but outside the Fermi sea when shifted by $-q$.   For sufficiently small $q$ it is simply
\begin{equation}
s_2(q) = \frac{\chi_F}{2\pi} |q|.
\label{sqd=1}
\end{equation}
Here $\chi_F$ is the Euler characteristic of the Fermi sea, which in 1D simply counts the number of components of the Fermi sea (or half the number of Fermi points).   This is not a Taylor expansion for small $q$.  It is exact for $q$ smaller than a fixed finite value that is determined by the smallest wavevector spanning the Fermi sea.   In the following sections we will see that this behavior has a generalization in higher dimensions. 

Fourier transforming, the $\abs{q}$ singularity in $s_2(q)$ determines the universal long distance limit of the equal-time correlations in real space, $s_2(x_A,x_B) = \langle \rho(x_A) \rho(x_B) \rangle_c$, given by
\begin{equation}
s_2(x_A,x_B) = \int \frac{dq}{2\pi} s(q) e^{i q(x_A-x_B)} = 
-\frac{\chi_F}{2\pi^2}\frac{1}{(x_A-x_B)^2}.
\end{equation}
Integrating $x_A$ over region $A$ ($-L<x_A<0$) and $x_B$ over region $B$ ($0<x_B<L$), gives the leading logarithmic divergence in the equal-time number correlation,
\begin{equation}
\langle Q_A Q_B \rangle_c = - \frac{\chi_F}{2\pi^2}\log \Lambda,
\label{qaqb log}
\end{equation}
where $\Lambda = k_F L$, and $k_F^{-1}$ is a non-universal short distance cutoff that depends on the size of the Fermi sea.   Using (\ref{S1AB}) along with $\zeta_2 = \pi^2/6$, this gives the Calabrese Cardy result
\begin{equation}
{\mathcal I}_2 = \frac{\chi_F}{3} \log \Lambda,
\end{equation}
where for free fermions, the central charge in conformal field theory is given by $c=\chi_F$.
Note that this calculation is for a geometry in which regions $A$ and $B$ meet at a single point $x=0$.   If instead region $A$ is surrounded by $B$, so that $A$ and $B$ meet at two points, then the result is doubled.  More generally, one could consider partitions in which region $A$ consists of multiple disconnected regions.   Then ${\mathcal I}_2$ counts the number of contact points between $A$ and $B$.
${\mathcal I}_2$ deserves the name ``topological mutual information" because it depends only on the topology of the Fermi sea, as well as the topology of the real space partition of $A$ and $B$.

\subsection{Topological mutual information in three dimensions}
\label{sec IIC}

As noted in the introduction, for $D>1$ the bipartite entanglement entropy exhibits a logarithmic divergence with area law coefficient that depends on the dimensions of the Fermi surface, and can be expressed in terms of the Widom formula.   Unlike the $D=1$ result, this coefficient is not quantized and does not reflect topological information about the state.   Here we show that the mutual information characterizing four regions that meet at a single point is quantized and reflects the topology of the filled Fermi sea.

We therefore consider free fermions with a momentum space dispersion $E_{\bf k}$ defined on a three dimensional region of size $L^3$ with open boundary conditions.   We partition the region into four sub regions $A$, $B$, $C$ and $D$ that meet at a single point, as shown in Fig. \ref{Fig1}(c), and consider the mutual information ${\mathcal I}_4$ defined in (\ref{i4 def}).  Since the entire system, $ABCD$ is in a pure state, we have $S_{ABCD}=0$ and $S_{ABC}=S_D$ and $S_{AB}=S_{CD}$.   Therefore, (\ref{i4 def}) simplifies to
\begin{equation}
{\mathcal I}_4 = 2(S_A + S_B + S_C + S_D - S_{AB} - S_{AC} - S_{AD}).
\label{mutual info}
\end{equation}
Note that 
this combination of entropies has a structure similar to the mutual information characterizing three regions $A$, $B$ and $C$ (along with their complement), which was introduced in Ref. \onlinecite{Kitaev2006} to isolate the topological entanglement entropy.  ${\mathcal I}_4$ characterizes the entanglement correlations that involve all four regions, and is insensitive to the local area-law contributions between pairs of regions.

We now use (\ref{s cumulant}) to evaluate ${\mathcal I}_4$.   It is straightforward to see that the first term in (\ref{s cumulant}), which involves $\langle Q^2 \rangle_c$ cancels.   We will argue that the leading divergence is dominated by the next term, involving $\langle Q^4 \rangle_c$,   
\begin{align}
{\mathcal I}_4 = &4 \zeta_4 \Bigl\langle Q_A^4 + Q_B^4 + Q_C^4 + Q_D^4 - (Q_A+Q_B)^4 \nonumber \\
&- (Q_A + Q_C)^4 -(Q_A + Q_D)^4 \Bigr\rangle_c.
\end{align}
Noting that $Q_A + Q_B + Q_C + Q_D$ is a constant and that the $Q$'s commute, this reduces to
\begin{equation}
{\mathcal I}_4 =  -48 \zeta_4 \langle Q_A Q_B Q_C Q_D \rangle_c.
\label{qaqbqcqd}
\end{equation}

Evaluating (\ref{qaqbqcqd}) is a bit more involved than it was for the 1D case, Eq. \ref{S1AB}, but the strategy is exactly the same.   We first compute the equal-time fourth order connected density correlation function in momentum space, which reveals a universal and quantized small $q$ singularity.   We then Fourier transform to get the fourth order equal-time density correlations in real space, followed by integrating the four positions over the four regions $A$, $B$, $C$ and $D$ (see Fig. \ref{Fig1}(c)).   Since the calculation is rather long, here we will summarize the results.   The momentum space density correlation function will be evaluated in Section \ref{sec III} and Appendix \ref{sec A}, while the Fourier transform and real space integrals will be described in Appendices \ref{sec B} and \ref{sec C}.

The momentum space fourth order equal-time density correlations, given by
\begin{equation}
s_4({\bf q}_1,{\bf q}_2,{\bf q}_3) = 
\int\frac{d^3{\bf q}_4}{(2\pi)^3} \langle \rho({\bf q}_1) \rho({\bf q}_2) \rho({\bf q}_3) \rho({\bf q}_4) \rangle_c
\label{s 3D}
\end{equation}
are the subject of Section \ref{sec IIIB}.   In (\ref{s 3D}) it is understood that due to translation invariance the constraint ${\bf q}_4 = -{\bf q}_1 - {\bf q}_2 - {\bf q}_3$ is enforced by a $\delta$ function, so only three of $({\bf q}_1,{\bf q}_2,{\bf q}_3,{\bf q}_4)$ are independent.  
We will show that $s_4$ has a universal small ${\bf q}$ behavior given by
\begin{equation}
s_4({\bf q}_1,{\bf q}_2,{\bf q}_3) = \frac{\chi_F}{(2\pi)^3} |{\bf q}_1 \cdot ({\bf q}_2 \times {\bf q}_3)|.
\label{sqd=3}
\end{equation}
Note that $s_3$ could equally well be described in terms of any three of the four ${\bf q}$'s, and that the triple product is the same in each case.   It describes the momentum space volume of the tetrahedron formed by the four ${\bf q}$'s.   The analytic derivation of (\ref{sqd=3}) presented in Section \ref{sec III} is valid in the ${\bf q} \rightarrow 0$ limit, and the leading divergence of ${\mathcal I}_4$ only depends on that limit.   However, we have checked by numerically evaluating the integrals that like (\ref{sqd=1}) this formula is exact for $|{\bf q}|$ smaller than a finite cutoff, which depends in a complicated way on the shape of the Fermi surface.

The next step is to evaluate the Fourier transform of (\ref{sqd=3}) to determine the fourth order density correlations in real space.   This calculation is described in Appendix \ref{sec B}, where we show that
\begin{equation}
s_4({\bf r}_A,{\bf r}_B,{\bf r}_C,{\bf r}_D) = \langle \rho({\bf r}_A) \rho({\bf r}_B) \rho({\bf r}_D) \rho({\bf r}_D) \rangle_c
\end{equation} 
can be written as,
\begin{equation}
s_4(\{{\bf r}_K\}) = \frac{2 \chi_F}{(2\pi)^6} \nabla_A \cdot ( \nabla_B \times  \nabla_C) \nabla_D \cdot {\bf F}(\{{\bf r}_K\}),
\end{equation}
where
\begin{equation}
{\bf F}(\{{\bf r}_K\}) = -({\bf r}_{BA} \times \nabla_C) \delta^2_\perp({\bf r}_{CA}\times {\bf r}_{BA}) \delta^2_\perp({\bf r}_{DA} \times {\bf r}_{BA}).
\end{equation}
Here we write ${\bf r}_{KK'} = {\bf r}_K - {\bf r}_{K'}$, and the two dimensional $\delta$-functions are evaluated in the plane perpendicular to ${\bf r}_{BA}$.  The $\delta$-functions require 
${\bf r}_{BA} \parallel {\bf r}_{CA} \parallel {\bf r}_{DA}$, so the connected fourth order correlations arise only when ${\bf r}_{A,B,C,D}$ are oriented along a straight line. 

We next need to integrate ${\bf r}_{A,B,C,D}$ over the four regions $A$, $B$, $C$, $D$ (see Fig. \ref{Fig1}(c)).   The calculation involves summing over sets of straight lines that visit all four regions.  Due to the derivatives on all four variables, the integrals can be evaluated on the boundaries of each of the regions. We find that the leading logarithmic divergence can be viewed as arising from the lines close to the ``triple contact lines",  along which three regions meet. For example, ${\bf r}_A$ could be on the boundary of $A$, near the point where all four regions meet, while ${\bf r}_{B,C,D}$ are distributed on the boundaries of $B$, $C$ and $D$ close to the line where $B$, $C$ and $D$ meet.   The calculation is a bit involved, and is described in Appendix \ref{sec C}.   Here we quote the result:
\begin{align}
\langle Q_A Q_B Q_C Q_D\rangle_c &= \int_{{\bf r}_K \in K} d^{12}{\bf r}_K s(\{{\bf r}_K\})\\
& = 
- \frac{ 3\chi_F}{8 \pi^6} \log^3 \Lambda.
\end{align}
Then, using (\ref{qaqbqcqd}) with $\zeta_4 = \pi^4/90$ we obtain the leading logarithmic divergence of the quadripartite mutual information,
\begin{equation}\label{i4 final}
{\mathcal I}_4 = \frac{\chi_F}{5 \pi^2} \log^3\Lambda.
\end{equation}
This is our central result for three dimensions, which shows that the mutual information $\mathcal{I}_4$ is topological in that it encodes the topology of 3D Fermi sea. While derived above for the case of free fermions, this result is actually robust in the presence of Fermi-liquid interaction. This is explained in Sec. \ref{sec V}.

\subsection{A particle-hole odd entanglement measure in two dimensions}
\label{sec IID}

\subsubsection{Charge-weighted bipartite entanglement entropy}
\label{sec IID1}

In this section we consider the the case of two dimensions.   As discussed in the introduction, even and odd dimensions behave differently.   This difference comes into sharp focus when one considers the effect of a particle-hole transformation on a system of fermions defined on a lattice.   This transformation replaces the Fermi sea by its complement.   The Euler characteristics of the Fermi sea $F$ and its complement $\bar F$ are related by $\chi_{\bar F} = (-1)^{D+1} \chi_F$, so in even dimensions, $\chi_F$ is odd under a particle-hole transformation.   Of course, the particle-hole transformation has no effect on the Fermi {\it surface}.    In general, the Euler characteristic of the Fermi surface $\chi_{\partial F}$ and the Euler characteristic $\chi_F$ of its interior, the Fermi sea, are related by \cite{Dieck2008}
\begin{equation}
\chi_{\partial F} = \left\{\begin{array}{ll} 2\chi_F & D \ {\rm odd} \\ 0 & D \ {\rm even}.\end{array}\right.
\end{equation}

Since the particle-hole transformation is just a change of basis it has no effect on the entanglement entropy.   Therefore, in odd dimensions, $\chi_F$ and $\chi_{\partial F}$ contain the same information, so 
one can view the topological mutual information defined in (\ref{i2 def}) and (\ref{i4 def}) as a property of the Fermi {\it surface}.   
In even dimensions, the vanishing of $\chi_{\partial F}$ is consistent with the vanishing of the attempted definition of the mutual information in Eq. \ref{i3 def}.   However, $\chi_F$ can still be non zero, and contains non-trivial topological information about the Fermi sea, but it can only be reflected in an entanglement measure that is odd under a particle-hole transformation.

This motivates us to define the {\it charge-weighted entanglement entropy},
\begin{equation}
S^Q_A = -{\rm Tr}[ (Q_A - \langle Q_A \rangle) \rho_A \log \rho_A ].
\end{equation}
Writing the trace in a basis of eigenstates of $Q_A$, this can be interpreted as the entanglement entropy weighted by the charge fluctuation in region $A$.  Under a particle-hole transformation $Q_A \rightarrow N_A - Q_A$ (where for a system defined on a lattice $N_A$ is the number of sites in region $A$).  Thus it is clear that $S^Q_A \rightarrow - S^Q_A$, \textit{i.e.} it is a particle-hole odd entanglement measure. 

We will first show how to compute $S^Q_A$ using the replica method, and then we will show that given a tripartition consisting of regions $A$, $B$ and $C$ that meet at a point we can define a charge-weighted topological mutual information, ${\mathcal I}^Q_2$, that has a universal $\log^2\Lambda$ divergence and probes the Euler characteristic $\chi_F$ of the 2D Fermi sea.   

\subsubsection{Replica analysis}
\label{sec IID2}

We first apply the replica analysis of Section \ref{sec IIA} to $S^Q_A$.   To this end, we introduce the quantity
\begin{equation}
S^{Q,n}_A = \frac{{\rm Tr}[Q_A \rho_A^n]}{{\rm Tr}[\rho_A^n]}.
\end{equation}
Noting that ${\rm Tr}[Q_A \rho_A] = \langle Q_A \rangle$ and ${\rm Tr}[\rho_A] = 1$ it can be seen that
\begin{equation}\label{replica limit of SQ}
S^Q_A = - \left. \frac{\partial S^{Q,n}_A}{\partial n} \right|_{n=1}.
\end{equation}

To evaluate $S^{Q,n}_A$ we introduce replicas $a = 0, ..., n-1$, along with the twist operator $T_{A,n}$ given in (\ref{twist}).  Then
\begin{equation}
S^{Q,n}_A = \frac{\langle Q_{A,a=0} T^A_n \rangle}{\langle T^A_n \rangle}.
\end{equation}
In the replica-momentum basis, this then has the form,
\begin{equation}
S^{Q,n}_A = \frac{1}{n}\sum_p\frac{\langle  Q_{A,p} \prod_{p'} e^{\frac{2\pi i p'}{n}Q_{A,p'}} \rangle}
{\langle \prod_{p'} e^{\frac{2\pi i p'}{n}Q_{A,p'}} \rangle},
\end{equation}
where the sums and products over $p$ and $p'$ range from $-(n-1)/2$ to $(n-1)/2$.  Since the replica momentum channels decouple, the terms with $p'\ne p$ cancel, and for the remaining $p'=p$ term $Q_{A,p}$ can be generated by differentiating with respect to $p$,
\begin{equation}
S^{Q,n}_A = \frac{1}{n}\sum_p \frac{n}{2\pi i} \frac{\partial}{\partial p} \log \langle e^{\frac{2\pi i p}{n} Q_{A,p}} \rangle.
\end{equation}
Performing the cumulant expansion for the expectation value of the exponent then gives
\begin{equation}
S^{Q,n}_A = \frac{1}{n}\sum_p \frac{n}{2\pi i} \frac{\partial}{\partial p} 
\sum_{M = 1}^\infty \frac{1}{M!} \left(\frac{2\pi i p}{n}\right)^M \langle Q_{A,p}^M \rangle_c.
\end{equation}
Since the expectation value is independent of $p$ we can perform the differentiation with respect to $p$ and evaluate the sum on $p$ using (\ref{harmonic number}).   After shifting $M\rightarrow M+1$ we obtain,
\begin{equation}
S^{Q,n}_A = \frac{1}{n}\sum_{M=0}^\infty \frac{1}{M!} \left(\frac{2\pi i}{n}\right)^M C_{n,M} \langle Q_A^{M+1} \rangle_c.
\end{equation}
Finally, using the same analysis as Eqs. \ref{h to zeta}-\ref{s cumulant} and differentiating with respect to $n$ in the limit $n\rightarrow 1$ we obtain,
\begin{equation}
S^Q_A = \sum_{l=1}^\infty  2 \zeta_{2l} \langle Q_A^{2l+1} \rangle_c.
\end{equation}
Thus, the charge-weighted bipartite entanglement entropy has a similar structure to (\ref{s cumulant}), except it picks out the odd cumulants of the charge fluctuations rather than the even cumulants.  This implies $S^Q_{A} = -S^Q_{\bar{A}}$, since $Q_A+Q_{\bar{A}}$ is a constant.

\subsubsection{Charge-weighted topological mutual information}
\label{sec IID3}

Following the procedure in Section \ref{sec IIB}, we consider a two dimensional system of free fermions with dispersion $E({\bf k})$ defined on a finite system of size $L^2$ with open boundary conditions.  We partition the system into three regions $A$, $B$ and $C$ that meet at a point, see Fig. \ref{Fig1}(b), and define the charge-weighted topological mutual information as
\begin{equation}
{\mathcal I}^Q_3 = S^Q_A + S^Q_B + S^Q_C - S^Q_{AB} - S^Q_{BC} - S^Q_{CA} +S^Q_{ABC}.
\end{equation}
Since $ABC$ is the entire system with a conserved total charge, $S^Q_{ABC}=0$.  Moreover, $S^Q_{AB} = - S^Q_C$, so that unlike in (\ref{i3 def}), the above terms do not cancel.
We anticipate that ${\mathcal I}^Q_3$ is dominated by the $l=1$ term in the cumulant expansion,
\begin{align}
{\mathcal I}^Q_3 = 2 \zeta_2 & \Bigl\langle Q_A^3 +  Q_B^3 + Q_C^3
-  (Q_A + Q_B)^3\nonumber\\- & (Q_B+Q_C)^3 -  (Q_C+Q_A)^3\Bigr\rangle_c
.
\end{align}
Using the fact that the total charge $Q_A+Q_B+Q_C$ is constant and the $Q$'s commute we obtain,
\begin{equation}
{\mathcal I}^Q_3 = 12 \zeta_2 \langle Q_A Q_B Q_C \rangle_c.
\end{equation}

We evaluate this following the same procedure as Sections \ref{sec IIB} and \ref{sec IIC}.   We first consider the momentum space correlator, defined on an infinite plane,
\begin{equation}
s_3({\bf q}_1,{\bf q}_2) = \int\frac{d^2{\bf q}_3}{(2\pi)^2}\langle \rho({\bf q}_1) \rho({\bf q}_2) \rho({\bf q}_3) \rangle_c,
\end{equation}
where translation symmetry fixes ${\bf q}_3 = - {\bf q}_1 - {\bf q}_2$.  
In Section \ref{sec IIIA}, we establish that like the $D=1$ and $D=3$ cases, this has a universal small ${\bf q}$ behavior, which is exact for $|{\bf q}_{1,2}|$ smaller than a finite cutoff that depends on the shape of the Fermi surface,
\begin{equation}
s_3({\bf q}_1,{\bf q}_2) = \frac{\chi_F}{(2\pi)^2} |{\bf q}_1 \times {\bf q}_2|,
\end{equation}
where we use the 2D (scalar) cross product.

We next Fourier transform to obtain the real space density correlations,
\begin{equation}
s_3({\bf r}_A,{\bf r}_B,{\bf r}_C) = \langle \rho({\bf r}_A) \rho({\bf r}_B)\rho({\bf r}_C) \rangle_c.
\end{equation}
In Appendix \ref{sec B} we show that this has the form
\begin{equation}
s_3({\bf r}_A,{\bf r}_B,{\bf r}_C) = \frac{2\chi_F}{(2\pi)^4} (\nabla_A \times \nabla_B) (\nabla_C \times {\bf F}(\{{\bf r}_K\})
\end{equation}
with
\begin{equation}
{\bf F}({\bf r}_A,{\bf r}_B,{\bf r}_C) = \frac{{\bf r}_{BA}}{|{\bf r}_{BA}|^2} \delta({\bf r}_{BA} \times {\bf r}_{CA}).
\end{equation}
Thus, as in the 3D case, the density correlations are finite when ${\bf r}_{A,B,C}$ lie along a straight line.

We next integrate ${\bf r}_{A,B,C}$ over regions $A$, $B$ and $C$ (see Fig. \ref{Fig1}(b)).   This calculation is described in Appendix \ref{sec C}.   The leading logarithmic divergence comes from straight lines that begin at the triple contact point and straddle one of the three lines that separate two regions.   The result is
\begin{equation}\label{QQQ_c}
\langle Q_A Q_B Q_C \rangle_c = \int_{{\bf r}_K \in K} d^6{\bf r}_K s(\{{\bf r}_K\}) = 
\frac{3\chi_F}{8\pi^4} \log^2 \Lambda,
\end{equation}
where $\Lambda = k_F L$, and $k_F^{-1}$ is a non-universal short distance cutoff that depends on the dimensions of the Fermi surface.    Then, using $\zeta_2 = \pi^2/6$ we obtain,
\begin{equation}
{\mathcal I}^Q_3 =  \frac{3\chi_F}{4 \pi^2} \log^2 \Lambda.
\label{iq3 final}
\end{equation}
This is the central result of this section.  The charge-weighted topological mutual information exhibits a universal $\log^2 \Lambda$ divergence that probes the topology of the two dimensional Fermi sea.  In Sec. \ref{sec IV}, we present numerical simulations that demonstrate consistency with this analytic result.

\section{Universal Density Correlations}
\label{sec III}

In this section we consider the $M$'th order equal-time correlation function of the density for a free Fermi gas, defined by
\begin{equation}
s_{M}({\bf q}_1, ..., {\bf q}_{M-1}) = \int\frac{d^D{\bf q}_M}{(2\pi)^D}
\langle \rho_{{\bf q}_1} \rho_{{\bf q}_2} ... \rho_{{\bf q}_M} \label{sk def}
\rangle_c
\end{equation}
for small ${\bf q}$, where 
\begin{equation}
\rho_{\bf q} = \int \frac{d^D{\bf k}}{(2\pi)^D} c^\dagger_{{\bf k}} c_{{\bf k}+{\bf q}},
\label{rhoq def}
\end{equation}
and 
${\bf q}_M = - \sum_{i=1}^{M-1} {\bf q}_i$.  We evaluate (\ref{sk def}) for a free fermion Hamiltonian of the form
\begin{equation}
{\cal H}_0 = \int \frac{d^D{\bf k}}{(2\pi)^D} E_{\bf k} c_{\bf k}^\dagger c_{\bf k}.
\end{equation} 
at zero temperature, where the electronic states are filled to for $E_{\bf k} < E_F$.
In general there will be a form factor $f({\bf q}) = \langle u_{{\bf k}+{\bf q}}|u_{\bf k}\rangle$ in (\ref{rhoq def}) that depends on the Bloch wavefunctions $|u_{\bf k}\rangle$, but since $f({\bf q}\rightarrow 0) = 1$ , this will not affect the small ${\bf q}$ limit, so we take $f({\bf q}) = 1$.    

For $M=D+1$, where $D$ is the dimensionality of the Fermi gas, we will show that $s_{D+1}$ has a universal behavior,
\begin{equation}
s_{D+1}({\bf q}_1, ..., {\bf q}_D) = \frac{\chi_F}{(2\pi)^D} |\det {\mathbb Q}|,
\label{s general}
\end{equation}
where ${\mathbb Q}$ is the $D\times D$ matrix formed out of the $D$ vectors ${\bf q}_i$.   Thus $|\det {\mathbb Q}|$ describes the volume of the $D$ dimensional parallelepiped formed by ${\bf q}_i$.   

For $D=1$ this result is trivial, and was explained in Section \ref{sec IIB}, where we argued that it is exact for $q$ smaller than a fixed finite cutoff that is determined by the size of the Fermi sea.   We have verified that the same is true for $D=2$ and $D=3$ by numerically evaluating the integrals that are obtained by evaluating (\ref{sk def}) using Wick's theorem.   We will describe the Wick's theorem analysis in some detail for $D=2$ in \ref{sec IIIA}, and then generalize to $D=3$ in \ref{sec IIIB}.   While straightforward in principle, this calculation does not provide any insight into why (\ref{s general}) is true.   Therefore, in Appendix \ref{sec A} we develop a different approach based on evaluating the closed fermion loop Feynman diagrams, which explains naturally the asymptotic behavior in the small $q$ limit.
Our proof of (\ref{s general}) will be given for the cases $D=2$ and $D=3$.   While we have not done the general case, we suspect it is valid for all $D$. 

Finally in Section \ref{sec IIIC} we will show that for a general value of $M$
\begin{equation}
s_{M}({\bf q}_1, ..., {\bf q}_{M-1}) \propto k_F^{D+1-M} q^{M-1}.
\label{s asymptotic}
\end{equation}
This result was used in Section \ref{sec IIA} to determine the order of divergence of the $M$'th order cumulant $\langle Q_A^M \rangle_c$.

\subsection{D=2}
\label{sec IIIA}

It is straightforward to evaluate (\ref{sk def}) using Wick's theorem.  Defining the Fermi occupation factor $f_{\bf k} = \langle c^\dagger_{\bf k} c_{\bf k}\rangle = \theta(E_F - E_{\bf k})$ and $\bar f_{\bf k} = 1 - f_{\bf k}$, we find for $M=3$,
\begin{equation}
s_3({\bf q}_1,{\bf q}_2) =\int\frac{d^2{\bf k}}{(2\pi)^2} \bar f_{{\bf k}+{\bf q}_1} f_{\bf k}  (\bar f_{{\bf k}-{\bf q}_3} - f_{{\bf k}-{\bf q}_2} ).
\label{s3 wick}
\end{equation}
This integral can be evaluated numerically, for a given Fermi surface specified by $E_{\bf k}=E_F$, and we have checked that (\ref{s general}) is numerically exact for sufficiently small (but finite) ${\bf q}$.  We have also established this exactness using a geometric argument, discussed in Appendix \ref{geoproof}, that is analogous to the $D=1$ argument explained in Sec. \ref{sec IIB}.   While this analysis provides more justification for  (\ref{s general}) for $D=2$, it is formidable to apply a similar reasoning to $D=3$. Hence, we choose to focus on analytic arguments that allow for a unified treatment of both the $D=2$ and $D=3$ cases.    Here, we will show that in the small ${\bf q}$ limit the integral may be expressed as a sum over critical points in the dispersion $E_{\bf k}$, where ${\bf v}_{\bf k} = \nabla_{\bf k} E_{\bf k} = 0$, which then can be related to the Euler characteristic $\chi_F$ by (\ref{chif morse}).   In this section, we will present a series of straightforward manipulations of (\ref{s3 wick}) that accomplishes this.   In Appendix \ref{sec A} we will present an alternative derivation of this result, which makes it clearer why this result is true, and shows how it can be generalized to higher order correlation functions.

Our strategy is to identify derivatives with respect to ${\bf q}$.  
We manipulate (\ref{s3 wick}) using the identity
\begin{equation}
\bar f_{{\bf k}+{\bf q}} f_{{\bf k}} = -
 \left( \Delta^{{\bf k}+{\bf q}}_{{\bf k}} f_{\bf k} \right)\Theta^{{\bf k}+{\bf q}}_{{\bf k}},
\label{f identity}
\end{equation}
where we introduce the notations
\begin{align}
&\Delta^{{\bf k}+{\bf q}}_{{\bf k}} f_{\bf k} \equiv f_{{\bf k}+{\bf q}} - f_{{\bf k}} 
\label{delta notation}\\
&\Theta^{{\bf k}+{\bf q}}_{{\bf k}} \equiv \theta(E_{{\bf k}+{\bf q}}-E_{{\bf k}}).
\end{align}
We then obtain
\begin{equation}
s_3 = \int\frac{d^2{\bf k}}{(2\pi)^2} \Theta^{{\bf k}+{\bf q}_1}_{\bf k} 
(f_{\bf k}-f_{{\bf k}+{\bf q}_1})(\bar f_{{\bf k}-{\bf q}_3} - f_{{\bf k}-{\bf q}_2}).
\end{equation}
Replacing $f_{{\bf k}(+{\bf q}_1)}=1-\bar f_{{\bf k}(+{\bf q}_1)}$ in the terms multiplying $f_{{\bf k}-{\bf q}_2}$ this becomes
\begin{align}
s_3 = \int\frac{d^2{\bf k}}{(2\pi)^2} \Theta^{{\bf k}+{\bf q}_1}_{\bf k}
\Bigl[&-\Delta_{\bf k}^{{\bf k}-{\bf q}_2} \left(\bar f_{{\bf k}-{\bf q}_3}f_{\bf k} \right)\\
&- \Delta_{\bf k}^{{\bf k}-{\bf q}_3} \left( \bar f_{\bf k} f_{{\bf k}-{\bf q}_2}\right) \Bigr].
\end{align}
Applying (\ref{f identity}) again leads to
\begin{align}
s_3 = \int\frac{d^2{\bf k}}{(2\pi)^2} \Theta_{\bf k}^{{\bf k}+{\bf q}_1} 
\Bigl[ &\Delta_{\bf k}^{{\bf k}-{\bf q}_2} \left(\Theta^{{\bf k}-{\bf q}_3}_{\bf k} 
\Delta_{\bf k}^{{\bf k}-{\bf q}_3} f_{\bf k}\right) \nonumber \\
- & \Delta_{\bf k}^{{\bf k}-{\bf q}_3} \left(\Theta_{{\bf k}-{\bf q}_2}^{\bf k} 
\Delta_{\bf k}^{{\bf k}-{\bf q}_2} f_{\bf k}\right) \Bigr].
\end{align}
We next perform a ``discrete integration by parts" using the identity
\begin{equation}
\int d^2{\bf k} F({\bf k})\left( \Delta_{\bf k}^{{\bf k}-{\bf q}} G({\bf k})\right) = 
\int d^2{\bf k} G({\bf k})\left( \Delta_{\bf k}^{{\bf k}+{\bf q}} F({\bf k}) \right)
\label{int by parts}
\end{equation}
and obtain
\begin{align}
s_3 = \int\frac{d^2{\bf k}}{(2\pi)^2} f_{\bf k} \Bigl[ & 
\Delta_{\bf k}^{{\bf k}+{\bf q}_3} \left(
\Theta^{{\bf k}-{\bf q}_3}_{\bf k} \Delta_{\bf k}^{{\bf k}+{\bf q}_2} 
\Theta^{{\bf k}+{\bf q}_1}_{\bf k} \right)\nonumber\\
- & 
\Delta_{\bf k}^{{\bf k}+{\bf q}_2} \left(
\Theta_{{\bf k}-{\bf q}_2}^{\bf k} \Delta_{\bf k}^{{\bf k}+{\bf q}_3} 
\Theta^{{\bf k}+{\bf q}_1}_{\bf k} \right) \Bigr].
\end{align}

We now take the limit ${\bf q}\rightarrow 0$, so $\Delta_{\bf k}^{{\bf k}+{\bf q}}F({\bf k})={\bf q}\cdot \nabla_{\bf k}F({\bf k})$ and $\Theta^{{\bf k}+{\bf q}}_{\bf k} = \theta({\bf v}_{\bf k}\cdot {\bf q})$.   Writing $\bar\theta(x) = \theta(-x)$ then obtain
\begin{align}
s_3 = \int &\frac{d^2{\bf k}}{(2\pi)^2} f_{\bf k} \Bigl[ 
{\bf q}_3\cdot\nabla_{\bf k}\Bigl(\bar\theta({\bf v}_{\bf k}\cdot{\bf q}_3) 
{\bf q}_2\cdot\nabla_{\bf k} \theta({\bf v}_{\bf k}\cdot{\bf q}_1)\Bigr) \nonumber\\
- & 
{\bf q}_2\cdot\nabla_{\bf k} \Bigl(\theta({\bf v}_{\bf k}\cdot{\bf q}_2)
{\bf q}_3\cdot\nabla_{\bf k} \theta({\bf v}_{\bf k}\cdot{\bf q}_1) \Bigr)\Bigl].
\end{align}
We next replace ${\bf q}_3 = - {\bf q}_1 -{\bf q}_2$ and note that since we can integrate the first derivative (outside the parentheses) by parts, so that it acts on $f_{\bf k}$, the second derivative inside the parenthesis will be proportional to $\delta({\bf v}_{\bf k}\cdot {\bf q}_1)$ and fixes  ${\bf v}_{\bf k}\cdot {\bf q}_1 = 0$.  Thus, we may replace 
$\bar\theta({\bf v}_{\bf k}\cdot{\bf q}_3) = \theta({\bf v}_{\bf k}\cdot{\bf q}_2)$.  Then, it can be seen that the terms in which all four derivatives are $({\bf q}_2\cdot\nabla_{\bf k})$, as well as the terms involving second derivatives $({\bf q}_1\cdot\nabla_{\bf k})({\bf q}_2\cdot\nabla_{\bf k})\theta({\bf v}_{\bf k}\cdot{\bf q}_1)$ all cancel.   This leaves
\begin{align}
\label{s3 detform}
s_3 = \int&\frac{d^2{\bf k}}{(2\pi)^2} f_{\bf k} \Bigl[ \\
- &\Bigl({\bf q}_1\cdot\nabla_{\bf k}\theta({\bf v}_{\bf k}\cdot{\bf q}_2) \Bigr)
\Bigl({\bf q}_2\cdot\nabla_{\bf k} \theta({\bf v}_{\bf k}\cdot{\bf q}_1)\Bigr) \nonumber\\
+ & 
\Bigl({\bf q}_2\cdot\nabla_{\bf k} \theta({\bf v}_{\bf k}\cdot{\bf q}_2)\Bigr) 
\Bigl({\bf q}_1\cdot\nabla_{\bf k} \theta({\bf v}_{\bf k}\cdot{\bf q}_1) \Bigr)\Bigl] \nonumber
\end{align}
Note that every term in the sum on ${\bf k}$ is proportional to $\delta({\bf v}_{\bf k}\cdot {\bf q}_1)\delta({\bf v}_{\bf k}\cdot {\bf q}_2)$.  Provided ${\bf q}_1$ and ${\bf q}_2$ are linearly independent, the sum will be restricted to {\it critical points} in the dispersion $E_{\bf k}$ where ${\bf v}_{\bf k} = 0$.   

We will now show that the integral evaluates the signature of each critical point.
Consider a critical point ${\bf k}_c$ near which
\begin{equation}
E_{\bf k} = E_c + \frac{1}{2} ({\bf k}-{\bf k_c})\cdot {\mathbb H}_c \cdot ({\bf k}-{\bf k}_c),
\end{equation}
where the Hessian ${\mathbb H}_c$ is the matrix of second derivatives of $E_{\bf k}$, and $v_{\bf k} = {\mathbb H}_c \cdot({\bf k}-{\bf k}_c)$.   We can then write (\ref{s3 detform}) as a sum over critical points inside the Fermi sea $s_3 = \sum_{c} f_{{\bf k}_c} s_3^c$, with
\begin{equation}
s^c_3 = \int\frac{d^2{\bf k}}{(2\pi)^2}
  \epsilon_{ij} ({\bf q}_i \cdot {\mathbb H}_c \cdot {\bf q}_1)( {\bf q}_j \cdot {\mathbb H}_c \cdot {\bf q}_2 )\delta({\bf v}\cdot {\bf q}_1) \delta({\bf v}\cdot{\bf q}_2).
\end{equation}
The coefficient of the $\delta$-functions in the integrand can be recognized as $\det[{\mathbb Q}^T {\mathbb H}_c {\mathbb Q}]$, where ${\mathbb Q}$ is the $2\times 2$ matrix built from the column vectors ${\bf q}_1$ and ${\bf q}_2$.  To evaluate the integral,  define new variables $X_j = {\bf v}_{\bf k} \cdot {\bf q}_j = ({\bf k}-{\bf k}_c)\cdot {\mathbb H}_c \cdot {\bf q}_j$.   The Jacobian of the transformation is, 
\begin{equation}
\frac{\partial(X_1,X_2)}{\partial(k_1,k_2)} = \det[ {\mathbb H}_c {\mathbb Q}].
\end{equation}
We then get
\begin{align}
s^c_3 &= \int \frac{dX_1dX_2}{(2\pi)^2} \frac{\det[{\mathbb Q}^T {\mathbb H}_c {\mathbb Q}]}{|\det[ {\mathbb H}_c {\mathbb Q}]|} \delta(X_1)\delta(X_2)\\
  &=  \eta_c \frac{|\det {\mathbb Q}|}{(2\pi)^2} ,
\end{align}
where $\eta_c = {\rm sgn} \det[{\mathbb H}_c]$ is the signature of the critical point $c$.

We thus obtain Eq. (\ref{s general}), where the Euler characteristic is expressed in terms of the critical points inside the  Fermi sea as
\begin{equation}
\chi_F = \sum_{c \in S_F} \eta_c.
\end{equation}

\subsection{D=3}
\label{sec IIIB}

We now analyze the case $s_4({\bf q}_1,{\bf q}_2,{\bf q}_3)$ in three dimensions.   It is straightforward to evaluate the equal-time expectation value using Wick's theorem, but now there are 6 contractions,
\begin{align}
\label{s4 wick}
s_4({\bf q}_1,&{\bf q}_2,{\bf q}_3) = \int\frac{d^3{\bf k}}{(2\pi)^3}
\bar  f_{{\bf k}+{\bf q}_1} f_{\bf k} \Bigl( \\
& \bar f_{{\bf k} + {\bf q}_1 + {\bf q}_2} \bar f_{{\bf k} - {\bf q}_4} 
+  f_{{\bf k} - {\bf q}_2}  f_{{\bf k}-{\bf q}_2 - {\bf q}_3} \nonumber \\
- &  \bar f_{{\bf k} - {\bf q}_3}  f_{{\bf k} - {\bf q}_2 - {\bf q}_3}  
-  \bar f_{{\bf k}-{\bf q}_4} f_{{\bf k}+ {\bf q}_1 + {\bf q}_3} \nonumber \\
- & \bar f_{{\bf k}+ {\bf q}_1 + {\bf q}_2}f_{{\bf k} - {\bf q}_3} 
- \bar f_{{\bf k}+{\bf q}_1 + {\bf q}_3}f_{{\bf k}-{\bf q}_2} \Bigr). \nonumber
\end{align}

We have evaluated Eq.  \ref{s4 wick} numerically for a series of 3 dimensional Fermi seas, which are specified by the function $f_{\bf k}$.   We find that Eq. \ref{s general} is  numerically exact for sufficiently small ${\bf q}$.   The momentum scale where (\ref{s general}) breaks down is set by the size and maximum curvature of the Fermi surface.   This mirrors a similar behavior for $s_2(q)$ in $D=1$, discussed in Section \ref{sec IIB} and $s_3({\bf q}_1,{\bf q}_2)$ in $D=2$, discussed in Section \ref{sec IIIA}.

While is possible to manipulate (\ref{s3 wick}) and (\ref{s4 wick}) into a form where the small $q$ behavior is more apparent, the algebra is quite complicated, and it is far from obvious how to proceed. 
We therefore seek a more systematic approach for extracting the limiting small $q$ behavior of (\ref{sk def}).  That will be developed in Appendix \ref{sec A}.  The basis for that approach is an observation in Ref. \onlinecite{Kane2022} that higher order {\it response} functions for a ballistic Fermi gas are related to solutions to the Boltzmann equation, which can be straightforwardly solved order by order in the external fields.   This suggests that a similar simplification should occur for the correlation function.  In Appendix \ref{sec A} we will show that the $M$'th order density correlation function computed in an imaginary time formalism can be computed to all orders with the aid of a Ward identity.   This then leads to a formulation in which (\ref{s4 wick}) can be expressed in the form,
\begin{align}
s_4({\bf q}_1&,{\bf q}_2,{\bf q}_3) = \int\frac{d^3{\bf k}}{(2\pi)^3}
 f_{\bf k} \sum_{abc=1}^3 \epsilon_{abc}
\Bigl({\bf q}_a\cdot\nabla_{\bf k}\theta({\bf v}_{\bf k}\cdot{\bf q}_1) \Bigr)\nonumber\\
&\Bigl({\bf q}_b\cdot\nabla_{\bf k}\theta({\bf v}_{\bf k}\cdot{\bf q}_2) \Bigr)
\Bigl({\bf q}_c\cdot\nabla_{\bf k}\theta({\bf v}_{\bf k}\cdot{\bf q}_3)\Bigr) .
\label{s4 epsilon}
\end{align}

Following the same analysis as the $D=2$ case, this integral will be dominated by critical points ${\bf k}_c$ where ${\bf v}_{\bf k} = 0$.   We introduce variables $X_a = {\bf v}_{\bf k}\cdot {\bf q}_a = ({\bf k}-{\bf k}_c)\cdot {\mathbb H}_c \cdot {\bf q}_a$ for $a = 1,2,3$, where ${\mathbb H}_c$ is the Hessian of $E_{\bf k}$ at ${\bf k}_c$.   Then,
\begin{equation}
s_4 = \frac{|\det {\mathbb Q}|}{(2\pi)^3} \sum_c f_{{\bf k}_c} {\rm sgn}[\det {\mathbb H}_c].
\label{s4 critical}
\end{equation}
Noting that $\chi_F = \sum_c f_{{\bf k}_c} {\rm sgn}[\det {\mathbb H}_c]$, we thus obtain Eq. (\ref{s general}) for $D=3$.

\subsection{$M$'th order correlator}
\label{sec IIIC}

We now briefly discuss the asymptotic behavior of $s_{M}({\bf q}_1,...,{\bf q}_{M-1})$ for general values of $M$ for a given dimension $D$.   As shown in Appendix \ref{sec A}, the $M$'th order frequency dependent correlator has a simple expression, which involves $M-1$ discrete derivatives of the form
$\Delta_{\bf k}^{{\bf k}-{\bf q}_i} \sim {\bf q}_i \cdot \nabla_{\bf k}$.   It is straightforward to perform the Matsubara frequency integrals to obtain the equal-time correlator, and as in the case for $s_{D+1}$ computed in Appendix \ref{sec A}, this will involve $\theta$ functions of the form $\theta(\Delta_{\bf k}^{{\bf k}+{\bf q}}E_{\bf k}) \sim \theta({\bf v}_{\bf k}\cdot \hat{\bf q})$.   Thus, the ${\bf q}\rightarrow 0$ limit is straightforward to take, and it will involve $M-1$ powers of ${\bf q}$, along with $D -(M-1)$ powers of $k_F$, which come from the momentum sum over ${\bf k}$, which involves $M-1$ powers of $\nabla_{\bf k}$.   This establishes Eq. \ref{s asymptotic}.

\section{\label{sec IV}Numerical study for $D=2$}
In this section, we present numerical evidence to support our prediction for the topological scaling of the charge-weighted tripartite mutual information in two dimensions, namely
\begin{equation}\label{prediction}
\mathcal{I}^Q_3= \frac{3\chi_F}{4\pi^2}\log^2(\frac{L}{a})+\mathcal{O}(L^0)
\end{equation}
with $L$ the linear size of the system and $a$ the short-distance cut-off. We focus on free-fermion lattice models, which can be simulated efficiently using the \textit{correlation matrix} method \cite{Peschel2001,Peschel2003,Henley2004,Peschel2009}. In particular, we study tight-binding models on triangular and square lattices with various topology of the Fermi sea. For free theories, the complete information of the ground state is encoded in two-point correlators $C_{ij} = \langle c^\dagger_i c_j\rangle$, which has a simple relation to the charge-weighted bipartite entanglement entropy $S^Q_A$. We first derive this relation, and describe our setup for the numerics, in Sec. \ref{sec IVA}. Then, by diagonalizing the $N \times N$ correlation matrix, with $N \sim L^2$ being the number of lattice sites, we obtain $\mathcal{I}^Q_3(L)$ up to $L \sim 300$. In Sec. \ref{sec IVB}, we present a quadratic-fitting analysis to extract the coefficient of $\log^2 L$ and demonstrate consistency with $\chi_F=\pm1$ for the case with one electron/hole-like Fermi surface. In Sec. \ref{sec IVC}, we adopt a ``ratio analysis" by calculating $\mathcal{I}^Q_3/\mathcal{I}^{Q*}_3$, and verify that this ratio is consistent with the quantization suggested by $\chi_F/\chi_F^*$. This analysis allows us to draw support from scenarios with more than one Fermi pockets (i.e. $\abs{\chi_F}>1$), while the performance of fitting is limited by finite-size error.

\subsection{\label{sec IVA}Correlation matrix method\\ and general setup}
Let us first review how to relate the von Neumann entanglement entropy for a subsystem $A$ to the two-point correlation matrix $(C_A)_{ij}= \langle c^\dagger_i c_j\rangle$, where $i,j \in A$. Following Ref. \cite{Peschel2003}, the key is to express the reduced density matrix $\rho_A$ in an exponential form,
\begin{equation}
\rho_A = \frac{e^{-\mathcal{H}_A}}{Z_A}
\end{equation}
with $Z_A = \Tr[e^{-\mathcal{H}_A}]$, and the entanglement Hamiltonian $\mathcal{H}_A$ chosen as a free-fermion operator
\begin{equation}\label{EntHam}
\mathcal{H}_A = \sum_{i,j \in A}(h_A)_{ij}c^\dagger_i c_j.
\end{equation}
As such, $n$-point correlation functions would factorize due to Wick's theorem, in accordance with our ground state (i.e. the filled Fermi sea) being a Slater determinant. Matrices $h_{A}$ and $C_{A}$ are related as follows,
\begin{equation}
(C_A)_{ij}  = \Tr[\rho_A c^\dagger_i c_j] = \Big(\frac{1}{1+e^{h_A}}\Big)_{ji}, 
\end{equation}
which can be shown easily by first transforming to the basis that diagonalizes $h_A$. Next, we define a generating function
\begin{equation}
\begin{split}
Z_A(\beta) &\equiv \Tr[e^{-\beta \mathcal{H}_A}]\\
&= \det[1+(C_A^{-1}-1)^{-\beta}], 
\end{split}
\end{equation} 
which relates to the von Neumann entanglement entropy as
\begin{equation}
\begin{split}
S_A &= (1-\partial_\beta) \log Z_A(\beta) \rvert_{\beta=1} \\
&= -\Tr[C_A\log C_A+(1-C_A)\log (1-C_A)].
\end{split}
\end{equation}
Therefore, instead of dealing with a $2^{N_A} \times 2^{N_A}$ density matrix, one  simply diagonalizes a much smaller $N_A \times N_A$ correlation matrix to obtain $S_A$. This is a standard result that has been used to simulate von-Neumann entanglement entropy in various non-interacting systems \cite{Barthel2006, Li2006, Peschel2009}. 

In this work, the \textit{charge-weighted} entanglement entropy has been introduced,
\begin{equation}
S^Q_A = - \Tr[(Q_A-\langle Q_A \rangle)\rho_A \log \rho_A]
\end{equation}
with $Q_A = \sum_{i\in A}c^\dagger_i c_i$ being the total charge in region $A$. By the same token, let us define a generating function
\begin{equation}
\begin{split}
\mathcal{Z}_A(\beta,\mu) &\equiv \Tr[e^{-\beta(\mathcal{H}_A - \mu Q_A)}] \\
& = \det[1+e^{\beta\mu}(C_A^{-1}-1)^{-\beta}],
\end{split}
\end{equation}
which generates the charge-weighted entanglement entropy by
\begin{equation}
\begin{split}
S^Q_A &= \partial_\mu[(1-\partial_\beta)\log \mathcal{Z}_A(\beta,\mu)]\rvert_{\beta=1,\mu=0} \\
&= \Tr[(1-C_A)C_A\log(C_A^{-1}-1)].
\end{split}
\end{equation}
This formula allows us to efficiently compute the charge-weighted tripartite mutual information $\mathcal{I}^Q_3$. Following the definition in Eq. (\ref{i3q def}), and noting $S^Q_{AB}= -S^Q_C$ and $S^Q_{ABC}=0$, we have $\mathcal{I}^Q_3 = 2(S^Q_A+S^Q_B+S^Q_C)$. 

\begin{figure}[t]
   \includegraphics[width=0.95\columnwidth]{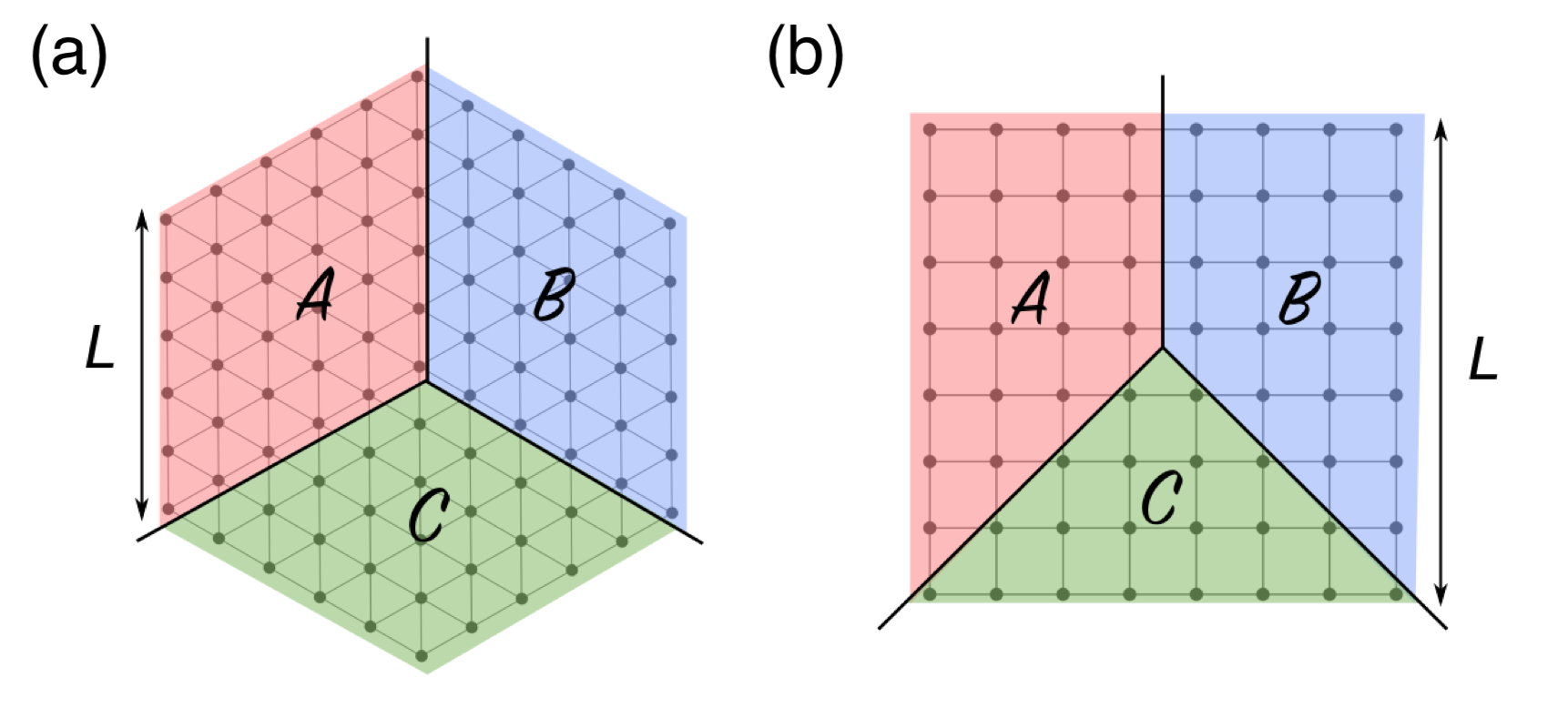}\centering
  \caption{\small{Tight-binding models and the respective tripartition for $D=2$ numerical study. (a) Triangular lattice with symmetric tripartition. Every angle at the triple contact is $2\pi/3$. (b) Square lattice with a different geometry of tripartition. Angles at the triple contact are $3\pi/4$, $3\pi/4$ and $\pi/2$. Open-boundary condition is implemented for simulations.}}
  \label{partition}
\end{figure}

Our numerical study focuses on two families of tight-binding models, with only spinless electrons for simplicity. The first setup is on the triangular lattice, with isotropic hopping among nearest-neighbors ($t_1$) and next-nearest-neighbors ($t_2$), 
\begin{equation}\label{triH}
H_{\text{tri}} = - t_1\sum_{\langle i,j \rangle} c^\dagger_i c_j - t_2 \sum_{\langle\langle i,j \rangle\rangle}  c^\dagger_i c_j .
\end{equation}
The second setup is on the square lattice, with hopping $t^x_{\ell}$ ($t^y_{\ell}$) to the $\ell$-th nearest-neighbor in the $x$ ($y$) direction, 
\begin{equation}\label{sqH}
H_{\text{sq}} = -\sum_{\ell = 1}^3 \sum_i (t^x_\ell c^\dagger_i c_{i+\ell\hat{x}}+t^y_\ell c^\dagger_i c_{i+\ell\hat{y}} + \text{H.c.}).
\end{equation}
By considering up to third-nearest-neighbor hopping in each direction, we can conveniently generate various Fermi-sea topology with $\abs{\chi_F} \leq 4$. The precise geometry of real-space tripartition for each setup is illustrated in Figure \ref{partition}.

In our simulation the open boundary condition is implemented, for the benefit that only one triple contact is present in the system. As each triple contact is associated with a $\chi_F \log^2 L$-divergence, implementing a periodic boundary condition should multiply this divergence by an appropriate factor that counts the number of contacts due to periodicity, but this also means that multiple triple contacts can \textit{interfere} with each other. This would lead to finite-size correction to our prediction in (\ref{prediction}), which is hard to control. We thus adopt the open boundary condition as a cleaner way to extract $\chi_F$ from the $\log^2L$-scaling. Next, we present two approaches of analysis to demonstrate consistency between numerics and our theoretical prediction. 

\begin{figure}[b!]
   \includegraphics[width=0.98\columnwidth]{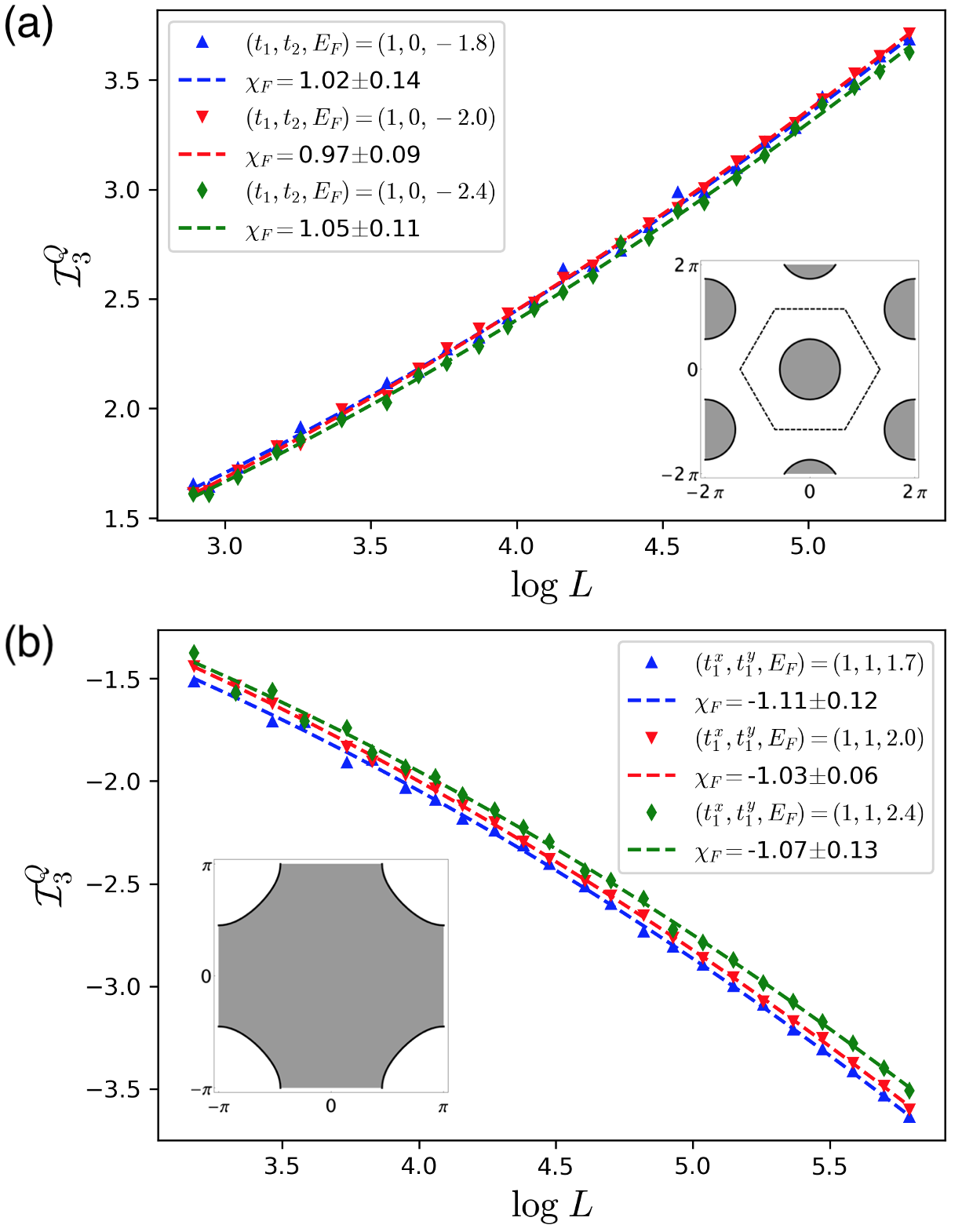}\centering
  \caption{\small{Fitting analysis for (a) triangular lattice tight-binding model with one electron-like Fermi surface, thus $\chi_F = 1$ is expected; (b) for square lattice with one hole-like Fermi surface, thus $\chi_F = -1$ is expected. Our results show collapse of data points for Fermi sea with varying geometry but the same topology. Fitting demonstrates that $\mathcal{I}^Q_3$ scales as $\log^2 L$, with a nearly quantized coefficient that encodes the Fermi sea topology, and is insensitive to the geometry of tripartition. Insets show the filled Fermi sea (shaded in grey) in corresponding setups. Details on optimizing the fitting analysis are presented in Appendix \ref{sec D}.}}
  \label{fitting}
\end{figure}

\subsection{\label{sec IVB}Fitting analysis}
Based on (\ref{prediction}), we consider a fitting function with three parameters $\{p,q,r\}$,
\begin{equation}
\mathcal{I}^Q_3(L) = p\log^2L +q\log L +r.
\end{equation}
Performing least-squares fit on $\mathcal{I}^Q_3$ as a polynomial of $\log L$, we obtain the quadratic coefficient $p$, and its uncertainty $\delta p $ (from the covariance matrix), which translate into
\begin{equation}
\chi_F \pm \delta\chi_F = \frac{4\pi^2}{3}(p\pm \delta p).
\end{equation}
For the triangular lattice with linear size $L$ (see Fig. \ref{partition}), we have computed $\mathcal{I}^Q_3$ from $L=18$ to $L=210$, with 25 data points sampled evenly in the scale of $\log L$. For $(t_1,t_2, E_F)=(1,0,-2)$, there is one \textit{electron-like} Fermi surface, and hence $\chi_F=1$. Our fitting result suggests $\chi_F = 0.97(9)$. By varying the Fermi energy $E_F$ to obtain Fermi sea of the same topology but different sizes, we consistently find $\chi_F \approx 1$, see Fig. \ref{fitting}(a). For the square lattice with linear size $L$, we have computed $\mathcal{I}^Q_3$ from $L=24$ to $L=327$, with 24 data points sampled evenly in the scale of $\log L$. For $(t^x_1,t^y_1, E_F)=(1,1,2)$, there is one \textit{hole-like} Fermi surface, and hence $\chi_F=-1$. Our fitting result suggests $\chi_F = -1.03(6)$. Again, we have varied $E_F$ to demonstrate that $\chi_F \approx -1$ is consistently obtained, see Fig. \ref{fitting}(b). Taking both the triangular and square lattices together, and noting that two different geometries of tripartition have been implemented, our numerical results support our prediction that the $\log^2L$-divergence of $\mathcal{I}^Q_3$ is \textit{universal} and \textit{topological}. 

\begin{figure*}[t!]
   \includegraphics[width=\textwidth]{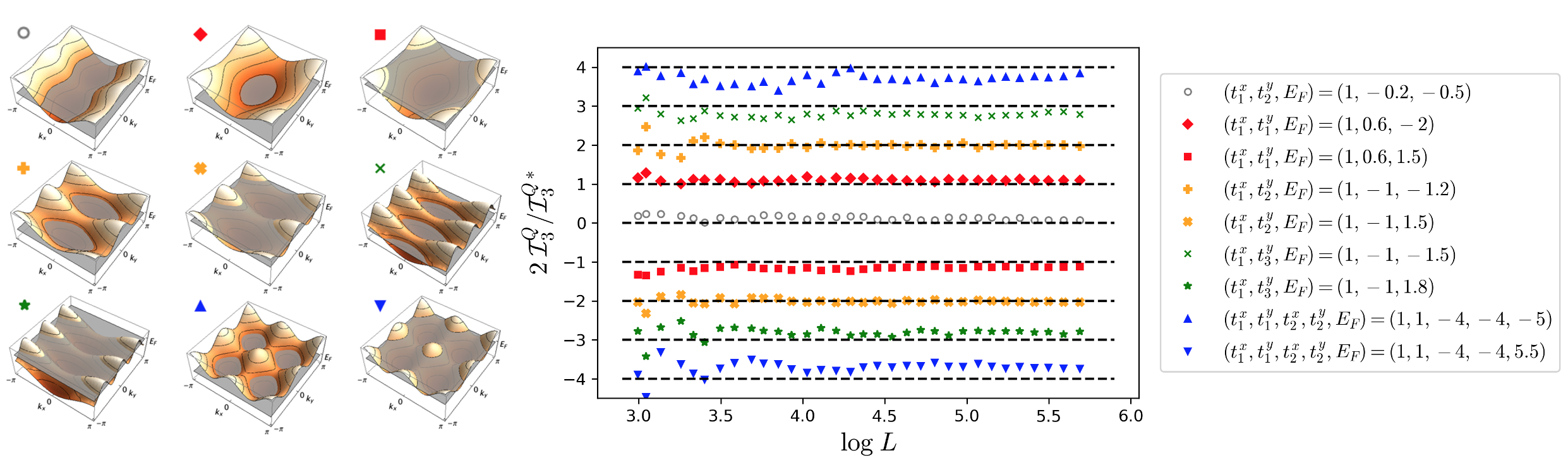}\centering
  \caption{\small{Ratio analysis for nine sets of Fermi-sea topology, with $\chi_F$ ranging from $-4$ to $+4$. Left: equi-energy contours for the nine sets of dispersion in the square lattice tight-binding model. Shaded regions correspond to the filled Fermi sea. Middle: ratio $\chi^*_F \mathcal{I}^Q_3/\mathcal{I}^{Q*}_3$ as a function of linear system size $L$. The standard ruler $\{\mathcal{I}^{Q*}_3, \chi^*_F\}$ is defined by (\ref{standard}), corresponding to two electron-like Fermi pockets, hence $\chi^*_F=2$. Right: specifications of hopping integrals and Fermi energy that generate the nine sets of Fermi sea under study. Unspecified hoppings are set to 0.}}
  \label{ratioplt}
\end{figure*}

In our data, the best-fit value deviates by roughly $5\%$ from the true $\chi_F$, and the uncertainty of fitting is about $10\%$. We attribute this deviation to finite-size error, after all, our prediction in (\ref{prediction}) is derived in the thermodynamic limit $L/a \rightarrow \infty$. Being incapable of simulating an infinite system (in fact our storage limits us to $\log L \lesssim 6$), we are confronted with fluctuations caused by the lattice discreteness and boundary effects. In fact, the best-fit values presented above are obtained after optimizing the raw data (the approaches are explained in Appendix \ref{sec D}) to help us sidestep the finite-size errors. Moreover, the effective UV cut-off ($a$) is determined not only by the lattice constant (which has been set to 1 already), but also by the size of Fermi sea and distance between Fermi surfaces in the momentum space. If one of these momentum scales happen to be too small, the UV cut-off would be rather large, and we are forced to study much larger system sizes in order to extract the topological scaling behavior. However, our storage resource limits us to $\log L \lesssim 6$, and thus severely limits the performance of quadratic-fitting for scenarios with multiple Fermi pockets packed in the first Brillouin zone. 

\subsection{\label{sec IVC}Ratio analysis}
In order to further test (\ref{prediction}) for larger $\chi_F$, let us consider the following ratio,
\begin{equation}\label{ratio}
\frac{\mathcal{I}^Q_3(L)}{\mathcal{I}^{Q*}_3(L)} = \frac{\chi_F \log^2(L/a)+\mathcal{O}(L^0)}{\chi^*_F \log^2(L/a^*)+\mathcal{O}(L^0)} \approx \frac{\chi_F}{\chi^*_F}.
\end{equation}
Here $\{\mathcal{I}^{Q*}_3,\chi^*_F\}$ defines a ``standard ruler", to which we use to compare with other systems. Admittedly, the ratio $\mathcal{I}^Q_3/\mathcal{I}^{Q*}_3$ equals to $\chi_F/\chi^*_F$ only in the thermodynamic limit, which we cannot reach. However, if we judiciously choose the size of Fermi sea and distance between Fermi surfaces to be comparable in the two scenarios that we are taking ratio of, i.e. ensure $a \approx a^*$, then we still expect (\ref{ratio}) to hold for sufficiently large $L$. In this way, we can circumvent the difficulty encountered in the fitting analysis.

The results of ratio analysis are summarized in Fig. \ref{ratioplt}. We adopt the square lattice tight-binding model in (\ref{sqH}), and first generate the ``standard ruler" by
\begin{equation}\label{standard}
(t^x_1, t^y_2, E_F) = (1,-1,-2.5), 
\end{equation}
with unspecified hoppings $t^{x/y}_\ell$ all set to zero. This gives two electron-like Fermi surfaces ($\chi^*_F = 2$), like the one labeled by the ``plus" symbol in Fig. \ref{ratioplt}, but with more negative Fermi energy such that the Fermi surfaces are smaller. Choosing this as our standard ruler allows us to better approach the expected quantization when comparing to scenarios with  three or four Fermi surfaces. We have then generated nine sets of Fermi-sea topology, ranging from $\chi_F = -4$ to $\chi_F=+4$, whose equi-energy contours are displayed in the left panel of Fig. \ref{ratioplt}. As shown in the plot, $\chi^*_F \mathcal{I}^Q_3/\mathcal{I}^{Q*}_3 $ approaches the true $\chi_F$ for each type of Fermi-sea topology as $L$ approaches the thermodynamic limit. All cases have demonstrated consistency with (\ref{ratio}). 

\begin{figure*}[t!]
   \includegraphics[width=\textwidth]{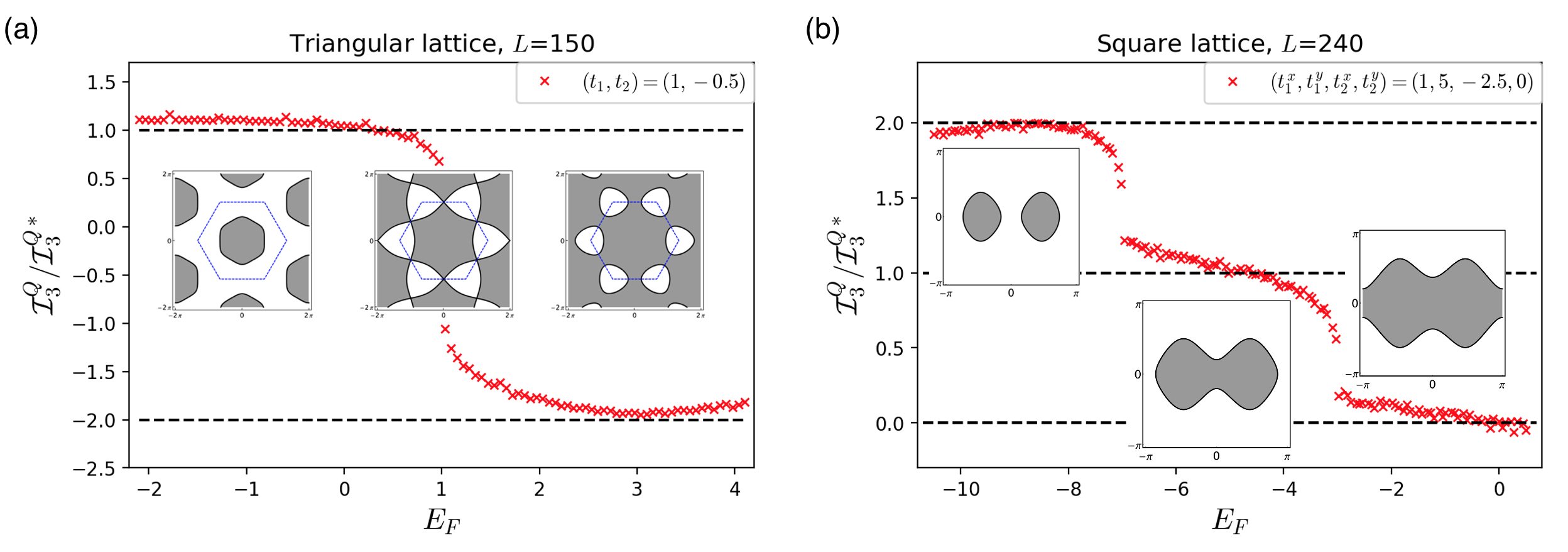}\centering
  \caption{\small{Lifshitz transition as probed by the charge-weighted tripartite mutual information $\mathcal{I}^Q_3$. (a) shows a triangular lattice undergoing a topological transition from $\chi_F = 1$ to $\chi_F =-2$ at $E_F=1.0$. The reference $\mathcal{I}^{Q*}_3$ is calculated at $E_F = 0.5$, corresponding to $\chi^*_F = 1$; (b) shows a square lattice undergoing a topological transition from $\chi_F = 2$ to $\chi_F =1$ at $E_F=-7.0$, and then from $\chi_F =1$ to $\chi_F=0$ at $E_F=-3.0$. The reference $\mathcal{I}^{Q*}_3$ is calculated at $E_F = -5.0$, with $\chi^*_F = 1$. Insets display the filled Fermi sea (shaded in grey) at different fillings.}}
  \label{transition}
\end{figure*}

In light of this, we have also used $\mathcal{I}^Q_3$ to probe Lifshitz transitions in a given band-structure, where the topology of Fermi sea changes as $E_F$ is varied. Two examples are illustrated in Fig. \ref{transition}, with the respective standard ruler $\{\mathcal{I}^{Q*}_3,\chi^*_F\}$ specified in the caption. One can clearly observe quantization in accordance with the topology of Fermi sea from the ratio $\mathcal{I}^Q_3/\mathcal{I}^{Q*}_3$, and see a drastic change at a Lifshitz transition. While one may notice that the quantization is not exact, the way it deviates is completely expected from our prediction in (\ref{prediction}). For example, in Fig. \ref{transition}(a) we see a curve (instead of a plateau in the ideal case) bending upward around $E_{F}=3.0$. This is because, as $E_F$ goes below $3.0$, the hole-like Fermi pockets are getting closer to each other; while as $E_F$ goes above $3.0$, the Fermi pockets are getting smaller. Either of these effects end up increases the UV cut-off $a$, and hence decreases the magnitude of $\mathcal{I}^Q_3$. The same reasoning can be applied to understand the deviation from quantization in Fig. \ref{transition}(b) as well. 

The numerics for free-fermion lattice models thus support our prediction in Eq. (\ref{prediction}). These results serve as a non-trivial check for the replica analysis presented in previous sections. In the next section, we return to the replica theory and examine the effect of interaction. 

\section{Effect of Interactions}
\label{sec V}

In the preceding sections we have established that the multipartite mutual information defined in (\ref{i2 def}), (\ref{i3q def}) and (\ref{i4 def}) provide a universal and topological characterization of the Fermi sea for a system of non-interacting fermions.   This was established by relating the mutual information to the universal and quantized number correlations.   It is important to address whether this characterization is robust in the presence of short-ranged electron-electron interactions.   In $D=1$, the non-interacting electron gas becomes a Luttinger liquid, while in higher dimensions it becomes a Fermi liquid.   In order to explain what the issues are, let us first review the well known situation in one dimension.

In a one dimensional single channel Luttinger liquid, the number correlations are {\it not} quantized.   A Luttinger liquid can be described by the Euclidean Lagrangian,
\begin{equation}
{\cal L} = \frac{1}{2\pi K v} [(\partial_\tau\theta)^2 + v^2 (\partial_x\theta)^2 ],
\end{equation}
where the charge density is given by $\rho(x) = \partial_x\theta/\pi$ and $K$ is the Luttinger parameter \cite{Giamarchi2004}.  It follows that (\ref{sqd=1}) and (\ref{qaqb log}) become
\begin{equation}
s_2(q)  = \frac{K}{2\pi}|q|
\label{sq luttinger}
\end{equation}
and
\begin{equation}
\langle Q_A Q_B \rangle_c = -\frac{K}{2\pi^2}\log\Lambda.
\end{equation}
Thus, for interacting fermions with $K \ne 1$ the coefficient of the logarithmic divergence of the number correlations is no longer topologically quantized.   Nonetheless, it is known that the coefficient of the logarithmic divergence of the entanglement entropy, $(c/6) \log \Lambda$, reflects the central charge $c$ of the conformal field theory, which is equal to $1$ even when $K \ne 1$.   

In the following we will show how to reconcile this discrepancy within the replica theory by studying the effect of interactions perturbatively.   We will then use that analysis to argue that in three dimensions the topological mutual information remains robustly quantized in the presence of interactions, even though the density correlations are no longer quantized.   Thus, the topological mutual information provides a robust topological characterization of the Fermi liquid phase.   In contrast, we will show that interactions destroy the quantization of the the charge-weighted entanglement entropy defined for a two dimensional system.   Thus, ${\mathcal I}^Q_3$ only provides a topological characterization of non-interacting (or weakly interacting) systems.

We will begin by recovering the known result in one dimension and using the understanding developed there proceed to three and then two dimensions.

\subsection{D=1 : Luttinger Liquid}
\label{sec VA}

\begin{figure}
\includegraphics[width=\columnwidth]{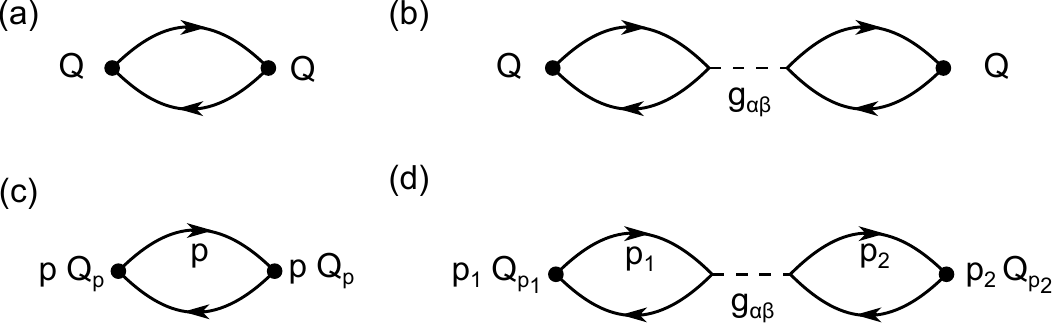}
\caption{(a,b): Representative Feynman diagrams for calculating $\langle Q_A^2 \rangle_c$, with (a) the non-interacting contribution and (b) the first-order correction from interaction. Small circles represent the number operator $Q_A$. (c,d) Diagrams involved in calculating $S_{A,n}$ in the replica analysis, focusing on the $\langle Q^2_A \rangle_c$ term which contributes to the logarithmic divergence in $\mathcal{I}_2$. Small circles here correspond to $pQ_{A,p}$, where $p$ is the replica momentum that is to be summed over. The summation over $p$ cancels all the interacting contributions, such as (d), leaving the non-interacting diagram in (c) as the only contribution. The quantization in $\mathcal{I}_2$ thus survives the interaction.}
\label{Fig8}
\end{figure}

Consider a single channel gas of spinless fermions, with a Hamiltonian ${\cal H} = {\cal H}_0 + V$, where ${\cal H}_0$ is a free fermion Hamiltonian and $V$ is a four fermion interaction.   
To address the effect of $V$ on the large $L$ behavior of the number correlations and the entanglement entropy it is sufficient to consider a low energy theory that focuses on the left ($L$) and right ($R$) moving Fermi points, and discards irrelevant interactions that flow to zero in the low energy limit.  This has the form of the Luttinger model, with
\begin{equation}
{\cal H}_0 = -i \hbar v_F (\psi_R^\dagger \partial_x \psi_R - \psi_L^\dagger \partial_x \psi_L).
\label{h0fermion}
\end{equation}
At low energy, the most relevant interactions are the forward scattering interactions between the densities at the right and left moving Fermi points, 
\begin{equation}
V = \sum_{\alpha, \beta} g_{\alpha\beta} \psi_\alpha^\dagger \psi_\alpha \psi_\beta^\dagger \psi_\beta,
\end{equation}
where $\alpha$ and $\beta$ are summed over $L$ and $R$, and we define $g_1 = g_{LL}=g_{RR}$ and 
$g_2 = g_{LR} = g_{RL}$.   
The effect of this interaction on the density-density correlation function (and hence the number correlations) can be studied perturbatively by summing the RPA (random phase approximation) diagrams in Fig. \ref{Fig8}(a,b).   To lowest order in $g_{\alpha\beta}$ this leads to Eq. \ref{sq luttinger} with
\begin{equation}
K = 1 - \frac{g_2}{\pi v_F}.
\end{equation}

In order to determine the interaction corrections to the entanglement entropy we must consider the replica theory, which involves $n$ copies of the interacting theory.  
The entanglement entropy in the interacting theory is then determined by writing 
\begin{equation}
S_{A,n} = \frac{{\rm Tr}[ T_{A,n} e^{-S_0 + \delta S}]}{{\rm Tr}[e^{-S_0 + \delta S}]},
\end{equation}
where $S_0 = \sum_a S_0[\psi_{\alpha,a}^\dagger,\psi_{\alpha,a}]$ is the replicated free fermion action corresponding to (\ref{h0fermion}), with $a$ the replica index and $\alpha=R,L$. $\delta S = \int dx d\tau V$ describes the interaction,   
\begin{equation}
V = \sum_{\alpha,\beta, a} g_{\alpha\beta} \psi^\dagger_{\alpha,a} \psi_{\alpha,a} \psi_{\beta,a}^\dagger \psi_{\beta,a}.
\end{equation}
Notice there is no interaction between replicas in the ``real replica space".   However,
to proceed with the analysis, we need to transform to the replica momentum basis, where we can write
\begin{equation}
S_{A,n} = \frac{{\rm Tr}[ \prod_p e^{\frac{2\pi i p}{n} Q_{A,p}}
e^{-S_0 + \delta S}]}{{\rm Tr}[e^{-S_0 + \delta S}]}
\label{san perturbation}
\end{equation}
with
\begin{equation}
V = \sum_{\alpha, \beta,\{p_i\}} g_{\alpha\beta} \delta_{p_1-p_2+p_3-p_4} 
\psi^\dagger_{\alpha,p_1} \psi_{\alpha,p_2} \psi_{\beta,p_3}^\dagger \psi_{\beta,p_4}.
\end{equation}
Thus, the different replica-momentum channels are no longer independent.   However, since the interaction is diagonal in replica space and invariant under replica translations, it is independent of replica momentum, and obeys ``replica momentum conservation".    

The first order correction to $S_{A,n}$ is determined by expanding (\ref{san perturbation}) to first order in $g$.   If in addition we formally expand $\exp[(2\pi i p/n) Q_{A,p}]$, then the resulting terms will involve connected Feynman diagrams containing a single interaction line along with $k$ $p Q_p$-vertices.   In particular, for $M=2$ the diagram (which corresponds to the same diagram which led to the modification of $\langle Q_A^2 \rangle_c$) is shown in Fig. \ref{Fig8}(d).   Note that since the interaction is independent of replica momentum, this involves two independent sums 
\begin{equation}
\left(\sum_{p_1 = -\frac{n-1}{2}}^{\frac{n-1}{2}} p_1 \right)
\left(\sum_{p_2 = -\frac{n-1}{2}}^{\frac{n-1}{2}} p_2 \right) = 0.
\label{d=1 cancel}
\end{equation}
Thus, this diagram does not contribute to $S_{A,n}$ and does not lead to a modification of the topological mutual information ${\mathcal I}_2$.   It is clear that higher order diagrams will behave similarly.   This confirms that ${\mathcal I}_2$ probes the central charge $c$, and is not perturbed by interactions.   Note also, that the cancellation in Eq. \ref{d=1 cancel} occurs for any value of $n$.   This means that in addition to the von Neumann entropy, the R\'enyi entropies for $D=1$ also exhibit a universal logarithmic divergence \cite{Calabrese2004}. 

\subsection{D=3 : Fermi Liquid}
\label{sec VB}

We now apply the same analysis to a three dimensional interacting Fermi liquid.   We will make the key assumption that the Fermi liquid phase can be described by a low energy theory, analogous to the Luttinger model, that describes excitations close to the Fermi surface, and which discards irrelevant interactions.   The remaining marginal interactions involve short-ranged forward scattering interactions between the densities associated with different points on the Fermi surface.  These determine the Fermi liquid parameters, which are central to Fermi liquid theory.   We thus consider a free fermion Hamiltonian of the form
\begin{equation}
{\cal H}_0 = \int \frac{d^3{\bf k}}{(2\pi)^3} E_{\bf k} c_{\bf k}^\dagger c_{\bf k}
\end{equation}
defined on a shell in momentum space straddling the Fermi surface $E_{\bf k} = E_F$.  We take the interaction to be
\begin{equation}
V = \int\frac{d^3{\bf k} d^3{\bf k}'}{(2\pi)^6} f_{{\bf k},{\bf k}'} \delta n_{\bf k} \delta n_{{\bf k}'}
\end{equation}
where $\delta n_{\bf k} = c_{\bf k}^\dagger c_{\bf k} - \langle c_{\bf k}^\dagger c_{\bf k} \rangle$.

\begin{figure}
\includegraphics[width=3 in]{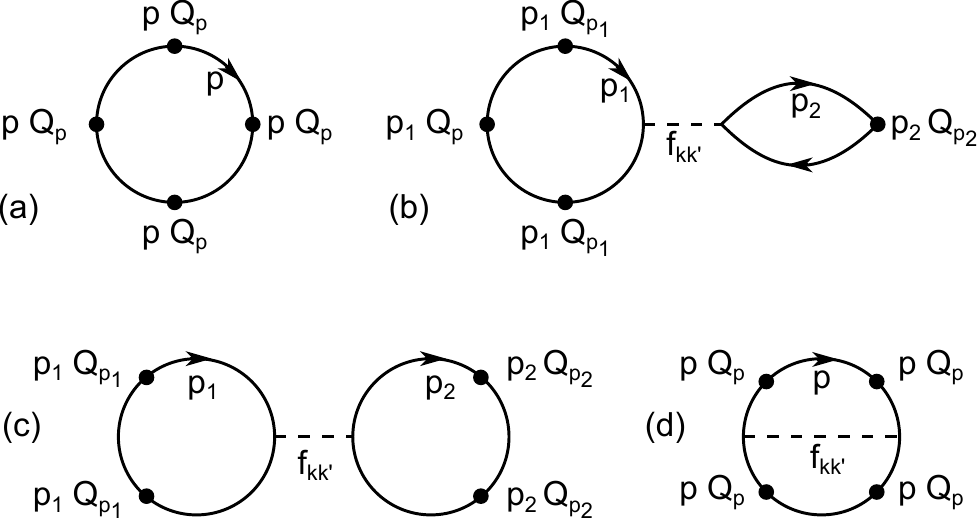}
\caption{Feynman diagrams for calculating $S_{A,n}$ in the replica analysis, focusing on the $\langle Q^4_A \rangle_c$ term which contributes to the logarithmic divergence in $\mathcal{I}_4$. Interacting correction of the type in (b) vanishes as the independent sums over replica momenta $p_i$ render it zero. Correction from (c) vanishes particularly in the replica limit $n\rightarrow 1$. Non-RPA contribution, such as (d), do not contribute in the Fermi liquid theory. This leaves the non-interacting diagram in (a) as the only contribution to $\mathcal{I}_4$, hence its quantization survives the interaction.}
\label{Fig9}
\end{figure}

As in $D=1$, interactions will lead to a correction to the leading divergence of the number correlations.   In particular, the fourth order fluctuation $\langle Q_A^4 \rangle_c$ will involve a correction to first order in $f_{{\bf k},{\bf k}'}$ described by the Feynman diagram in Fig. \ref{Fig9}(b) (without the replica indices).    As argued in Ref. \onlinecite{Polchinski1992,Shankar1994}, Fermi liquid theory amounts to summing all RPA-like diagrams involving the low energy interaction $f_{{\bf k},{\bf k}'}$, at order $(f_{{\bf k},{\bf k}'})^n$ which involve $n+1$ fermion loops and independent sums on ${\bf k}$.  Diagrams with fewer fermion loops (like the self energy correction to the single particle propagator in Fig \ref{Fig9}(d)) are suppressed by phase space constraints in the low energy renormalized theory, and do not contribute.
Examination of the simplest allowed diagram in Fig. \ref{Fig9}(b)  shows that the coefficient of $\log^3 \Lambda$ includes a correction at first order.    Thus, the coefficient of the $\log^3 \Lambda$ divergence in $\langle Q_A^4\rangle_c$ is not topologically quantized in the presence of interactions.

To compute the entanglement entropy, we again introduce the replica theory, where now the interaction $f_{{\bf k},{\bf k}'}$ is independent of replica-momentum and obeys replica momentum conservation.   Thus, each closed fermion loop in the diagrams of Fig. \ref{Fig9} will include an independent sum over replica momentum.   The diagram in Fig. \ref{Fig9}(b) vanishes for the same reason as the 1D case because it involves
\begin{equation}
\left(\sum_{p_1 = -\frac{n-1}{2}}^{\frac{n-1}{2}} p_1^3 \right)
\left(\sum_{p_2 = -\frac{n-1}{2}}^{\frac{n-1}{2}} p_2 \right) = 0.
\end{equation}
The diagram in Fig. \ref{Fig9}(c) is not equal to zero.   Rather, it involves
\begin{equation}
\left(\sum_{p_1 = -\frac{n-1}{2}}^{\frac{n-1}{2}} p_1^2 \right)
\left(\sum_{p_2 = -\frac{n-1}{2}}^{\frac{n-1}{2}} p_2^2 \right) = \left( C_{n,2}\right)^2.
\label{d=3 vanish}
\end{equation}
However, since $\lim_{n\rightarrow 1} C_{n,2} = (n-1)/6$ the contribution of this diagram will be of order $(n-1)^2$.   Therefore, this contribution vanishes in (\ref{replica limit}).

It is clear that all allowed diagrams describing interaction corrections to $S_{A,n}$ will vanish upon taking the replica limit.   This shows that, like in $D=1$, the $\log^3 \Lambda$ divergence of the mutual information $\mathcal{I}_4$ remains quantized in the presence of interactions. Equation (\ref{i4 final}) therefore provides a robust topological characterization of the interacting Fermi liquid phase.   We can now identify distinct classes of {\it topological Fermi liquids}, which are distinguished by the Euler characteristic of the Fermi sea (or equivalently of the Fermi surface, since $\chi_{\partial F} = 2\chi_F$).

It is interesting to note that since the vanishing of (\ref{d=3 vanish}) only occurs in the limit $n \rightarrow 1$, there is potential for the R\'enyi entropies to be perturbed by interactions.  We further remark that the topological mutual information $\mathcal{I}_4$, as introduced in Sec. \ref{sec IIC}, can also be constructed from R\'enyi entropies instead of the von Neumann entropy. The resulting $\mathcal{I}^{(n)}_{4}$, in the non-interacting case, exhibits a $\log^3 \Lambda$ divergence whose coefficient is again proportional to $\chi_F$, but with a modified proportionality factor.  However, we have not established that the quantization is robust in the presence of interaction when $n \rightarrow 1$ limit is \textit{not} taken. Nevertheless, since R\'enyi entropies are easier to simulate numerically \cite{Hastings2010, Humeniuk2012, Grover2013, Broecker2014, Wang2014}, and also more convenient to probe experimentally \cite{Pichler2013, Islam2015, Brydges2019, Cornfeld2019}, it is worthwhile to understand precisely how interactions renormalize the $\log^3 \Lambda$ divergence in those cases. This will depend on the correction contributed by the diagram in Fig. \ref{Fig9}(c). We will leave the explicit evaluation of that diagram to future work.

\subsection{D=2}
\label{sec VC}

\begin{figure}
\includegraphics[width=3 in]{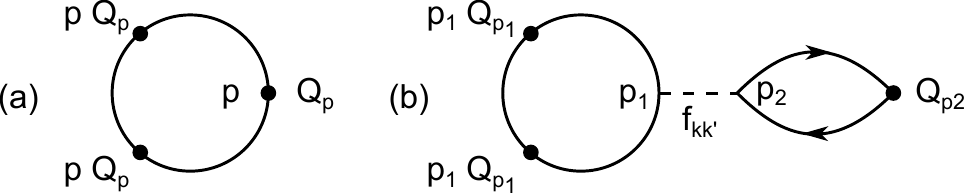}
\caption{Feynman diagrams for calculating $S^Q_{A,n}$ in the replica analysis, focusing on the $\langle Q^3_A \rangle_c$ term which contributes to the logarithmic divergence in $\mathcal{I}^Q_3$. While the non-interacting diagram in (a) contributes a quantized logarithmic divergence in $\mathcal{I}^Q_3$, Fermi-liquid interaction gives rise to additional diagrams, such as (b), which modify the coefficient of divergence. }
\label{Fig10}
\end{figure}

We now consider the interaction corrections to the charge-weighted entanglement entropy that was defined for $D=2$.   In two dimensions, the $\log^2\Lambda$ divergence of the third order number correlations, $\langle Q_A^3 \rangle_c$, will be modified by RPA-like corrections to the charge vertex, just as in $D=3$.   However, the charge-weighted entanglement entropy behaves differently because it involves computing
\begin{equation}
S^Q_{A,n} = \frac{{\rm Tr}[ Q_{a=0} T_{A,n}
e^{-S_0 + \delta S}]}{{\rm Tr}[e^{-S_0 + \delta S}]}.
\label{sanq perturbation}
\end{equation}
In the replica momentum representation, the $Q_{a=0}$ inside the trace becomes $\sum_p Q_p/n$, so that when evaluating the third order cumulant, only two of the three $Q_p$'s come with a factor of $p$.   It follows that the RPA correction to the $Q_{a=0}$ vertex, shown in Fig. \ref{Fig10} does not vanish in the replica limit.   Rather, it involves
\begin{equation}
\left(\sum_{p_1 = -\frac{n-1}{2}}^{\frac{n-1}{2}} p_1^2 \right)
\left(\sum_{p_2 = -\frac{n-1}{2}}^{\frac{n-1}{2}} \frac{1}{n}\right) = C_{n,2}.
\end{equation}
For $n\rightarrow 1$ this is $(n-1)/6$, and since it is linear in $n-1$ it survives in Eq. (\ref{replica limit of SQ}).   Therefore, assuming this diagram indeed contributes to the $\log^2\Lambda$ divergence, the topological quantization of ${\mathcal I}^Q_3$ found in Eq. (\ref{iq3 final}) is only a property of non-interacting Fermi gas, and does not persist in a two dimensional Fermi liquid.

\section{Discussion and Conclusion}
\label{sec VI}

In this paper we have established that the topology of the Fermi sea in a $D$-dimensional Fermi gas, characterized by the Euler characteristic $\chi_F$, is reflected in the multipartite entanglement of $D+1$ regions that meet at a point.   For odd $D$, the ($D+1$)-partite mutual information ${\mathcal I}_{D+1}$ provides a robust characterization of the Fermi gas, even in the presence of interactions.   Thus, our result serves as a generalization of Calabrese and Cardy's calculation of the bipartite entanglement entropy of a 1+1D conformal field theory, applied to an interacting Fermi gas.  For $D=3$, ${\mathcal I}_4$ provides a universal characterization of the entanglement that distinguishes distinct topological Fermi liquid phases, see Eq. (\ref{i4 final}).  For even dimensions (specifically $D=2$), the multipartite mutual information does not probe $\chi_F$, but we introduced a modified ``charge-weighted" mutual information ${\mathcal I}_3^Q$ that does, see Eq. (\ref{iq3 final}). Unlike the ${\mathcal I}_4$, however, ${\mathcal I}_3^Q$ only provides a topological characterization of the non-interacting Fermi gas, and is perturbed by interactions.

Our results motivate several further questions.   What is the significance of our inability to define an entanglement measure reflecting the topology of the Fermi sea for a two dimensional interacting Fermi liquid?   Is there a fundamental obstruction to doing that, or might there be some other entanglement measure?  We have found that Fermi liquids in $D=3$ fall into distinct topological classes indexed by $\chi_F$.   
One of the grand themes in topological band theory has been the interplay between topology and symmetry \cite{Hasan2010, Qi2011}.  Does the presence of symmetries, like crystal symmetries, time reversal or Bogoliubov-de Gennes particle-hole symmetry in a superconductor, refine the topological classes of Fermi liquids?   

It will also be interesting to clarify the connection between the topological mutual information that we have introduced and characterizations of the entanglement in higher dimensional conformal field theories.   For example, for a $D=3+1$ CFT, the bipartite entanglement entropy for a spherical region contains a universal logarithmic term proportional to the quantity ``$a$", which like ``$c$" in $D=1+1$, counts the low energy degrees of freedom \cite{Ryu2006, Solodukhin2008, Casini2009, Myers2011, Casini2011, Liu2013}.  The Euler characteristic $\chi_F$, even for $D=3$, however, is more like $c$ than $a$.  It characterizes not just the Fermi sea but also the Fermi \textit{surface} (since $\chi_{\partial F}=2\chi_F$), which defines a family of $1+1$ dimensional chiral fermions. In this sense, a 3D Fermi liquid is more closely related to a $1+1$D CFT than to $3+1$D relativistic fermions.  It will be interesting to clarify the connection between our theory of topological entanglement and a recently developed effective field theory of the Fermi surface \cite{Son2022} that accounts for non-linear effects, and generalizes earlier theories of bosonization of the Fermi surface \cite{CastroNeto1994,Houghton2000} which have been applied to study entanglement \cite{Ding2012}.

It will be interesting to examine the scaling behavior of ${\mathcal I}_4$ at a Lifshitz transition and to ask what happens in a Weyl semimetal, where the Fermi energy is at a Weyl point at $E=0$, and the Fermi surface shrinks to zero.   It is easy to check that $\chi_F$ is the same when the Fermi energy is above or below the Weyl point.   However, when $E_F=0$, our calculation of ${\mathcal I}_4$ breaks down, as Eq. (\ref{sd formula0}) for $s_{4}({\bf q}_1,{\bf q}_2,{\bf q}_3)$ is only valid for $\abs{\bf q} \lesssim k_F$.   When $k_F \rightarrow 0$ the expression is no longer valid.  It follows that the limits $L \rightarrow \infty$ and $E_F \rightarrow 0$ do not commute, and for $E_F = 0$ we do {\it not} expect a $\log^3 L$ divergence of ${\mathcal I}_4$.    We speculate that, for $E_F = 0$, ${\mathcal I}_4$ exhibits a weaker $\log L$ divergence that probes a topological property of the $3+1$D free Dirac fermion conformal field theory.  Indeed, for smooth entangling surfaces, the bipartite entanglement entropy for relativistic fermions exhibits a $\log L$ term whose coefficient probes both the intrinsic and extrinsic curvature of the surface \cite{Solodukhin2008}.   However, extending this result to entangling surfaces with corners - which are necessarily present in our theory - remains an open problem \cite{Myers2012, Bednik2019, Bueno2019}.  

It will be of interest to devise methods for measuring ${\mathcal I}_{D+1}$ - either numerically or experimentally.   While we have shown that numerical calculation is straightforward for free fermions in $D=2$, generalizing to $D=3$, and in particular to {\it interacting} systems will require innovation.   As discussed in Sec. \ref{sec V}, the robustness of $\mathcal{I}_4$ in the presence of interaction is guaranteed only when it is constructed from the von Neumann entropy, see Eq. (\ref{i4 def}). Fermionic projected entangled pair states, and their Gaussian variants, can in principle grant access to the von Neumann entropy in higher dimensions \cite{Kraus2010, Mortier2020}, but for low bond dimensions they do not capture the requisite logarithmic corrections to the entanglement area law.  The same issue also plagues the fermionic multiscale entanglement renormalization ansatz (MERA) for $D>1$ \cite{Corboz2009, Corboz2010, Barthel2009, Pineda2010, Barthel2010}. Nevertheless, a generalized version known as the \textit{branching} MERA has been proposed to display logarithmic correction to the area law, given a suitably chosen holographic tree that characterizes the branching structure \cite{Evenbly2014, Evenbly2014B, Haegeman2018}. We thus expect the branching MERA to properly account for the mutual information that characterizes Fermi sea topology in higher dimensions. The connection between optimal branching network structures and the topology of Fermi sea is an interesting question for future studies. R\'enyi entropies are easier to access, but it remains to be investigated whether, and how, interactions would renormalize the mutual information $\mathcal{I}^{(n)}_{4}$ constructed from R\'enyi entropies.

While it is difficult to measure entanglement directly in experiment, there have been proposals for how to measure it indirectly through charge fluctuations \cite{Klich2009, Song2011, Song2012}, which can be probed by existing experimental techniques \cite{Reulet2003, Sukhorukov2007, Gershon2008, Flindt2009}. In particular, the statistics of charge fluctuations at a quantum point contact as it is opened and closed contains a signature from which the 1D entanglement can be extracted \cite{Klich2009}.   It is tempting to ask whether such measurement setups could be generalized to two dimensions for measuring the third-order cumulant $\langle Q_A Q_B Q_C\rangle_c$ in Eq. (\ref{QQQ_c}).  One obvious complication is that, as we have shown, the connection between number correlations and the entanglement entropy is corrupted by electron-electron interactions.  But perhaps for a sufficiently weakly interacting system, measurement of the number correlations could still probe the topology of Fermi sea. In addition, for a weakly interacting 2D system it would be interesting to devise a method for directly measuring $s_3({\bf q}_1,{\bf q}_2)$ in Eq. (\ref{sd formula0}), by scattering experiments.

In one dimension, the central charge $c$ can be probed experimentally by the {\it thermal} response \cite{Nightingale1986, Kane1997, Read2000, Cappelli2002}, which is also related to the gravitational response \cite{Luttinger1964, Ryu2012_anomaly, Stone2012_anomaly}.   Importantly, this quantized response is insensitive to interactions.   For $D=3$ we have found that the quantization of ${\mathcal I}_4$ in terms of $\chi_F$ is similarly robust in the presence of interactions.   Does this define a quantized gravitational response?   If so, is there a thermal analog that is accessible experimentally?

Finally, one of the fascinating things about $c$ in $D=1+1$ is that it can be fractional \cite{Ginsparg1988, BYB}.  In a superconductor, chiral Majorana edge modes have $c=1/2$ \cite{Read2000}, and more exotic strongly correlated states can have other fractions.   It will be interesting to generalize our analysis to describe strongly interacting non-Fermi liquid phases \cite{Else2021}, and explore whether $\chi_F$ in three dimensions has an interpretation as a higher-order anomaly for the emergent Fermi surface. 

$$ \\ $$
\acknowledgments
This work was supported by a Simons Investigator Grant to C.L.K. from the Simons Foundation.

\appendix

\section{Matsubara Correlation Function}
\label{sec A}

Here we compute the correlation function of the density as a function of Matsubara frequency in a zero temperature imaginary time formalism.   
\begin{align}
s_{M}(\{{\bf q}_a& , i\omega_a \}|_{a=1}^{M-1}) = \int d^{M-1}\tau  e^{i \sum_i \omega_i(\tau_i
-\tau_M)}\nonumber \\
&\int \frac{d^D{\bf q}_M}{(2\pi)^D} 
 \langle T_\tau[\rho_{{\bf q}_1}(\tau_1) ... \rho_{{\bf q}_{M}}(\tau_{M})] \rangle_c,
\end{align}
where $\rho_{\bf q}(\tau) = e^{{\cal H}\tau} \rho_{\bf q} e^{-{\cal H}\tau}$ and 
$T_\tau$ indicates a time ordered product, with $\tau$ increasing from right to left.  $s_{M}$ is a function of $M-1$ frequencies and momenta, constrained by $\sum_i {\bf q}_i =0$ and $\sum_i \omega_i =0 $.   Since it depends only on time differences, we can fix $\tau_M=0$ in the integral.   $s_{M}$ can be evaluated using the standard diagrammatic technique by evaluating a single Fermion loop with $M$ density vertices placed in all possible orders.   For small $M$ this is rather simple, but the calculation gets increasingly cumbersome for larger $M$.  We therefore seek a more systematic approach.   We will employ a Ward identity that allows us to formulate a simple recursion relation that allows us to determine $s_{M}$ for all $M$.

\subsection{Ward Identity}
\label{sec A1}

Here we derive a recursive formula that allows us to determine $s_{M}$ for all $M$ by exploiting a Ward identity.   We begin by defining the the vertex function for $M$ density operators,
\begin{equation}
\Lambda_M({\bf k},\tau,\{{\bf q}_a,\tau_a\}_{a=1}^M)
= \langle T_\tau[c^\dagger_{{\bf k} + {\bf q}_\Sigma}(\tau) c_{{\bf k}}(\tau) \prod_{a = 1}^M \rho_{{\bf q}_a}(\tau_a)] \rangle
\end{equation}
along with its Fourier transform,
\begin{align}
\Lambda_M({\bf k},&\{{\bf q}_a,i\omega_a\}_{a=1}^M) = \nonumber\\ &
\int d^M \tau_a e^{i \sum_a \omega_a (\tau_a - \tau)} \Lambda_M({\bf k},\tau,\{{\bf q}_a,\tau_a\}).
\end{align}
Here ${\bf q}_\Sigma = \sum_a {\bf q}_a$.   Clearly,
\begin{equation}
s_{M}(\{{\bf q}_a,\omega_a\}_{a=1}^{M-1}) = \int \frac{d^D{\bf k}}{(2\pi)^D} \Lambda_{M-1}({\bf k},\{{\bf q}_a,\omega_a\}).
\end{equation}
We derive a Ward identity for $\Lambda_M$ by differentiating with respect to $\tau$, taking into account both the time dependence of the Heisenberg operators and the discontinuity at $\tau = \tau_a$ due to the time ordering.  Using the facts that
\begin{equation}
 [{\cal H},c^\dagger_{{\bf k} + {\bf q}_\Sigma} c_{{\bf k}}] 
 = \left(\Delta_{\bf k}^{{\bf k} + {\bf q}_\Sigma} E_{\bf k}\right)  c^\dagger_{{\bf k} + {\bf q}_\Sigma} c_{{\bf k}}
\end{equation}
and
\begin{equation}
[c^\dagger_{{\bf k} + {\bf q}_\Sigma} c_{{\bf k}} , \rho_{{\bf q}_b}] = \Delta_{\bf k}^{{\bf k}+{\bf q}_b} \left(c^\dagger_{{\bf k} + {\bf q}_\Sigma - {\bf q}_b} c_{\bf k}\right)
\end{equation}
($\Delta_{\bf k}^{{\bf k}+{\bf q}_b}$ is defined in \ref{delta notation}) we obtain
\begin{align}
\Bigl(\partial_\tau - &\Delta_{\bf k}^{{\bf k} + {\bf q}_\Sigma} E_{\bf k}\Bigr)
\Lambda_M({\bf k},\tau,\{{\bf q}_a,\tau_a\})
= \\
&\sum_{b=1}^M \Delta_{\bf k}^{{\bf k}+{\bf q}_b} 
\Lambda_{M-1}({\bf k},\tau,\left.\{{\bf q}_a,\tau_a\}\right|_{a \ne b} )\delta(\tau-\tau_b).
\nonumber
\end{align}
Fourier transforming on $\tau_a$, this leads to a recursion relation obeyed by 
$\Lambda_M({\bf k},\{{\bf q}_a,\omega_a\})$:
\begin{align}
\Lambda_M({\bf k},\{{\bf q}_a&,i\omega_a\}) = \sum_{b=1}^M \frac{1}{i\omega_\Sigma - \Delta_{\bf k}^{{\bf k}+{\bf q}_\Sigma} E_{\bf k}}\\
&  \Delta_{\bf k}^{{\bf k}+{\bf q}_b} \Lambda_{M-1}({\bf k},\{{\bf q}_a,i\omega_a\}|_{a \ne b}).
\nonumber
\end{align}
where $\omega_\Sigma = \sum_a \omega_a$

This allows us to determine $\Lambda_M$ for all $M$, and after integrating over ${\bf k}$ we obtain $s_{M}$.  For $M=1$ we have simply
\begin{equation}
s_1 = \int\frac{d^D{\bf k}}{(2\pi)^D} f_{\bf k}.
\label{s1 formula}
\end{equation}
Then,
\begin{equation}
s_2({\bf q},i\omega) = \int\frac{d^D{\bf k}}{(2\pi)^D} \frac{1}{i\omega - \Delta_{\bf k}^{{\bf k}+{\bf q}} E_{\bf k}}
 \Delta_{\bf k}^{{\bf k}+{\bf q}} f_{\bf k},
\label{s2 formula}
\end{equation}
and
\begin{align}
s_3({\bf q}_{1,2},i\omega_{1,2})& = \int\frac{d^D{\bf k}}{(2\pi)^D}
\sum_{a\ne b=1}^2 
\frac{1}{i\omega_a + i\omega_b - \Delta_{\bf k}^{{\bf k}+{\bf q}_a + {\bf q}_b} E_{\bf k}} \nonumber\\
&\Delta_{\bf k}^{{\bf k}+{\bf q}_b} \left(
\frac{1}{i\omega_a - \Delta_{\bf k}^{{\bf k}+{\bf q}_a} E_{\bf k}}
\Delta_{\bf k}^{{\bf k}+{\bf q}_a} f_{\bf k}\right).
\label{s3 formula}
\end{align}
The case $M=3$ was also studied in Ref. \onlinecite{Kane2022}, where the retarded response function was computed.  That result is related to this by analytic continuation $i\omega \rightarrow \omega + i\eta$.
For $M=4$ we have
\begin{align}
s_4(&\{{\bf q}_a,i\omega_a\}) = \int\frac{d^D{\bf k}}{(2\pi)^D}\sum_{a\ne b \ne c = 1}^3 \nonumber\\
&\frac{1}{i\omega_a+i\omega_b+i\omega_c - \Delta_{\bf k}^{{\bf k}+{\bf q}_a + {\bf q}_b + {\bf q}_c}E_{\bf k}}\nonumber \\
&\Delta_{\bf k}^{{\bf k}+{\bf q}_c} \Bigl[
\frac{1}{i\omega_a+i\omega_b-\Delta_{\bf k}^{{\bf k}+{\bf q}_a + {\bf q}_b}E_{\bf k}}  \nonumber\\ 
&\Delta_{\bf k}^{{\bf k}+{\bf q}_b} \Bigl(
\frac{1}{i\omega_a-\Delta_{\bf k}^{{\bf k}+{\bf q}_a}E_{\bf k}} 
\Delta_{\bf k}^{{\bf k}+{\bf q}_a} f_{\bf k} \Bigr)\Bigr].
\label{s4 formula}
\end{align}
Clearly this pattern extends to all orders.

\subsection{Equal-Time Correlation Function}
\label{sec A2}

To compute the equal-time correlation function, we integrate over the Matsubara frequencies.   Here we will do the analysis for the cases $D=2$ and $D=3$ separately.

\subsubsection{D=2}
\label{sec A2a}

For a given $a$ and $b$ in the sum in (\ref{s3 formula}) it is useful to define new frequencies, $\Omega_1 = \omega_a $ and $\Omega_2 = \omega_a + \omega_b$ .   Then
\begin{equation}
e^{-i(\omega_a \tau_{a3} + \omega_b \tau_{b3})} = e^{-i(\Omega_1 \tau_{ab} + \Omega_2 \tau_{b3})},
\end{equation}
and we can evaluate the sums over $\Omega_1$ and $\Omega_2$ independently.  
We evaluate the sums in the equal-time limit, but we must account for the operator ordering in
$\langle \rho_{{\bf q}_1}\rho_{{\bf q}_2}\rho_{{\bf q}_3}\rangle$ by keeping an infinitesimal time difference $\tau_{ab} = \eta {\rm sgn}(b-a)$, that serves as a convergence factor for the integration.   The sums are then evaluated by contour integration.   For example, 
\begin{equation}
\sum_{i\Omega_1} \frac{e^{-i \eta \Omega_1 {\rm sgn}(b-a)}}
{i\Omega_1 - \Delta_{\bf k}^{{\bf k}+{\bf q}_a}E_{\bf k}} = 
\left\{\begin{array}{ll} -\theta_a & a<b \\ \bar\theta_a & a > b
\end{array}\right.
\label{freqint1}
\end{equation}
and since $b<3$
\begin{equation}
\sum_{i\Omega_2} \frac{e^{-i \eta\Omega_2 {\rm sgn}(3-b)}}
{i\Omega_2 - \Delta_{\bf k}^{{\bf k}+{\bf q}_a + {\bf q}_b}E_{\bf k}} = 
 -\theta_{ab},
\label{freqint2}
\end{equation}
Here we adopt the shorthand notation
\begin{equation}
\theta_{a(b)} = 1-\bar\theta_{a(b)} = \theta(\Delta_{\bf k}^{{\bf k}+{\bf q}_a (+ {\bf q}_b)}E_{\bf k}).
\end{equation}
This leads to
\begin{align}
s_3({\bf q}_1,{\bf q}_2) = \int\frac{d^2{\bf k}}{(2\pi)^2} \Bigl[
&\theta_{12} \Delta_{\bf k}^{{\bf k}+{\bf q}_2} \left(\theta_1 \Delta_{\bf k}^{{\bf k}+{\bf q}_1} f_{\bf k} \right)\nonumber\\
-&\theta_{21} \Delta_{\bf k}^{{\bf k}+{\bf q}_1} \left( \bar\theta_2 \Delta_{\bf k}^{{\bf k}+{\bf q}_2} f_{\bf k}\right)\Bigr].
\end{align}

We now consider the limit ${\bf q}\rightarrow 0$, so that 
$\Delta_{\bf k}^{{\bf k}+{\bf q}_1} \rightarrow {\bf q}_1\cdot\nabla_{\bf k}$ and 
$\theta_1 \rightarrow \theta({\bf v}_{\bf k}\cdot {\bf q}_1)$.  We will keep the discrete derivative notation for a few more steps because it makes the formulas more compact, but now it should be understood that $\Delta_{\bf k}^{{\bf k}+{\bf q}} = \Delta_{{\bf k}-{\bf q}}^{\bf k}$.   In addition, we
use ${\bf q}_2 = -{\bf q}_1 - {\bf q}_3$ to write $s_3$ in terms of ${\bf q}_1$ and ${\bf q}_3$.  (We are free to do this, since $s_3$ can be expressed in terms of any pair of ${\bf q}$'s.)  It follows that $\theta_{12} \rightarrow \bar\theta_3$.
We also integrate by parts on ${\bf k}$ to obtain
\begin{align}
s_3({\bf q}_1,{\bf q}_3) = \int\frac{d^2{\bf k}}{(2\pi)^2}f_{\bf k} \Bigl[
&\Delta_{\bf k}^{{\bf k}-{\bf q}_1}\left(\theta_1 \Delta_{\bf k}^{{\bf k}+{\bf q}_1 + {\bf q}_3}
\bar\theta_3 \right)\nonumber\\
-& \Delta_{\bf k}^{{\bf k}+{\bf q}_1+{\bf q}_3}\left( \theta_{13} \Delta_{\bf k}^{{\bf k}-{\bf q}_1 }\bar\theta_3\right)\Bigr].
\end{align}
Since the derivatives on the left (outside the parentheses) can be integrated by parts so that they act on $f_{\bf k}$, the derivatives inside the parentheses are proportional to $\delta({\bf v}_{\bf k}\cdot {\bf q}_3)$.   We can therefore set ${\bf v}\cdot {\bf q}_3 = 0$ in the rest of the integral, so $\theta_{13} \rightarrow \theta_1$.   It can then be checked that the terms involving second derivatives $\Delta_{\bf k}^{{\bf k}-{\bf q}_1 }\Delta_{\bf k}^{{\bf k}+{\bf q}_1 + {\bf q}_3} \bar\theta_3$ cancel.  Using $\nabla_{\bf k} \bar\theta_3 = -\nabla_{\bf k}\theta_3$ we write
\begin{align}
s_3({\bf q}_1,{\bf q}_3) = \int\frac{d^2{\bf k}}{(2\pi)^2} &f_{\bf k} \Bigl[
-\left(\Delta_{\bf k}^{{\bf k}-{\bf q}_1}\theta_1\right)\left(\Delta_{\bf k}^{{\bf k}+{\bf q}_1 + {\bf q}_3}
\theta_3 \right)\nonumber\\
+& \left(\Delta_{\bf k}^{{\bf k}+{\bf q}_1+{\bf q}_3} \theta_{1} \right)\left(\Delta_{\bf k}^{{\bf k}-{\bf q}_1 }\theta_3\right)\Bigr].
\label{s3 simplified}
\end{align}
Now it can be observed that the terms in which 
 all four derivatives involve ${\bf q}_1\cdot \nabla_{\bf k}$ cancel, so
\begin{align}
s_3({\bf q}_1,{\bf q}_3) = \int\frac{d^2{\bf k}}{(2\pi)^2} &f_{\bf k} \Bigl[
\left({\bf q}_1 \cdot \nabla_{\bf k} \theta_1\right)
\left({\bf q}_3 \cdot \nabla_{\bf k} \theta_3\right)\nonumber\\
-&\left({\bf q}_3 \cdot \nabla_{\bf k} \theta_1\right)
\left({\bf q}_1 \cdot \nabla_{\bf k} \theta_3\right)\Bigr].
\end{align}
This has exactly the same form as (\ref{s3 detform}).  The only difference is that it is written as a function of ${\bf q}_1$ and ${\bf q}_3$ instead of ${\bf q}_1$ and ${\bf q}_2$.   But since $s_3$ can be expressed in terms of any pair of ${\bf q}$'s that does not matter.

\subsubsection{D=3}
\label{sec A2b}

We now evaluate the frequency integral for (\ref{s4 formula}).   The calculation is very similar to the $D=2$ calculation in the previous section.   It's just a bit more complicated.   For a given $a$, $b$ and $c$ in the sum in (\ref{s4 formula}) it is again useful to define new independent frequencies,
$\Omega_1 =\omega_a $, $\Omega_2 = \omega_a + \omega_b$ and $\Omega_3 = \omega_a+ \omega_b + \omega_c$.  Then, noting that
\begin{equation}
e^{-i(\omega_a\tau_{a4} + \omega_b\tau_{b4} + \omega_c\tau_{c4})} = e^{-i(\Omega_1 \tau_{ab} + \Omega_2 \tau_{bc} + \Omega_3 \tau_{c4})}
\end{equation}
the integrals over $\Omega_{1,2,3}$ can be performed independently, as in (\ref{freqint1}) and (\ref{freqint2}).  There are now 6 terms in the sum over $a,b,c$.  The result is
\begin{align}
s_4({\bf q}_1,&{\bf q}_2,{\bf q}_3) = \int\frac{d^3{\bf k}}{(2\pi)^3} \Bigl\{ \\
- &\theta_{123} \Delta_{\bf k}^{{\bf k}+{\bf q}_3} 
\left[ \theta_{12}\Delta_{\bf k}^{{\bf k}+{\bf q}_2} 
\left( \theta_1 \Delta_{\bf k}^{{\bf k}+{\bf q}_1} f_{\bf k}\right)\right] \nonumber \\
+ & \theta_{123} \Delta_{\bf k}^{{\bf k}+{\bf q}_3} 
\left[ \theta_{12}\Delta_{\bf k}^{{\bf k}+{\bf q}_1} 
\left( \bar\theta_2 \Delta_{\bf k}^{{\bf k}+{\bf q}_2} f_{\bf k}\right)\right]  \nonumber \\
+ & \theta_{123} \Delta_{\bf k}^{{\bf k}+{\bf q}_2} 
\left[ \bar\theta_{13}\Delta_{\bf k}^{{\bf k}+{\bf q}_3} 
\left( \theta_1 \Delta_{\bf k}^{{\bf k}+{\bf q}_1} f_{\bf k}\right)\right]  \nonumber \\
+ & \theta_{123} \Delta_{\bf k}^{{\bf k}+{\bf q}_2} 
\left[ \theta_{13}\Delta_{\bf k}^{{\bf k}+{\bf q}_1} 
\left( \bar\theta_3 \Delta_{\bf k}^{{\bf k}+{\bf q}_3} f_{\bf k}\right)\right]  \nonumber \\
+ & \theta_{123} \Delta_{\bf k}^{{\bf k}+{\bf q}_1} 
\left[ \bar\theta_{23}\Delta_{\bf k}^{{\bf k}+{\bf q}_3} 
\left( \theta_2 \Delta_{\bf k}^{{\bf k}+{\bf q}_2} f_{\bf k}\right)\right]  \nonumber \\
- & \theta_{123} \Delta_{\bf k}^{{\bf k}+{\bf q}_1} 
\left[ \bar\theta_{23}\Delta_{\bf k}^{{\bf k}+{\bf q}_2} 
\left( \bar\theta_3 \Delta_{\bf k}^{{\bf k}+{\bf q}_3} f_{\bf k}\right)\right]  \nonumber 
\Bigr\}.
\end{align}
We take the ${\bf q}\rightarrow 0$ limit, write ${\bf q}_1 = -{\bf q}_2 - {\bf q}_3 - {\bf q}_4$ and integrate by parts to obtain
\begin{align}
s_4({\bf q}_2,&{\bf q}_3,{\bf q}_4) = \int\frac{d^3{\bf k}}{(2\pi)^3}  f_{\bf k} \Bigl\{ \\
- &
\Delta_{\bf k}^{{\bf k}+{\bf q}_2+{\bf q}_3+{\bf q}_4} \left[ \bar\theta_{234}
\Delta_{\bf k}^{{\bf k}-{\bf q}_2} \left(  \bar\theta_{34} 
\Delta_{\bf k}^{{\bf k}-{\bf q}_3} \bar\theta_4 \right)\right] \nonumber \\
+ & 
\Delta_{\bf k}^{{\bf k}-{\bf q}_2} \left[ \bar\theta_{2}
\Delta_{\bf k}^{{\bf k}+{\bf q}_2+{\bf q}_3 + {\bf q}_4} \left(  \bar\theta_{34} 
\Delta_{\bf k}^{{\bf k}-{\bf q}_3} \bar\theta_4 \right)\right] \nonumber \\
+ & 
\Delta_{\bf k}^{{\bf k}+{\bf q}_2+{\bf q}_3 + {\bf q}_4} \left[ \bar\theta_{234}
\Delta_{\bf k}^{{\bf k}-{\bf q}_3 } \left(  \theta_{24} 
\Delta_{\bf k}^{{\bf k}-{\bf q}_2} \bar\theta_4 \right)\right] \nonumber \\
+ & 
\Delta_{\bf k}^{{\bf k}-{\bf q}_3} \left[ \bar\theta_{3}
\Delta_{\bf k}^{{\bf k}+{\bf q}_2+{\bf q}_3 + {\bf q}_4 } \left(  \bar\theta_{24} 
\Delta_{\bf k}^{{\bf k}-{\bf q}_2} \bar\theta_4 \right)\right] \nonumber \\
+ & 
\Delta_{\bf k}^{{\bf k}-{\bf q}_2} \left[ \theta_{2}
\Delta_{\bf k}^{{\bf k}-{\bf q}_3} \left(  \bar\theta_{23} 
\Delta_{\bf k}^{{\bf k}+{\bf q}_2+{\bf q}_3 + {\bf q}_4 } \bar\theta_4 \right)\right] \nonumber \\
- & 
\Delta_{\bf k}^{{\bf k}-{\bf q}_3} \left[ \bar\theta_{3}
\Delta_{\bf k}^{{\bf k}-{\bf q}_2} \left(  \bar\theta_{23} 
\Delta_{\bf k}^{{\bf k}+{\bf q}_2+{\bf q}_3 + {\bf q}_4 } \bar\theta_4 \right)\right] \nonumber
\Bigr\}.
\end{align}

This can be simplified similar to (\ref{s3 simplified}), except there are a few more steps.   First, since we can have the first three derivatives act to the left, the single derivative on $\bar\theta_4$ allows us to replace
$\bar\theta_{234}\rightarrow\bar\theta_{23}$ and $\bar\theta_{34}\rightarrow\bar\theta_3$ in the first term, along with similar replacements in the other terms.   Then it can be observed that all terms with more than one derivative acting to the right on $\bar\theta_4$ cancel.   This allows us to integrate by parts again to have a single derivative acting on $\bar\theta_3$, which allows us to replace $\bar\theta_{23} \rightarrow \bar\theta_2$ in the first term, along with similar replacements in the other terms.   Finally it can be observed that all terms with two derivatives acting on the middle $\theta$ function cancel.
Replacing $\nabla_{\bf k}\bar\theta\rightarrow -\nabla_{\bf k}\theta$ this becomes
\begin{align}
s_4({\bf q}_2,&{\bf q}_3,{\bf q}_4) = \int\frac{d^3{\bf k}}{(2\pi)^3}  f_{\bf k} \Bigl\{ \\
+&
\left(\Delta_{\bf k}^{{\bf k}+{\bf q}_2+{\bf q}_3+{\bf q}_4} \theta_{2} \right)\left(
\Delta_{\bf k}^{{\bf k}-{\bf q}_2}   \theta_{3}\right)\left( 
\Delta_{\bf k}^{{\bf k}-{\bf q}_3} \theta_4 \right) \nonumber \\
- & 
\left(\Delta_{\bf k}^{{\bf k}-{\bf q}_2}  \theta_{2}\right)\left(
\Delta_{\bf k}^{{\bf k}+{\bf q}_2+{\bf q}_3 + {\bf q}_4} \theta_{3} \right)\left(  
\Delta_{\bf k}^{{\bf k}-{\bf q}_3} \theta_4 \right) \nonumber \\
+ & 
\left(\Delta_{\bf k}^{{\bf k}-{\bf q}_3 }  \theta_{2} \right) \left(
\Delta_{\bf k}^{{\bf k}+{\bf q}_2+{\bf q}_3 + {\bf q}_4} \theta_{3}\right)\left(
\Delta_{\bf k}^{{\bf k}-{\bf q}_2} \theta_4 \right) \nonumber \\
- & 
\left(\Delta_{\bf k}^{{\bf k}+{\bf q}_2+{\bf q}_3 + {\bf q}_4 } \theta_{2} \right)
\left(\Delta_{\bf k}^{{\bf k}-{\bf q}_3}  \theta_{3}\right)\left(  
\Delta_{\bf k}^{{\bf k}-{\bf q}_2} \theta_4 \right) \nonumber \\
+ & 
\left(\Delta_{\bf k}^{{\bf k}-{\bf q}_2}  \theta_{2}\right)\left(
\Delta_{\bf k}^{{\bf k}-{\bf q}_3}  \theta_{3} \right) \left( 
\Delta_{\bf k}^{{\bf k}+{\bf q}_2+{\bf q}_3 + {\bf q}_4 } \theta_4 \right)\nonumber \\
- & 
\left(\Delta_{\bf k}^{{\bf k}-{\bf q}_3}  \theta_{2}\right)\left(
\Delta_{\bf k}^{{\bf k}-{\bf q}_2} \theta_{3} \right)\left(  
\Delta_{\bf k}^{{\bf k}+{\bf q}_2+{\bf q}_3 + {\bf q}_4 } \theta_4 \right) \nonumber
\Bigr\}.
\end{align}
Finally, terms with repeated $q$'s cancel, allowing us to write
\begin{align}
s_4({\bf q}_2,&{\bf q}_3,{\bf q}_4) = \int\frac{d^3{\bf k}}{(2\pi)^3}  f_{\bf k} \Bigl\{ \\
+&
\left(
{\bf q}_4 \cdot\nabla_{\bf k} \theta_{2} \right)\left(
{\bf q}_2 \cdot\nabla_{\bf k} \theta_{3}\right)\left( 
{\bf q}_3 \cdot\nabla_{\bf k} \theta_4 \right) \nonumber \\
- & 
\left(
{\bf q}_2 \cdot\nabla_{\bf k} \theta_{2}\right)\left(
{\bf q}_4 \cdot\nabla_{\bf k} \theta_{3} \right)\left(  
{\bf q}_3 \cdot\nabla_{\bf k} \theta_4 \right) \nonumber \\
+ & 
\left(
{\bf q}_3 \cdot\nabla_{\bf k} \theta_{2} \right) \left(
{\bf q}_4 \cdot\nabla_{\bf k} \theta_{3}\right)\left(
{\bf q}_2 \cdot\nabla_{\bf k} \theta_4 \right) \nonumber \\
-& 
\left(
{\bf q}_4 \cdot\nabla_{\bf k} \theta_{2} \right)\left(
{\bf q}_3 \cdot\nabla_{\bf k}  \theta_{3}\right)\left(  
{\bf q}_2 \cdot\nabla_{\bf k} \theta_4 \right) \nonumber \\
+ & 
\left(
{\bf q}_2 \cdot\nabla_{\bf k}  \theta_{2}\right)\left(
{\bf q}_3 \cdot\nabla_{\bf k} \theta_{3} \right) \left( 
{\bf q}_4 \cdot\nabla_{\bf k} \theta_4 \right)\nonumber \\
- & 
\left(
{\bf q}_3 \cdot\nabla_{\bf k} \theta_{2}\right)\left(
{\bf q}_2 \cdot\nabla_{\bf k} \theta_{3} \right)\left(  
{\bf q}_4 \cdot\nabla_{\bf k} \theta_4 \right) \nonumber
\Bigr\}.
\end{align}
This can then be expressed in the form of Eq. \ref{s4 epsilon}.   Following the same analysis as the $D=2$ case, this allows us to express the result in terms of the critical points in $E({\bf k})$ using (\ref{s4 critical}).

\section{\label{geoproof} Geometric Proof of Universal Density Correlations for $D=2$}

In Sec. \ref{sec III} and Appendix \ref{sec A}, we have argued that the $(D+1)$'th order equal-time density correlation function for a free Fermi gas in $D$-dimension obeys a universal behavior, namely
\begin{equation}\label{geoproof1}
\begin{split}
s_{D+1}({\bf q}_1, ..., {\bf q}_{D}) &\equiv \int\frac{d^D{\bf q}_{D+1}}{(2\pi)^D}
\langle \rho_{{\bf q}_1} \rho_{{\bf q}_2} ... \rho_{{\bf q}_{D+1}} 
\rangle_c \\
& = \frac{\chi_F}{(2\pi)^D} \abs{\det {\mathbb Q}}, 
\end{split}
\end{equation}
where $\rho_{\bf q}$ is defined Eq. (\ref{rhoq def}), $\abs{\det {\mathbb Q}}$ is the volume of the $D$-dimensional parallelepiped formed by ${\bf q}_i$, and $\chi_F$ is the Euler characteristic of the Fermi sea. The $D=1$ result is elementary, while for $D=2$ and $D=3$ we have presented analytic proofs in the limit ${\bf q} \rightarrow 0$. Notably, numerical evaluation of the integral in Eq. (\ref{geoproof1}) suggests that this result holds \textit{exactly} even away from the limit ${\bf q} \rightarrow 0$, as long as ${\bf q}$ is smaller than a finite momentum cut-off that characterizes the shape of Fermi sea. In this appendix, we present a geometric proof for $D=2$, in spirit of the $D=1$ proof, to establish this stronger statement.

\begin{figure}[t!]
   \includegraphics[width=0.9\columnwidth]{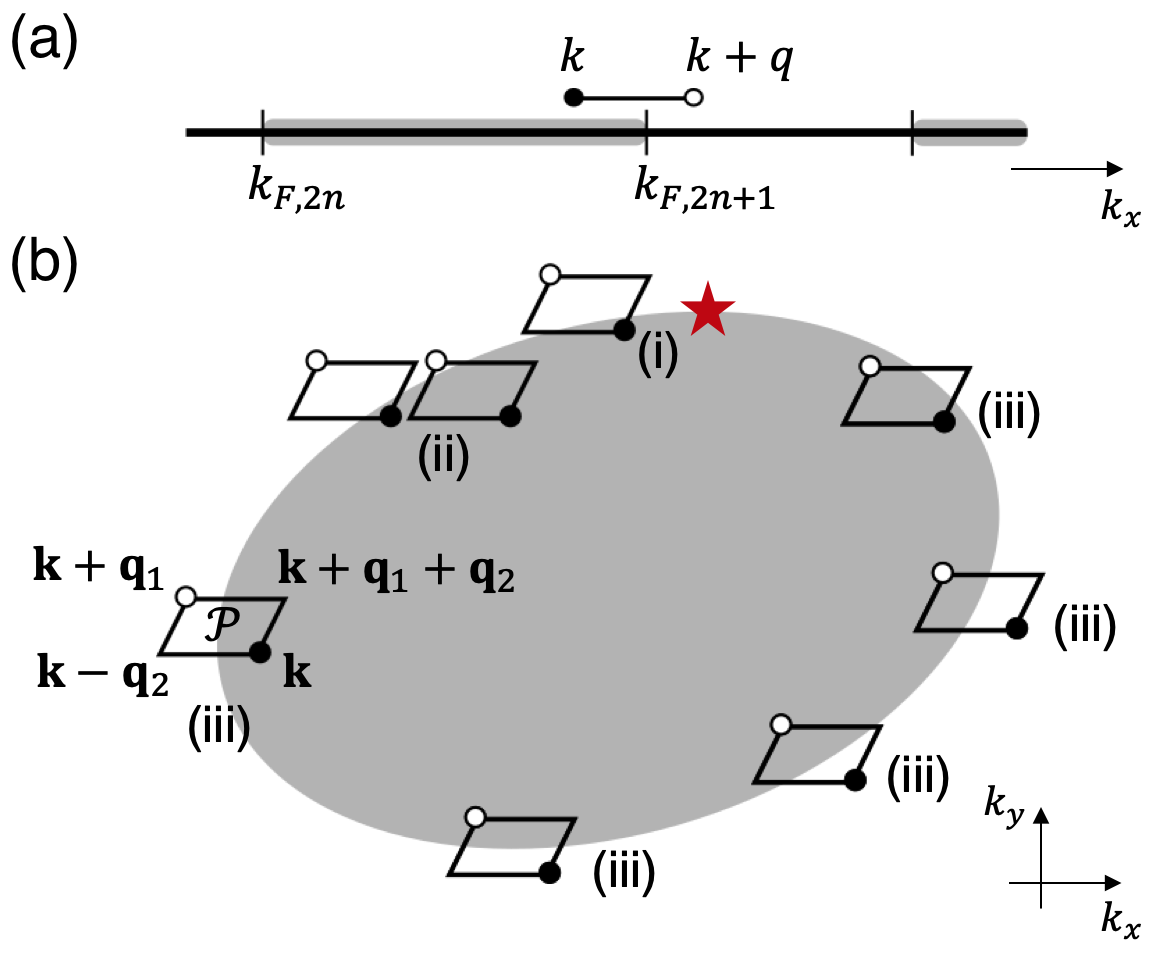}\centering
  \caption{\small{Geometric proof for the universal behavior of density correlation in (a) $D=1$ and in (b) $D=2$. The shaded region represents the filled Fermi sea. The integral involved in $s_{D+1}$ can be evaluated pictorially. For $D=1$, it is evaluated from the overlap between the interval $(k,k+q)$ and the 1D Fermi sea, with net contribution arising near Fermi points. For $D=2$, the integral is evaluated from the overlap between the parallelogram $\mathcal{P}$ and the 2D Fermi sea. By considering various placements of  $\mathcal{P}$, labeled by (i-iii), we conclude that the net contribution arises near the Fermi surface critical point marked by the red star. See the main text for details of the proof.}}
  \label{geopic1}
\end{figure}

For $D=1$, the universal relation reads
\begin{equation}\label{geo1D}
\int dk\; f_k \bar{f}_{k+q} = \chi_F \abs{q}, 
\end{equation}
where $f_k = \theta(E_F - E_k)$ is the Fermi occupation factor and $\bar{f}_k \equiv 1-f_k$. This can be easily understood by noticing that the integrand $f_k \bar{f}_{k+q}$ is 1 only when the momentum point $k$ lies within the Fermi sea while $k+q$ lies outside. The integrand is 0 otherwise. If we visualize $f_k \bar{f}_{k+q}$ as an \textit{interval}, see Fig. \ref{geopic1}(a), then the integral measures the totality of configurations for putting this interval around the boundary of Fermi sea (here the Fermi points) such that one end (with $f_k$) is inside the Fermi sea while the other end (with $\bar{f}_{k+q}$) is outside. Clearly, as long as $\abs{q}$ is smaller than the distance between any two Fermi points, the result is simply $\abs{q}$ (i.e. the length of the interval) times the number of pairs of Fermi points (i.e. $\chi_F$ for 1D Fermi sea). This establishes Eq. (\ref{geo1D}). 

For $D=2$, as explained in Sec. \ref{sec IIIA}, the universal relation asserts that
\begin{equation}\label{geoproof2}
\int d^2{\bf k} f_{\bf k} \bar{f}_{{\bf k} +{\bf q}_1} (\bar{f}_{{\bf k}+{\bf q}_1+{\bf q}_2} - f_{{\bf k}-{\bf q}_2})  = \chi_F\abs{{\bf q}_1 \times {\bf q}_2}.
\end{equation}
In spirit of the geometric argument for $D=1$, let us visualize the integrand $f_{\bf k} \bar{f}_{{\bf k} +{\bf q}_1} (\bar{f}_{{\bf k}+{\bf q}_1+{\bf q}_2} - f_{{\bf k}-{\bf q}_2})$ as a parallelogram (call it $\mathcal{P}$), see Fig. \ref{geopic1}(b). The integral measures the totality of configurations for overlapping $\mathcal{P}$ with the Fermi sea such that certain corners are inside the Fermi sea while others are outside. There are only two scenarios where the integrand $\mathcal{P}$ can be non-zero: 
\begin{equation*}
\begin{split}
\text{(A)   } &\text{when ${\bf k}$ is the only corner of $\mathcal{P}$ located \textit{inside}}\\
&\text{the Fermi sea, then $\mathcal{P} = +1$;}\\
\text{(B)   } &\text{when ${\bf k}+{\bf q}_1$ is the only corner located \textit{outside}}\\
&\text{the Fermi sea, then $\mathcal{P} = -1$.}
\end{split}
\end{equation*}
To help distinguishing these two scenarios, we label the ${\bf k}$-corner by a solid circle and the $({\bf k}+{\bf q}_1)$-corner by an open circle.

Given these two scenarios, let us look at a schematic Fermi sea depicted in Fig. \ref{geopic1}(b). For concreteness, we consider an electron-like Fermi pocket. We put the parallelogram $\mathcal{P}$ close to the Fermi surface at various locations labeled by (i-iii).  It is clear that when $\mathcal{P}$ is placed at location (iii), \textit{neither} scenario (A) nor (B) can be realized. These are locations far away from the critical point on the Fermi surface where $v_x=0$ and $\text{sgn}(v_y)=\text{sgn}(q_{1y})$ (the coordinate system is set up such that ${\bf q}_2 \parallel \hat{x}$), which is marked by the red star in the figure and shall be denoted as $\mathcal{C}$. For those configurations, $\mathcal{P}=0$. When the parallelogram is placed at location (ii), \textit{both} scenarios (A) and (B) can be realized. Nevertheless, by shifting the parallelogram along the ${\bf q}_2$-direction, one can easily check that scenario (A) is realized just \textit{as much as} (B) is, and hence the net contribution there is again zero. We are left with location (i) to examine.

\begin{figure}[t!]
   \includegraphics[width=0.95\columnwidth]{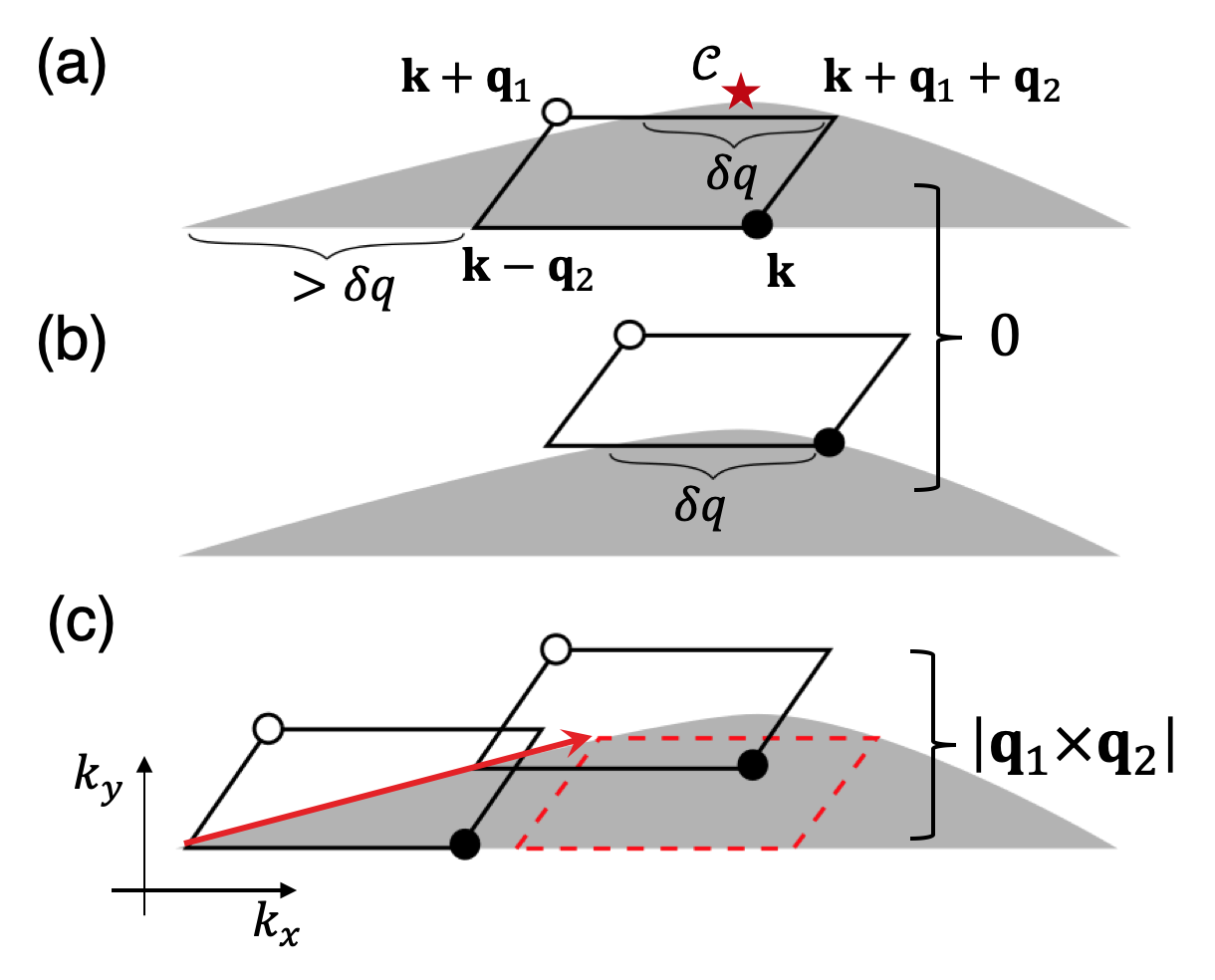}\centering
  \caption{\small{Configurations of the integrand, represented as a parallelogram $\mathcal{P}$, contributing to the density correlation when placed near a Fermi surface critical point $\mathcal{C}$. As argued in the text, each convex/concave critical point (with $v_x=0$ and $\text{sgn}(v_y) = \text{sgn}(q_{1y})$) contributes $\pm \abs{{\bf q}_1 \times {\bf q}_2}$ to the integral, leading to the $\chi_F  \abs{{\bf q}_1 \times {\bf q}_2}$ dependence of $s_3({\bf q}_1, {\bf q}_2)$.}}
  \label{geopic2}
\end{figure}

We have illustrated in Fig. \ref{geopic2} several situations in which the parallelogram $\mathcal{P}$ is placed close to a Fermi surface critical point $\mathcal{C}$, at which $v_x=0$ and $\text{sgn}(v_y)=\text{sgn}(q_{1y})$. Here what we mean by ``close" is that it is \textit{impossible} to shift the parallelogram in the $\hat{x}$-direction such that all four corners are entirely inside the Fermi sea. The dashed parallelogram correspond to the critical situation, so that if $\mathcal{P}$ is placed below it, then $\mathcal{P}$ is located at region (ii) shown in Fig. \ref{geopic1}. Thus, let us focus on the situation where $\mathcal{P}$ is located above the dashed parallelogram. Figure \ref{geopic2}(a) illustrates the situation when the top edge of $\mathcal{P}$ is slightly below the critical point $\mathcal{C}$, with ${\bf k}+{\bf q}_1+{\bf q}_2$ right at the Fermi surface and ${\bf k}+{\bf q}_1$ being the \textit{only} corner outside the Fermi sea. This realizes scenario (B) and gives $\mathcal{P}=-1$. This remains true when the parallelogram is shifted in the $-\hat{x}$-direction until ${\bf k}+{\bf q}_1+{\bf q}_2$ \textit{exits} the Fermi sea. Such a negative contribution, however, is \textit{completely canceled} by the situation depicted in Fig. \ref{geopic2}(b), which realizes scenario (A) to give $\mathcal{P}=+1$ for each such configuration.  Notice that a {\it crucial assumption} has been made: the radius of curvature on the Fermi surface near $\mathcal{C}$ has to be large enough (or ${\bf q}_{1,2}$ be small enough) such that the corner ${\bf k}+{\bf q}_1+{\bf q}_2$ exits the Fermi sea \textit{before} ${\bf k}-{\bf q}_2$ does, as indicated in Fig. \ref{geopic2}(a). More precisely, in addition to requiring $\abs{{\bf q}_{1,2}} < c R$, we also require $\abs{q_{1y}} > c' q^2/R$ (where $R$ is the radius of curvature, $c$ and $c'$ are some geometry-dependent parameters). It is this assumption that guarantees the aforementioned cancellation, and leads to the universal density correlation. Clearly, this assumption can always be satisfied for small enough yet finite $q$, as long as $q_{1y} \neq 0$ ($i.e.$ $\abs{{\bf q}_1 \times {\bf q}_2} \neq 0$). When this assumption is violated, there will be $\mathcal{O}(q^3)$ correction to Eq. (\ref{geoproof2}).

Finally, let us consider the non-vanishing contribution coming from the situation depicted in Fig. \ref{geopic2}(c). Let us first place the $({\bf k}-{\bf q}_2)$-corner right at the Fermi surface. The assumption made above guarantees that the ${\bf k}$-corner is the only corner lying inside the Fermi sea, thus realizing scenario (A) and gives $\mathcal{P}=+1$. This remains true when the parallelogram is shifted in the $-\hat{x}$-direction until the whole parallelogram exits the Fermi sea (the amount of shift is $\abs{{\bf q}_2}$). Notice that we can also shift the parallelogram in the vertical direction, as long as $\bf k$ lies between the top and bottom edges of the dashed parallelogram. We can thus slide the $({\bf k}-{\bf q}_2)$-corner along the Fermi surface, as indicated by the red arrow in the figure, and perform horizontal shift by amount up to $\abs{{\bf q}_2}$ to obtain all configurations that contribute to Eq. (\ref{geoproof2}). The integral is thus evaluated to be $\abs{{\bf q}_1 \times {\bf q}_2}$. 

In the specific case we just discussed, $\chi_F=1$, and there is just one critical point $\mathcal{C}$ on the Fermi surface, which is \textit{convex} in nature (i.e. $\partial_x v_x >0$). If there is a \textit{concave} critical point, the same reasoning presented above would conclude that there is a $\abs{{\bf q}_1 \times {\bf q}_2} \cdot (-1)$ contribution to the integral in Eq. (\ref{geoproof2}). Therefore, in the most generic case where the Fermi surface has $c$-convex critical points and $\tilde{c}$-concave critical points, and noting that $\chi_F= c-\tilde{c}$ \cite{Kane2022}, we arrive at Eq. (\ref{geoproof2}).

\section{Fourier Transform of $s_{D+1}({\bf q}_1, ..., {\bf q}_D)$}
\label{sec B}

In this section we compute the long wavelength correlation function $s_{D+1}({\bf r}_1, ...,{\bf r}_{D+1})$ by taking the Fourier transform of Eq. \ref{sd formula0},
\begin{equation}
s_{D+1} = \chi_F\int \frac{(d^D{\bf q})^D}{(2\pi)^{D(D+1)}}e^{i\sum_{a=1}^D {\bf q}_a ({\bf r}_a-{\bf r}_{D+1})} |\det {\mathbb Q}|.
\end{equation}
We will consider the cases $D=2$ and $D=3$ separately.  In each case it will be useful to write $|\det {\mathbb Q}| = (\det {\mathbb Q}) {\rm sgn}[\det {\mathbb Q}]$.   $\det {\mathbb Q}$ will act as derivatives (which will subsequently be integrated) acting on the Fourier transform of ${\rm sgn}[\det {\mathbb Q}]$.   Since $[\rho({\bf r}_a),\rho({\bf r}_b)]=0$ we expect $s_{D+1}({\bf r}_1, ...,{\bf r}_{D+1})$ to be invariant under permutations of ${\bf r}_a$.

\subsection{D=2}
\label{sec B1}

In $D=2$ we have $\det[{\mathbb Q}] = {\bf q}_1 \times {\bf q}_2$ (expressed as a 2D cross product).   It follows that we can write
\begin{equation}
s_3({\bf r}_1,{\bf r}_2,{\bf r}_3) = - \frac{\chi_F}{(2\pi)^6}(\nabla_1 \times \nabla_2) F_3({\bf r}_{13},{\bf r}_{23}),
\label{s3 total derivative}
\end{equation}
where ${\bf r}_{ab} = {\bf r}_a-{\bf r}_b$ and
\begin{equation}
F_3({\bf r}_1,{\bf r}_2,{\bf r}_3) = \int d^2{\bf q}_1 d^2{\bf q}_2 e^{i{\bf q}_1\cdot{\bf r}_{13} + {\bf q}_2 \cdot{\bf r}_{23}} {\rm sgn}( {\bf q}_1 \times {\bf q}_2 ).
\end{equation}
Let us write ${\bf q}_2 = q_{2\parallel} \hat {\bf q}_1 + q_{2\perp} \hat z \times \hat {\bf q}_1$.  Then the integral over $q_{2\parallel}$ gives $2\pi \delta(\hat{\bf q}_1 \cdot {\bf r}_{23})$, while the integral over $q_{2\perp}$ (which involves $\rm sgn q_{2\perp}$) gives $2 i/(\hat {\bf q}_1 \times {\bf r}_{23})$.   This can be written as
\begin{equation}
F_3({\bf r}_1,{\bf r}_2,{\bf r}_3) = 4\pi i \int d^2{\bf q}_1 e^{i {\bf q}_1\cdot{\bf r}_{13}}
 \frac{|{\bf q}_1|^2}{|{\bf r}_{23}|^2}
\frac{\delta({\bf q}_1 \cdot \hat {\bf r}_{23})}{{\bf q}_1 \times \hat{\bf r}_{23}}.
\end{equation}
Next write ${\bf q}_1 = q_{1\parallel} \hat {\bf r}_{23} + x_\perp {\bf r}_{23} \times \hat z$.  Then, the $\delta$-function sets $q_{1\parallel}=0$, so we obtain
\begin{equation}
F_3({\bf r}_1,{\bf r}_2,{\bf r}_3) =  4\pi i \int dx_\perp \ x_\perp e^{i x_\perp ({\bf r}_{13} \times{\bf r}_{23})} .
\end{equation}
This can be simplified by noting that without the $x_\perp$ outside the exponent the integral is $2\pi\delta(A_{123})$, where $A_{123} \equiv {\bf r}_{13}\times{\bf r}_{23} = {\bf r}_{21} \times {\bf r}_{31}$ is the area of the triangle formed by ${\bf r}_1$, ${\bf r}_2$ and ${\bf r}_3$.  The
$x_\perp$ can be generated by differentiating with respect to ${\bf r}_3$.  
Noting that $({\bf r}_{21}\times\nabla_3) A_{123} = {\bf r}_{21}\times(\hat z \times {\bf r}_{21}) = |{\bf r}_{21}|^2$ we obtain
\begin{equation}
F_3({\bf r}_1,{\bf r}_2,{\bf r}_3) = 8\pi^2 \frac{{\bf r}_{21}\times \nabla_3}{|{\bf r}_{21}|^2} \delta({\bf r}_{21} \times {\bf r}_{31}).
\end{equation}
Thus,
\begin{equation}
s_3({\bf r}_1,{\bf r}_2,{\bf r}_3) = -\frac{\chi_F}{8\pi^4} \nabla_1 \times \nabla_2
\frac{{\bf r}_{21} \times \nabla_3}{|{\bf r}_{21}|^2} \delta({\bf r}_{21} \times {\bf r}_{31}).
\label{s3(r) 2d final}
\end{equation}
Note that despite its appearance, $s_3$ is invariant under permutations of ${\bf r}_1$, ${\bf r}_2$ and ${\bf r}_3$.   It can also be written as $\chi_F \delta''(A_{123})/(8\pi^4)$.  However,  Eq. \ref{s3(r) 2d final} is a more useful form for integrating over ${\bf r}_a$.   Since the dominant contribution comes from when $A_{123}\sim 0$, ${\bf r}_1$, ${\bf r}_2$ and ${\bf r}_3$ must lie along a straight line.

\subsection{D=3}

For $D=3$ we have $\det[{\mathbb Q}] = ({\bf q}_1 \times {\bf q}_2)\cdot {\bf q}_3$, so we can write
\begin{equation}
s_4(\{{\bf r}_K\}) =  (\nabla_1 \times \nabla_2)\cdot\nabla_3 F_4(\{{\bf r}_K\}),
\label{s4 f4}
\end{equation}
where $K=1,2,3,4$ and
\begin{align}
F_4(&\{{\bf r}_K\}) = i \chi_F \int \frac{d^3{\bf q}_1 d^3{\bf q}_2 d^3{\bf q}_3}{(2\pi)^{12}} \\
&e^{i({\bf q}_1\cdot{\bf r}_{14}+{\bf q}_2\cdot{\bf r}_{24}+{\bf q}_3\cdot{\bf r}_{34})}
{\rm sgn}[({\bf q}_1 \times {\bf q}_2)\cdot{\bf q}_3] .
\nonumber
\end{align}

We now write
\begin{equation}
{\bf q}_3 = q_{3\parallel} \frac{{\bf q}_1\times{\bf q}_2}{|{\bf q}_1\times{\bf q}_2|} + {\bf q}_{3\perp},
\end{equation}
where ${\bf q}_{3\perp}$ is in the plane spanned by ${\bf q}_1$ and ${\bf q}_2$.   Then, ${\rm sgn}[({\bf q}_1\times{\bf q}_2)\cdot{\bf q}_3] = {\rm sgn} q_{3\parallel}$, so that the integral over ${\bf q}_{3\perp}$ gives $(2\pi)^2 \delta^2({\bf r}_{34} \times ({\bf q}_1\times{\bf q}_2)/|{\bf q}_1\times{\bf q}_2|)$.    Note that this is a two dimensional $\delta$-function that constrains ${\bf r}_{34} \parallel
{\bf q}_1 \times {\bf q}_2$, so that the two components of ${\bf r}_{34}$ in the plane spanned by ${\bf q}_1$ and ${\bf q}_2$ vanish.    The integral over $q_{3\parallel}$ gives $2i |{\bf q}_1\times{\bf q_2}|/ ({\bf q}_1\times{\bf q}_2)\cdot {\bf r}_{34}$.    The result of the ${\bf q}_3$ integration can thus be written as
\begin{align}
F_4(\{{\bf r}_K\}) =& \frac{-2\chi_F}{(2\pi)^{10}} \int d^3{\bf q}_1 d^3{\bf q}_2 e^{i({\bf q}_1\cdot{\bf r}_{14} + {\bf q}_2\cdot{\bf r}_{24})} \\
&\delta^2\left({\bf r}_{34} \times \frac{{\bf q}_1\times{\bf q}_2}{|{\bf q}_1\times{\bf q}_2|}\right)
\frac{|{\bf q}_1\times{\bf q}_2|}
{({\bf q}_1\times{\bf q}_2)\cdot {\bf r}_{34}}.
\nonumber
\end{align}

We next decompose ${\bf q}_1$ and ${\bf q}_2$ into components parallel and perpendicular to ${\bf r}_{34}$, by writing ${\bf q}_a = q_{a\parallel} \hat {\bf r}_{34} + {\bf q}_{a\perp}$, for $a=1,2$, where ${\bf q}_{a\perp} \cdot {\bf r}_{34} = 0$.    Using the fact that ${\bf r}_{34}\times ({\bf q}_1 \times{\bf q}_2) = {\bf q}_1 ({\bf r}_{34} \cdot {\bf q}_2) - {\bf q}_2 ({\bf r}_{34} \cdot {\bf q}_1)
= |{\bf r}_{34}| ({\bf q}_{1\perp} q_{2\parallel} - {\bf q}_{2\perp} q_{1\parallel})$, 
the $\delta$-function in the integral can be written
\begin{align}
\delta^2\left( \frac{{\bf r}_{34} \times ({\bf q}_1\times{\bf q}_2)}{|{\bf q}_1\times{\bf q}_2|}\right)
&= \frac{|{\bf q}_1\times{\bf q}_2|^2}{|{\bf r}_{34}|^2}
\delta^2({\bf q}_{1\perp} q_{2\parallel} - {\bf q}_{2\perp} q_{1\parallel})  \nonumber\\
= &\frac{|{\bf q}_{1\perp}\times{\bf q}_{2\perp}|}{|{\bf r}_{34}|^2} \delta(q_{1\parallel}) \delta(q_{2\parallel}),
\end{align}
where in the second equality we used the fact that the Jacobian of the argument of the $\delta$-function, when expressed as a function of $q_{1\parallel}$ and $q_{2\parallel}$ is
$|\partial({\bf q}_{1\perp} q_{2\parallel} - {\bf q}_{2\perp} q_{1\parallel})/\partial(q_{1\parallel},q_{2\parallel})| = |{\bf q}_{1\perp}\times{\bf q}_{2\perp}|$, and the $\delta$-function sets ${\bf q}_a = {\bf q}_{a\perp}$.

Now, using the fact that $|{\bf q}_{1\perp}\times {\bf q}_{2\perp}|^2/({\bf q}_{1\perp}\times {\bf q}_{2\perp})\cdot{\bf r}_{34} = ({\bf q}_{1\perp}\times {\bf q}_{2\perp})\cdot{\bf r}_{34}/|{\bf r}_{34}|^2$ we get
\begin{align}
F_4(\{{\bf r}_K\}) = \frac{-2\chi_F}{(2\pi)^{10}} \int& d^2{\bf q}_{1\perp} d^2{\bf q}_{2\perp}  
e^{i({\bf q}_{1\perp}\cdot{\bf r}_{14} + {\bf q}_{2\perp}\cdot{\bf r}_{24})}\nonumber\\
&\frac{{\bf r}_{34}\cdot({\bf q}_{1\perp}\times {\bf q}_{2\perp})}{|{\bf r}_{34}|^4 } ,
\end{align}
where ${\bf q}_{1\perp}$ and ${\bf q}_{2\perp}$ are integrated over the plane perpendicular to ${\bf r}_{34}$.   This can then be written as
\begin{equation}
F_4(\{{\bf r}_K\}) = \frac{2\chi_F}{(2\pi)^6}  \frac{{\bf r}_{34} \cdot (\nabla_1\times\nabla_2)}{|{\bf r}_{34}|^4}
\delta^2_\perp({\bf r}_{14}) \delta^2_\perp({\bf r}_{24}) ,
\end{equation}
where $\delta^2_\perp({\bf r}) = (2\pi)^{-2}\int d^2{\bf q}_\perp e^{i{\bf q}_\perp\cdot{\bf r}}$ is a 2D $\delta$-function for the components of ${\bf r}$ perpendicular to ${\bf r}_{34}$.  Using $\delta^2_\perp({\bf r} \times {\bf r}_{34}) = \delta^2_\perp({\bf r})/|{\bf r}_{34}|^2$ this can be written as
\begin{equation}
F_4(\{{\bf r}_K\}) = -\frac{2\chi_F}{(2\pi)^6}  \nabla_1 \cdot({\bf r}_{34} \times\nabla_2)
\delta^2_\perp({\bf r}_{14}\times{\bf r}_{34}) \delta^2_\perp({\bf r}_{24}\times{\bf r}_{34}) .
\label{f4 final}
\end{equation}

Finally, in (\ref{s4 f4}), we can substitute $\nabla_1 = -\nabla_2-\nabla_3-\nabla_4$ to trade $(\nabla_1\times\nabla_2)\cdot\nabla_3$ for $(\nabla_4\times\nabla_3)\cdot\nabla_2$, and after combining (\ref{s4 f4}) and (\ref{f4 final}) we obtain
\begin{align}
\label{s4 total derivative}
s_4(\{{\bf r}_K&\}) = -\frac{2\chi_F}{(2\pi)^6} \Bigl((\nabla_4\times\nabla_3)\cdot\nabla_2\Bigr)
\nabla_1 \cdot\Bigl( \\
&({\bf r}_{34}\times \nabla_2) \delta^2_\perp({\bf r}_{14}\times{\bf r}_{34}) \delta^2_\perp({\bf r}_{24}\times{\bf r}_{34})\Bigr) .
\nonumber
\end{align}
Note, that like (\ref{s3(r) 2d final}), this expression is invariant under permutations of ${\bf r}_1$, ${\bf r}_2$, ${\bf r}_3$ and ${\bf r}_4$.   In addition, it can be seen that ${\bf r}_1$, ${\bf r}_2$, ${\bf r}_3$ and ${\bf r}_4$ must all lie close to the same straight line.

\section{Real Space Integrals}
\label{sec C}

In this section we evaluate $\langle Q_A Q_B Q_C \rangle_c$ for $D=2$ and $\langle Q_A Q_B Q_C Q_D \rangle_c$ for $D=3$ by integrating $s_{D+1}({\bf r}_1, ...,{\bf r}_{D+1})$ over the $D+1$ regions.   We will treat the cases $D=2$ and $D=3$ separately.

\subsection{D=2}
\label{sec C1}

Here we evaluate
\begin{equation}
\langle Q_A Q_B Q_C\rangle_c = \int_{A,B,C} d^2{\bf r}_A d^2{\bf r}_B d^2{\bf r}_C s_3({\bf r}_A,{\bf r}_B,{\bf r}_C),
\end{equation}
where the regions $A$, $B$, and $C$ that partition the infinite plane meet at the origin and are separated by three rays, specified by unit vectors $\hat{\bf m}_{ab}= \hat{\bf m}_{ba}$, for $a \ne b = A,B,C$.   In addition the boundary rays define unit normals, $\hat{\bf n}_{ab} = - \hat{\bf n}_{ba}$, which points to region $a$ from region $b$.  We anticipate that the integral will diverge logarithmically with system size, so we cut off the integrals at a finite radius $r= L$.   We will see that the coefficient of the $\log^2 L$ divergence will be independent of the angles between the rays, provided all angles are less than $\pi$.   When one of the angles is larger than $\pi$ the result is modified (though it is still quantized).   A related modification of the third order response function was discussed in Ref. \onlinecite{Kane2022}.    For simplicity, here we will focus only on the case where all the angles are less than $\pi$.

Using Eq. (\ref{s3(r) 2d final}), which shows that $s_3(\{{\bf r}_K\})$ is a total derivative, we can write
\begin{equation}
\langle Q_A Q_B Q_C\rangle_c = \frac{\chi_F}{8\pi^4} I
\label{qaqbqc i}
\end{equation}
with
\begin{equation}
I = -\int_{\partial A,\partial B,\partial C} \frac{(d{\bf S}_A\times d{\bf S}_B)({\bf r}_{BA}\times d{\bf S}_C)}{|{\bf r}_{BA}|^2}
\delta({\bf r}_{BA}\times {\bf r}_{CA}),
\label{i formula}
\end{equation}
where the integrals are over the boundaries of regions $A$, $B$ and $C$, with normal length elements $d{\bf S}_{A,B,C}$.    

\begin{figure}
\includegraphics[width=2.5 in]{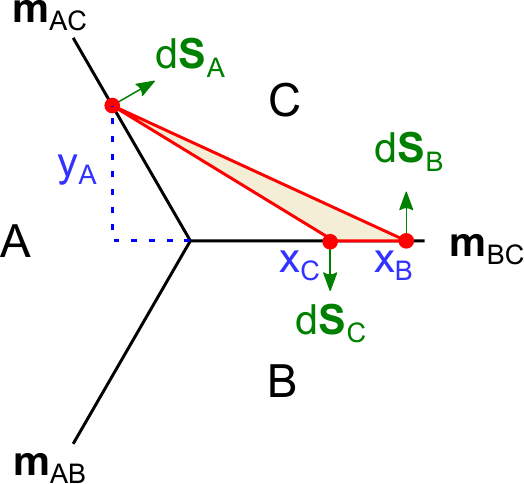}
\caption{Representative contribution to the $D=2$ integral $I$ in Eq. (\ref{i formula}), where ${\bf r}_B$ and ${\bf r}_C$ are integrated along ray $\hat{\bf m}_{BC}$, and ${\bf r}_A$ is integrated along rays $\hat{\bf m}_{AB}$ and $\hat{\bf m}_{AC}$. The $\delta$-function in the integrand requires ${\bf r}_A$, ${\bf r}_B$ and ${\bf r}_C$ to lie on the same straight line, hence placing ${\bf r}_A$ at the triple point. The remaining integrations over ${\bf r}_B$ and ${\bf r}_C$ lead to a $\log^2$-divergence.}
\label{Fig13}
\end{figure}

The argument of the $\delta$-function is the twice the area of the triangle formed by ${\bf r}_A$, ${\bf r}_B$ and ${\bf r}_C$.   It follows that provided ${\bf r}_A$, ${\bf r}_B$ and ${\bf r}_C$ are separated from each other, the integral will be dominated by straight lines in which two of the ${\bf r}'s$ are along one of the rays and the third ${\bf r}$ is at the triple point where the three regions meet.  There will be two cases: (1) ${\bf r}_B$ and ${\bf r}_C$ are integrated along ray $\hat{\bf m}_{BC}$, and ${\bf r}_A$ is integrated along rays $\hat{\bf m}_{AB}$ and $\hat{\bf m}_{AC}$, and (2) ${\bf r}_A$ and ${\bf r}_C$ are integrated along ray $\hat{\bf m}_{AC}$, and ${\bf r}_B$ is integrated along rays $\hat{\bf m}_{AB}$ and $\hat{\bf m}_{BC}$.  A third possibility, where ${\bf r}_A$ and ${\bf r}_B$ are along ray $\hat {\bf m}_{AB}$ is not present because $d{\bf S}_A \parallel d{\bf S}_B$.   Case (1) is shown in Fig. \ref{Fig13} and will be considered in detail.   Case (2) gives an identical result, so we will write $I = I_1 + I_2 = 2 I_1$.

To evaluate $I_1$, shown in Fig. \ref{Fig13}, consider Cartesian coordinates centered at the triple point with the $x$ axis along $\hat {\bf m}_{BC}$.   Then ${\bf r}_B = x_B \hat x$, ${\bf r}_C = x_C \hat x$ and ${\bf r}_A = x_A(y_A) \hat x + y_A \hat y$.   Here it is useful to parametrize the position along the boundary of region $A$ in terms of $y_A$.   $x_A(y_A)$ will depend on the angles of $\hat{\bf m}_{AB}$ and $\hat{\bf m}_{AC}$, but we will see that that dependence drops out of the answer.  The ingredients of the integral are then,
\begin{align}
&{\bf r}_{BA} = (x_B-x_A(y_A))\hat x - y_A \hat y \\
&{\bf r}_{BA} \times {\bf r}_{CA} = -(x_B - x_C) y_A.
\end{align}
The normal length elements are,
 \begin{align}
d{\bf S}_A &= \hat x dy_A - \hat y (dx_A/dy_A) dy_A \\
d{\bf S}_B &= \hat y dx_B \\
d{\bf S}_C &= - \hat y dx_C.
\end{align}
Note that $d{\bf S}_A$ depends on $x_A(y_A)$, but that dependence drops out in 
$d{\bf S}_A \times d{\bf S}_B$.   The integral then becomes
\begin{equation}
I_1 = \int_0^\infty dx_B dx_C \int_{-\infty}^\infty dy_A
\frac{x_{BA}}{x_{BA}^2 + y_A^2}\delta( x_{BC} y_A).
\label{i1 delta int}
\end{equation}
We do the integral over $y_A$ first.   The $\delta$-function fixes $y_A=0$, along with
$x_A(y_A)=0$.   We then obtain
\begin{equation}
I_1 = \int_0^\infty \frac{ dx_B dx_C }{x_B|x_B - x_C|}.
\label{i1 log int}
\end{equation}
Introducing upper cutoffs $x_B < L$ and $x_C <L$ and lower cutoffs $|x_{BC}|>a$ and
$x_B = |x_{BA}| > a$ for $a \sim k_F^{-1}$, we then obtain
\begin{align}
I_1 &= \int_a^L \frac{dx_B}{x_B} \left(\log\frac{x_B}{a} + \log\frac{L}{a}\right)\\
 &= \frac{3}{2} \log^2 \frac{L}{a}.
\label{i1 result}
\end{align}
Accounting for the multiplicative factor $2$ and (\ref{qaqbqc i}) we obtain
\begin{equation}
\langle Q_A Q_B Q_C\rangle_c = \frac{3\chi_F}{8\pi^4} \log^2\frac{L}{a}.
\label{qabc final}
\end{equation}

The reader may observe that if instead in (\ref{i1 delta int}) we do the integral over $x_C$ first, then we would obtain
\begin{align}
I_1 &= \int_0^\infty dx_B \int_{-\infty}^{\infty} dy_A \frac{x_{BA}}{x_{BA}^2+y_A^2}
\frac{1}{|y_A|}\nonumber\\
&\sim \int_{-\infty}^\infty \frac{dy_A}{|y_A|}\log \frac{L}{|y_A|}.
\label{i1 wrong int}
\end{align}
If we cut off the $y_A\rightarrow 0$ divergence by $a$, then the result would appear to disagree with (\ref{i1 result}).   The origin of this discrepancy is the choice for the lower cutoff of $y_A$.  Since (\ref{s general}) is valid for $|{\bf q}| < k_F$, Eq. \ref{s3(r) 2d final} is valid for $|{\bf r}_a - {\bf r}_b| > a \sim k_F^{-1}$.   Therefore, it is appropriate that the lower cutoffs for $x_B$ and $|x_B-x_C|$ in (\ref{i1 log int}) is $a$.   However, the lower cutoff for $y_A$ in (\ref{i1 wrong int}) is not set by $a$.   If we introduce a finite width to the junction region, so that $y_B=0$, $y_C \sim a$, then after integrating $x_C$ one finds the range of integration for $y_A$ is $a < y_A < L$ and $-L < y_A < - x_B a/L$, which when incorporated into (\ref{i1 wrong int}) reproduces (\ref{i1 result}).   This subtlety does not arise for the integration of $x_B$ and $x_C$ in (\ref{i1 log int}).

\subsection{D=3}
\label{sec C2}

We now evaluate
\begin{equation}
\langle Q_A Q_B Q_C Q_D\rangle_c = \int_{A,B,C,D} d^{12}{\bf r}_{A,B,C,D}  s_4(\{{\bf r}_K\}),
\end{equation}
where the four regions $K=A$, $B$, $C$ and $D$ that partition the 3D volume meet at the origin and are specified by four triple contact rays $\hat{\bf m}_{abc}$, symmetric in $a \ne b \ne c = A,B,C,D$.  The boundary plane separating region $a$ from region $b$ is then normal to $\hat {\bf m}_{abc} \times \hat{\bf m}_{abd}$, where $c\ne d\ne a,b$.   As in the previous section, we will assume for simplicity that all four regions are concave, and subtend a solid angle less than $2\pi$.   Given that condition, the result is independent of the directions of $\hat{\bf m}_{abc}$.

Eq. (\ref{s4 total derivative}) shows that $s_4(\{{\bf r}_K\})$ is a total derivative.  Identifying ${\bf r}_{A,B,C,D}$ with ${\bf r}_{4,3,2,1}$ in (\ref{s4 total derivative}) we can write
\begin{equation}
\langle Q_A Q_B Q_C Q_D\rangle_c = -\frac{\chi_F}{32\pi^6} I
\label{qaqbqcqd i}
\end{equation}
with
\begin{align}
\label{i4 formula}
I = \int \bigl[(d&{\bf S}_A \times d{\bf S}_B)\cdot d{\bf S}_C\bigr]\bigl[(d{\bf S}_D \times {\bf r}_{BA})\cdot \nabla_C \bigr]\nonumber\\
&\delta^2_\perp({\bf r}_{DA}\times {\bf r}_{BA})\delta^2_\perp({\bf r}_{CA}\times {\bf r}_{BA}).
\end{align}
Here ${\bf r}_a$ ($a=A,B,C,D$) is integrated over the boundary of region $a$ with outward normal area element $d{\bf S}_a$, and $\delta^2_\perp({\bf r})$ is the $2D$ delta-function for the two components of ${\bf r}$ perpendicular to ${\bf r}_{BA}$.

\begin{figure}
\includegraphics[width=3.2 in]{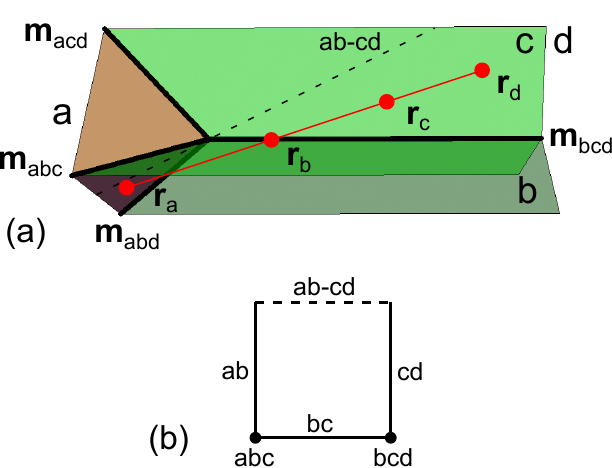}
\caption{(a) Representative contribution to the $D=3$ integral $I$ in Eq. (\ref{i4 formula}), with ${\bf r}_a$ ($a=A,B,C,D$) integrated over the boundary of region $a$. The $\delta$-functions in the integrand requires ${\bf r}_{A,B,C,D}$ to lie on a straight line (the red line in figure), hence two of the ${\bf r}$'s are the same boundary plane. Given a fixed ordering $abcd$, its contribution to $I$ consists of all possible orientation of the $abcd$-line such that it is parallel to one of the boundary. (b) Schematic diagram for a one-parameter family of lines with fixed distances between ${\bf r}_a$'s. Vertex $bcd$ represents the case when the $abcd$-line orients along the triple contact ${\hat{\bf m}}_{bcd}$; Edge $cd$ represents the case when the $abcd$-line rotates on the double contact plane $cd$ (shown in (a)); Dashed edge $ab-cd$ represents the case when the $abcd$-line lies on both the $ab$ and $cd$ plane.}
\label{Fig14}
\end{figure}

Due to the $\delta$-functions, the integrand will be nonzero when ${\bf r}_{A,B,C,D}$ are close to a straight line that visits all four boundaries.   Provided the points are separated from one another, this is only possible if two of ${\bf r}_{A,B,C,D}$ share the same boundary plane.   The configurations of such straight lines can be described by considering an ordering $abcd$ along the line that is a permutation of $ABCD$, and fixing the distances $r_{ab}$, $r_{bc}$ and $r_{cd}$ shown in Fig. \ref{Fig14}(a).   The resulting one parameter family of lines then has the structure shown in Fig. \ref{Fig14}(b).    Starting near the $bcd$ triple contact line in which ${\bf r}_a$ is at the origin, the line can continuously rotate into the $cd$ double contact plane.   (This configuration is shown in Fig. \ref{Fig14}(a).)  When the line reaches the $ab-cd$ line (which is the line common to the $ab$ and $cd$ planes, shown by the dashed line in Fig. \ref{Fig14}) the line can rotate no further because point ${\bf r}_b$ has reached the origin.  At that point, the line can slide along the $ab-cd$ line (keeping $r_{ab}$, $r_{bc}$ and $r_{cd}$ fixed) until ${\bf r}_c$ reaches the origin.   Then the line can tilt into the $ab$ plane and rotate until it reaches the $abc$ triple line (with ${\bf r}_d$ at the origin).   Then the line can tilt into the $bc$ plane and rotate until it gets back to the starting point at the $bcd$ triple line with ${\bf r}_a$ at the origin.

\begin{table}
\centering
\begin{ruledtabular}
\begin{tabular}{ccccccc}
$abcd$ & $c^{ab}_{abcd}$ &  $c^{abc}_{abcd}$ & $c^{bc}_{abcd}$ & $c^{bcd}_{abcd}$ &  $c^{cd}_{abcd}$ & $c^{\rm total}_{abcd}$  \\
\hline
$ABCD$ & $0$ & $0$ & $0$ & $1$ & $0$ & $1$ \\
$ABDC$ & $0$ & $0$ & $1$ & $0$ & $0$ & $1$ \\
$CABD$ & $0$ & $0$ & $0$ & $0$ & $1$ & $1$ \\
$CADB$ & $0$ & $1/6$ & $1/3$ & $0$ & $1/2$ & $1$ \\
$ACBD$ & $0$ & $0$ & $0$ & $1/3$ & $2/3$ & $1$ \\
$ACDB$ & $0$ & $1/3$ & $0$ & $0$ & $2/3$ & $1$
\end{tabular}
\end{ruledtabular}
\caption{Summary of logarithmic divergence $c\cdot \log^3\Lambda$, contributed from the straight line configuration with ordering $abcd$ (as permutation of $ABCD$). For instance, $c^{abc}_{abcd}$ is the coefficient of $\log^3$-divergence contributed by orienting the $abcd$-line along the triple contact $\hat{\bf m}_{abc}$. The superscript of $c$ labels a specific part of the one-parameter family in Fig. {\ref{Fig14}(b)}.} 
\label{itable}
\end{table}

Since the triple product $d{\bf S}_A \cdot (d{\bf S}_B \times d{\bf S}_C)$ vanishes when any pair of $d{\bf S}_{A,B,C}$ are parallel, the segments along the line $ab-cd$ (the dashed line in Fig. \ref{Fig14}(a,b)) will vanish.   Moreover, only the segments on the planes that include ${\bf r}_D$ will contribute to the integral.   For $d=D$ only the segment along plane $cD$ contributes, while for $c =D$, the $bD$ and the $Dd$ planes contribute.   Since $dcba$ describes the same line as $abcd$, the $a=D$ and $b=D$ cases are redundant.  
 
In the following we will show that integral in the double contact planes is an integral of the form $\int_R dx dy \delta'(x-y)$, where $R$ is a finite, bounded region.   Contribution to the integral from the interior of $R$ vanishes, but the contribution from the boundary $\partial R$ must be evaluated carefully and consistently.  The boundaries for the planar integrals will involve the $abc$ or $bcd$ triple contact lines and the $ab-cd$ lines.   Since the triple contact lines involve the intersection of multiple double contact planes, a separate calculation near those lines is necessary.   This will be done in Section \ref{sec C2A}, where the contribution from lines in which ${\bf r}_D$ is within a small parameter $\epsilon$ from the triple contact line.   This will then be matched with a second calculation (in Section \ref{sec C2B}) on the double contact planes (with ${\bf r}_D$ further than $\epsilon$ from the triple contact lines), in which all of the boundary terms can be correctly accounted for.

The results of those calculations are summarized in Table \ref{itable}, which for each permutation $abcd$ of $ABCD$ shows the contributions from each part of the square in Fig. \ref{Fig14}(b).   Since the lines $abcd$ and $dcba$ are equivalent there are only 12 independent permutations.   $6$ permutations are shown in Table \ref{itable}.  Since (\ref{i4 formula}) is invariant under the interchange of $A$ and $B$, the other $6$ are obtained by interchanging $A$ and $B$.  Each of the contributions to $I$ has a logarithmic divergence of the form $c\cdot \rm log^3 \Lambda$.   The coefficient $c$ for each term is shown, and for each ordering $abcd$ the sum $c^{\rm total}_{abcd}$ is shown in the right column.   It can be observed that for each ordering, $c^{\rm total}_{abcd}=1$, so that combining the 12 independent contributions leads to
\begin{equation}
I = 12 \log^3 \Lambda.
\end{equation}
It then follows that
\begin{equation}
\langle Q_A Q_B Q_C Q_D \rangle_c = -\frac{3 \chi_F}{8 \pi^6} \log^3 \Lambda.
\label{qabcd final}
\end{equation}
This is the final result.   We note that (\ref{qaqb log},\ref{qabc final},\ref{qabcd final}) exhibit a regular pattern, which suggests that in general dimensions
\begin{equation}
\langle Q_{A_1} ... Q_{A_{D+1}} \rangle_c = \chi_F \cdot (-1)^D  \frac{(D+1)!}{(2\pi)^{2D}} \log^{D} \Lambda.
\end{equation}

In the remainder of this section we will explain the calculations of the entries in Table \ref{itable}.   Section \ref{sec C2A} details the calculation near the triple contact lines, and Section \ref{sec C2B} explains the calculation in the double contact planes.

\subsubsection{Contribution from triple contact lines}
\label{sec C2A}

\begin{figure}
\includegraphics[width=2.5 in]{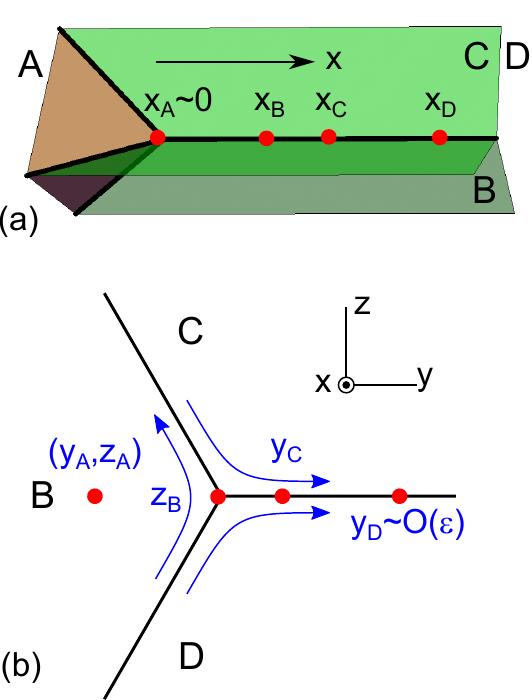}
\caption{Coordinate system for calculating the contribution to $I$ when ${\bf r}_{A,B,C,D}$ are very close to the triple contact line $BCD$ (where $y=z=0$).}
\label{Fig15}
\end{figure}

Here we consider the contribution to (\ref{i4 formula}) from configurations where ${\bf r}_{A,B,C,D}$ are very close to one of the four triple lines $BCD$, $ACD$, $ABD$ and $ABC$.   We will consider the $BCD$ line in detail and discuss the other three at the end.  For the $BCD$ line, as shown in Fig. \ref{Fig15}, ${\bf r}_A$ will be close to the origin, while ${\bf r}_{B,C,D}$ are distributed along the line at positions $x_{B,C,D}$.   We will consider all possible orderings of $x_{B,C,D}$ along the line, so this calculation will give the entries $c^{BCD}_{Abcd}$ in Table \ref{itable}, with $bcd$ accounting for the 6 permutations of $BCD$.

We organize the integral by writing
\begin{equation}
I^{BCD} = \int dx_B dx_C dx_D \tilde I(x_B,x_C,x_D),
\label{ibcd to tilde i}
\end{equation}
where $\tilde I(x_B, x_C,x_D)$ includes integrals over the remaining parameters, which describe deviations of ${\bf r}_{A,B,C,D}$ from the triple line.   These can be parametrized by the independent variables $y_A, z_A, z_B, y_C$ and $y_D$, as shown in Fig. \ref{Fig15}.     We thus write,
\begin{align}
{\bf r}_A &= x_A(y_A,z_A) \hat x + y_A \hat y + z_A \hat z \nonumber\\
{\bf r}_B &= x_B \hat x + y_B(z_B) \hat y + z_B \hat z \nonumber\\
{\bf r}_C &= x_C \hat x + y_C \hat y + z_C(y_C) \hat z
\label{rabcd parametrize}\\
{\bf r}_D &= x_D \hat x + y_D \hat y + z_D(y_D) \hat z,\nonumber
\end{align}
where $x_A(y_A,z_A)$, $y_B(z_B)$, $z_C(y_C)$ and $z_D(y_D)$ depend on the contact angles.
Note that the independent variables can all be positive or negative. For example, when $y_D<0$ the boundary of region $D$ deviates from the $z=0$ plane with a slope $dz_D/dy_D$.
Thus, all configurations close to the triple line are accounted for.
We will limit the integral to configurations close to the triple line by only integrating $y_D$ between $-\epsilon$ and $\epsilon$.   We integrate $y_A, z_A, z_B$ and $y_C$ from $-\infty$ to $\infty$, but since the integrand is only non-zero when all of the points are on a straight line, ${\bf r}_{A,B,C}$ will also be within $\epsilon$ of the triple line.   
We will consider $x_B$, $x_C$ and $x_C$ to be of order $1$ and separated by a finite distance larger than $k_F^{-1}$ and work to lowest order in $\epsilon$.   Divergent integrals when $x_{A,B,C,D}$ approach one another will be cut off by $k_F^{-1}$.   All possible orderings of $x_{B,C,D}$ will be considered.   The case $0<x_B<x_C<x_D$ is shown in Fig. \ref{Fig15}.

The $\delta$-functions in (\ref{i4 formula}) ensure that ${\bf r}_{A,B,C,D}$ lie along a straight line, so that the deviations are arranged proportionately, 
\begin{equation}
y_{BA}:y_{CA}:y_{DA} = z_{BA}:z_{CA}:z_{DA} = x_B:x_C:x_D.
\label{straight line}
\end{equation} 
It will be useful to express the $\delta$-functions in a slightly different form.     We can write,
\begin{align}
\delta^2_\perp({\bf r}_{BA} &\times {\bf r}_{CA}) \delta^2_\perp({\bf r}_{BA} \times {\bf r}_{DA})  \nonumber\\
\label{new delta fun}
&= \delta^2_\perp({\bf r}_{BA} \times {\bf r}_{CB}) \delta^2_\perp({\bf r}_{BA} \times {\bf r}_{DB}) \\
&= \frac{x_{DB}^2}{x_{B}^2}
\delta^2_\perp({\bf r}_{DB} \times {\bf r}_{CB}) \delta^2_\perp({\bf r}_{BA} \times {\bf r}_{DB}). \nonumber
\end{align} 
The first equality follows because ${\bf r}_{CA} = {\bf r}_{CB} + {\bf r}_{BA}$.  The second equality follows because the second $\delta$-function enforces ${\bf r}_{BA} \parallel {\bf r}_{DB}$, so that ${\bf r}_{BA} = {\bf r}_{DB} |{\bf r}_{BA}|/|{\bf r}_{DB}|$.  The constant factor can then be extracted from the $\delta$-function giving the factor $(|{\bf r}_{DB}|/|{\bf r}_{BA}|)^2 = x^2_{DB}/x^2_{B} + O(\epsilon)$.   The $\delta$-functions have the explicit form,
\begin{align}
\delta^2_\perp({\bf r}_{DB} \times {\bf r}_{CB}) = 
&\delta(x_{CD} y_B+ x_{DB} y_C+x_{BC} y_{D}) \nonumber\\
&\delta(x_{CD} z_B + x_{DB} z_C+x_{BC} z_D) \label{dbcb delta}\\
\delta^2_\perp({\bf r}_{BA} \times {\bf r}_{DB}) = 
&\delta(x_{DB}y_A - x_D y_B + x_B y_D)\nonumber \\
& \delta(x_{DB} z_A - x_D z_B + x_B z_D).  \label{badb delta}   
\end{align}

Given (\ref{rabcd parametrize}), the perpendicular area elements are,
\begin{align}
d{\bf S}_A &= dy_A dz_A (\hat x - \hat y \partial x_A/\partial y_A - \hat z \partial x_A/\partial z_A) \nonumber\\
d{\bf S}_B &= dx_B dz_B (\hat y - \hat z \partial y_B/\partial z_B) \nonumber\\
d{\bf S}_C &= - dx_C dy_C (\hat z - \hat y \partial z_C/\partial y_C) \\
d{\bf S}_D &= dx_D dy_D (\hat z - \hat y \partial z_D/\partial y_D).\nonumber
\end{align}
We can therefore write,
\begin{equation}
d{\bf S}_A \cdot (d{\bf S}_B \times d{\bf S}_C) =
- \Upsilon dy_A dz_A dx_B dz_B dx_C dy_C 
\label{triple volume}
\end{equation}
with
\begin{equation}
\Upsilon(z_B,y_C) \equiv
\left(1 - \frac{\partial y_B}{\partial z_B}\frac{\partial z_C}{\partial y_C}\right).
\label{upsilon}
\end{equation}
To leading order in $\epsilon$ we have
\begin{equation}
d{\bf S}_D \times {\bf r}_{BA} = x_B dx_D dy_D \left( \hat y + \hat z \frac{\partial z_D}{\partial y_D}\right).
\label{dsd times rba}
\end{equation}

The derivative $\nabla_C$ acting on the $\delta$-function will give
\begin{align}
\nabla_C &\delta^2_\perp({\bf r}_{DB} \times {\bf r}_{CB}) =  x_{DB} \Bigl( \label{nablacdelta}\\
&\hat y \delta' (x_{DB} y_{CB} - x_{CB} y_{DB}) \delta(x_{DB} z_{CB} - x_{CB} z_{DB}) \nonumber\\
 + &\hat z \delta (x_{DB} y_{CB} - x_{CB} y_{DB})  \delta'(x_{DB} z_{CB} - x_{CB} z_{DB}) \Bigr). \nonumber
\end{align}
Combining (\ref{nablacdelta}) and (\ref{dsd times rba}), it can be seen that the result can be expressed as a derivative with respect to the independent variable $y_D$:
\begin{align}
(d&{\bf S}_D \times {\bf r}_{BA})\cdot\nabla_C \delta^2_\perp({\bf r}_{DB} \times {\bf r}_{CB})=  \nonumber\\
& - dx_D dy_D\frac{x_B x_{DB}}{x_{CB}} \frac{\partial}{\partial y_D} 
\delta^2_\perp({\bf r}_{DB} \times {\bf r}_{CB}).
\label{dsd term}
\end{align}

We now have all of the ingredients for the $BCD$ triple line contribution to (\ref{i4 formula}).  Combining (\ref{ibcd to tilde i},\ref{new delta fun},\ref{triple volume},\ref{dsd term}) we obtain
\begin{align}
\tilde I(&x_B,x_C,x_D) = \int dy_A dz_A dz_B dy_C dy_D \Upsilon(z_B,y_C) \nonumber\\
&\frac{x_{DB}^3}{x_B x_{CB}}\left(\frac{\partial}{\partial y_D} \delta^2_\perp({\bf r}_{DB}\times{\bf r}_{CB})\right) 
\delta^2_\perp({\bf r}_{BA}\times{\bf r}_{DB}).
\label{ycyd integral}
\end{align}
We next note that the integrals over $y_A$ and $z_A$ can be evaluated using (\ref{badb delta}).   The result is then the integral of a total derivative with respect to $y_D$, which may be written as
\begin{equation}\label{tilde i initial}
\tilde I = \int_{-\epsilon}^\epsilon dy_D \frac{\partial K(y_D)}{\partial y_D} = K(y_D = +\epsilon) - K(y_D = -\epsilon)
\end{equation}
with
\begin{align}
K(x_B,x_C&,x_D; y_D) = \frac{x_{DB}}{x_B x_{CB}} \int dz_B dy_C \nonumber\\
&\Upsilon(z_B,y_C) \delta^2_\perp({\bf r}_{BA}\times{\bf r}_{DB}).
\end{align}
The result of the integral over $z_B$ and $y_C$ will depend on whether for a given $x_B, x_C, x_D$ and $y_D$ there exists a solution to the ``straight line condition", (\ref{straight line}).  This, in turn, will depend on ${\rm sgn} y_D$, as well as what order $x_B, x_C$ and $x_D$ are in.   If there is a solution $(z_B^*,y_C^*)$, then the integral can be evaluated using (\ref{dbcb delta}).  The Jacobian for the integration of (\ref{dbcb delta}) with respect to $y_C$ and $z_B$ is $x_{DB}x_{CD} \Upsilon(z_B,y_C)$, with $\Upsilon$ given in (\ref{upsilon}).  We then obtain
\begin{equation}
K(x_B,x_C,x_D; y_D) = 
\frac{{\rm sgn}\left[\Upsilon(z_B^*, y_C^*) x_B  x_{DB} x_{CB}\right] }{| x_B x_{CB} x_{DC} |}.
\end{equation}
If there is no solution, then the contribution is zero.

\begin{figure}
\includegraphics[width=3 in]{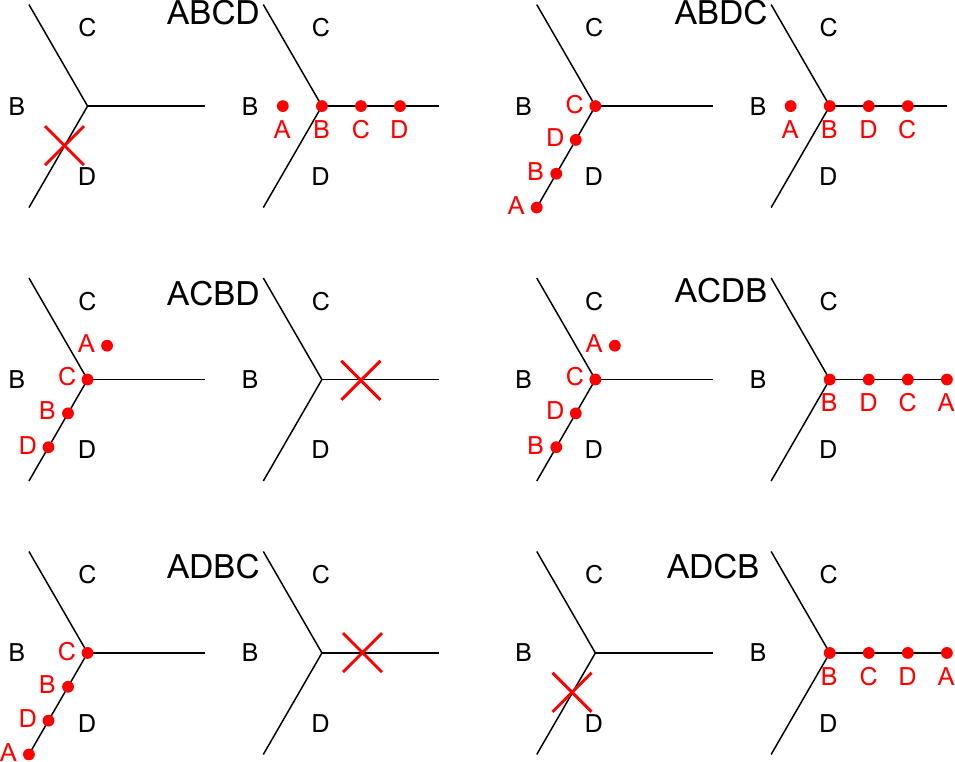}
\caption{Existence and non-existence of solution to the straight-line condition in (\ref{straight line}). For a given ordering $abcd$, the left panel shows a solution (if exists) with $y_D<0$ and the right panel shows a solution (if exists) with $y_D>0$. For instance, with $abcd=ABCD$, straight-line solutions are allowed only when $y_D>0$. This analysis leads us from (\ref{tilde i initial}) to (\ref{tilde i final})}. 
\label{Fig16}
\end{figure}

Fig. \ref{Fig16} shows the solutions for the six possible orderings of $x_B, x_C$ and $x_D$ for $y_D<0$ and $y_D>0$.   For example for the first entry, $ABCD$, which considers $x_A < x_B < x_C < x_D$, there is a solution for $y_D >0$, but there is no solution for $y_D <0$.   This ordering, along with $ADCB$ therefore contributes with a plus sign.  For $ACBD$ and $ADBC$, there is a solution for $y_D <0$, but no solution for $y_D>0$, so that ordering contributes with a minus sign.  However, note that for those (and only those) contributions ${\rm sgn}[x_{DB} x_{CB}]=-1$, which cancels the minus sign.
 For $ABDC$ and $ACDB$, there is a solution for both $y_D>0$ and $y_D<0$, so those contributions cancel.   In addition, for each of the contributions ${\rm sgn}\Upsilon(z_B^*,y_C^*) = +1$.
The result of this analysis is therefore,
\begin{equation}
\tilde I(x_B,x_C,x_D) = 
\frac{\theta_{ABCD} + \theta_{ACBD} + \theta_{ADBC} + \theta_{ADCB}}{| x_B x_{CB} x_{DC} |}.
\label{tilde i final}
\end{equation}
where $\theta_{abcd}$ specifies the ordering $x_a< x_b < x_c < x_d$.

We are now ready to perform the final integration over $x_B$, $x_C$ and $x_D$.  It is simplest to define new variables that for a given ordering describe the non-negative intervals between consecutive $x$'s.   For example, for $\theta_{ABCD}$ we define $u_1 = x_{BA} = x_B$, $u_2 = x_{CB}$ and $u_3 = x_{DC}$.   This analysis then gives
\begin{align}
I^{BCD}_{ABCD} &= \int_0^\infty \frac{du_1 du_2 du_3}{u_1 u_2 u_3}, \\
I^{BCD}_{ACBD} &= \int_0^\infty \frac{du_1 du_2 du_3}{(u_1+u_2)u_2(u_2+u_3)}, \\
I^{BCD}_{ADBC} &= \int_0^\infty \frac{du_1 du_2 du_3}{(u_1+u_2)u_3(u_2+u_3)}, \\
I^{BCD}_{ADCB} &= \int_0^\infty \frac{du_1 du_2 du_3}{(u_1+u_2+u_3)u_3 u_2}, \\
I^{BCD}_{ABDC} &= I^{BCD}_{ACDB} = 0.
\end{align}
To extract the leading logarithmic divergence of the integrals, we cut off the integrals at large distance by the system size $L$ and at short distance by $k_F^{-1}$.   This leads to
\begin{equation}
I^{BCD}_{abcd} = c^{BCD}_{abcd} \log^3 k_F L
\label{ibcdlog}
\end{equation}
with 
\begin{equation}
\begin{array}{ll}
c^{BCD}_{ABCD} = 1, \quad &c^{BCD}_{ABDC} = 0, \\
c^{BCD}_{ADBC} = 1/6, \quad  &c^{BCD}_{ADCB} = 1/3, \\
c^{BCD}_{ACBD} = 1/3, \quad  &c^{BCD}_{ACDB} = 0.  
\end{array}
\end{equation}

We now briefly mention the contributions from the other three triple contact lines and conclude the analysis of (\ref{i4 formula}).    Since (\ref{i4 formula}) is symmetric under the interchange of $A$ and $B$ the contribution from the $ACD$ triple line will be 
\begin{equation}
\begin{array}{ll}
c^{ACD}_{BACD} = 1, \quad & c^{ACD}_{BADC} = 0,  \\
c^{ACD}_{BDAC} = 1/6, \quad &c^{ACD}_{BDCA} = 1/3, \\
c^{ACD}_{BCAD} = 1/3, \quad &c^{ACD}_{BCDA} = 0. 
\end{array}
\end{equation}

For the $ABC$ triple line, the area elements $d{\bf S}_{A,B,C}$ will all lie in the plane perpendicular to ${\bf m}_{ABC}$, so $d{\bf S}_A \cdot (d{\bf S}_B \times d{\bf S}_C) = 0$.   Thus, 
\begin{equation}
c^{ABC}_{Dbcd} = 0.
\end{equation}

Finally, for $I_{ABD}$, ${\bf r}_C$, which will be at the end of the line, will have an unconstrained integral over the directions perpendicular to ${\bf m}_{ABD}$.  Due to the derivative $\nabla_C$, this will involve an unconstrained integral over the derivative of a $\delta$-function and will therefore also vanish.   We conclude that 
\begin{equation}
c^{ABD}_{Cbcd} = 0.
\end{equation}
This completes the evaluation of all of the triple contact line entries in Table \ref{itable}.

\subsubsection{Contribution from double contact planes}
\label{sec C2B}

\begin{figure}
\includegraphics[width=2.2 in]{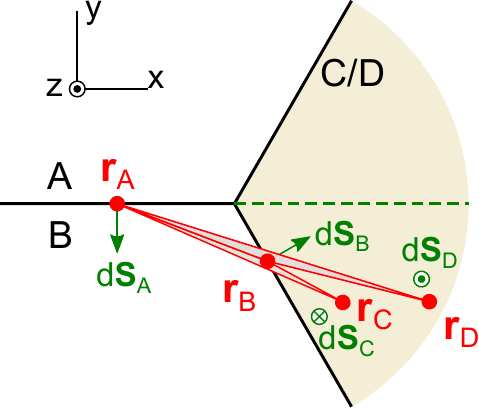}
\caption{Coordinate system for calculating the contribution to $I$ when ${\bf r}_{A,B,C,D}$ form a straight line on the $CD$ double contact plane, where points are ordered as $ABCD$. The origin is set at the point where the three rays (separating different regions) meet, and the $CD$ plane is at $z=0$. }
\label{Fig17}
\end{figure}

Here we evaluate (\ref{i4 formula}) when ${\bf r}_D$ and one other of ${\bf r}_{A,B,C}$ is in a double contact plane.  To be specific, we  consider the ordering $abcd = ABCD$, in the $CD$ contact plane, as shown in Fig. \ref{Fig17}.   This will give us the coefficient $c^{CD}_{ABCD}$ in Table \ref{itable}.   The other possible orderings will be discussed at the end.

We choose a coordinate system in which the $CD$ plane is in the plane $z=0$, and the $x$ axis points along the $AB-CD$  line.   Then we can parametrize the points on the boundary as
\begin{align}
{\bf r}_A &= x_A \hat x + y_A(z_A) \hat y + z_A \hat z \nonumber\\
{\bf r}_B &= x_B(y_B,z_B) \hat x + y_B \hat y + z_B \hat z \nonumber\\
{\bf r}_C &= x_C \hat x + y_C \hat y \label{rabcd def}  \\ 
{\bf r}_D &= x_D \hat x + y_D \hat y,\nonumber
\end{align}
where each surface is parametrized in terms of two coordinates, and $y_A(x_A)$ and $x_B(y_B,z_B)$ depend on the contact angles.   Here we have used the fact that when $x_B <0$ (so that ${\bf r}_A$ and ${\bf r}_B$ are both on the $AB-CD$ line) the area element $d{\bf S}_A \cdot (d{\bf S}_B \times d{\bf S}_C)$ will vanish.   This allows us to consider $y_B$ to be the independent variable that is subject to the constraint $y_B <0$.    This defines the boundary of the integration domain at the $AB-CD$ line.   The other boundary, near the $BCD$ triple contact line, is set by the analysis in the previous section:  we require that ${\bf r}_D$ be further than $\epsilon$ from the $BCD$ line.   Note that due to the straight line condition, the integrand will vanish at the boundaries of the ${\bf r}_C$ integral, so the $x_C$ and $y_C$ integrals will be unconstrained.  

The perpendicular area elements are,
\begin{align}
d{\bf S}_A &= - dx_A dz_A (\hat y - \hat z \partial y_A/\partial z_A ) \nonumber\\
d{\bf S}_B &= dy_B dz_B (\hat x - \hat y \partial x_B/\partial y_B - \hat z \partial x_B/\partial z_B) \nonumber\\
d{\bf S}_C &= - dx_C dy_C \hat z \\
d{\bf S}_D &= dx_D dy_D \hat z.\nonumber
\end{align}
It can be observed that $d{\bf S}_A \cdot (d{\bf S}_B \times d{\bf S}_C) =  - dx_A dz_A dy_B dz_B dx_C dy_C$ is independent of the contact angles.
In addition,
\begin{equation}
(d{\bf S}_D \times {\bf r}_{BA})\cdot \nabla_C
= d^2{\bf r}^\perp_D {\bf r}^\perp_{BA}\times \nabla_C,
\end{equation}
where ${\bf r}^\perp_{BA}$ gives the components of ${\bf r}_B - {\bf r}_A$ in the $xy$ plane, 
$d^2{\bf r}^\perp_D = dx_D dy_D$, and the right hand side uses the scalar 2D cross product.

The $\delta$-functions in (\ref{i4 formula}) are
\begin{align}
\delta^2_\perp({\bf r}_{BA} \times {\bf r}_{CA}) &= 
\delta({\bf r}_{BA}^\perp \times {\bf r}_{CA}^\perp) \delta(z_{BA} r_{CA} + z_A
r_{BA}) \nonumber\\
\delta^2_\perp({\bf r}_{BA} \times {\bf r}_{DA}) &= 
\delta({\bf r}_{BA}^\perp \times {\bf r}_{DA}^\perp) \delta(z_{BA} r_{DA} + z_A
r_{BA}) ,
\label{2d delta functions}
\end{align}
where $r_{BA} = |{\bf r}^\perp_{BA}|$.

We now combine (\ref{i4 formula},\ref{rabcd def}-\ref{2d delta functions}).   The $\delta$-functions involving $z_{A,B}$ fix $z_A=z_B=0$ with Jacobian $r_{BA}(r_{CA}-r_{DA})$.
Integrating $z_A$ and $z_B$ leads to an integral in the $xy$ plane,
\begin{align}
I = \int dy_A dx_B &d^2{\bf r}^\perp_C d^2{\bf r}^\perp_D
\frac{{\bf r}^\perp_{BA} \times \nabla_C}{|r_{BA}(r_{CA}-r_{DA})|} \nonumber\\
&\delta({\bf r}_{BA}^\perp \times {\bf r}_{CA}^\perp) \delta({\bf r}_{BA}^\perp \times {\bf r}_{DA}^\perp) 
\end{align}
It is useful to consider the integrals in polar coordinates with origin ${\bf r}^\perp_B$.   Let
\begin{align}
{\bf r}^\perp_C &= {\bf r}^\perp_B + r_{CB} (\cos \theta_{CB},\sin \theta_{CB}) ,\\
{\bf r}^\perp_D &= {\bf r}^\perp_B + r_{CB} (\cos \theta_{DB},\sin \theta_{DB}), \\
{\bf r}^\perp_{BA} &\equiv (-x_A,y_B) = r_{BA} (\cos \theta_{BA},\sin \theta_{BA}).
\end{align}
The integration limit defined by the $AB-CD$ line boundary will be $\theta_{AB} < 0$.   The integration limit for the $BCD$ line boundary will be $\theta_{DB} > \theta_0 \equiv \theta_{BCD} + \epsilon/r_{DB}$, where $\theta_{BCD}$ is the angle of the $BCD$ contact line relative to the x axis.  Due to the straight line condition, $\theta_{CB}$ will be unconstrained.

In terms of these variables, 
${\bf r}_{BA}^\perp \times {\bf r}_{CA}^\perp ={\bf r}_{BA}^\perp \times {\bf r}_{CB}^\perp = r_{BA} r_{CB} \sin(\theta_{CB}-\theta_{BA})$,
and ${\bf r}^\perp_{BA} \times \nabla_C = (r_{BA}/r_{CB}) \partial/\partial \theta_{CB}$ (plus a term that vanishes for $\theta_{CB}=\theta_{BA}$).   
The integral then becomes 
\begin{equation}
I^{CD}_{ABCD} = I_r I_\theta
\end{equation}
with
\begin{equation}
I_r = \int_0^\infty dr_{BA} dr_{CB}\int_{r_{CB}}^\infty dr_{DB} \frac{1 }{r_{BA}r_{CB}|r_{CB}-r_{DB}|}
\end{equation}
and
\begin{align}
I_\theta = \int_{-\infty}^\infty d\theta_{CB}&\int_{-\infty}^0 d\theta_{DB} \int_{\theta_0}^\infty d\theta_{BA}
\\
&\delta'(\theta_{CB}-\theta_{BA}) \delta(\theta_{DB}-\theta_{BA}).
\nonumber\end{align}
$I_r$ will involve a $\log^3\Lambda$ divergence.   However, due to the unconstrained integral over $\theta_{CB}$ of $\delta'(\theta_{CB}-\theta_{BA})$, $I_\theta$ will vanish.  
We conclude that  $c^{CD}_{ABCD}  =0$.

A similar analysis can be applied to all of the possible orderings $abcd$, on planes $ab$, $bc$ and $cd$.   There are several cases, so we will not repeat the details.   However, in some cases, the analog of $I_\theta$ does not vanish.   For example, we find $I^{BD}_{CABD} = I_r I_\theta$ with
\begin{equation}
I_r = \int_0^\infty dr_{BA} dr_{AC} \int_{r_{BA}}^\infty dr_{DA} \frac{1}{r_{BA} r_{AC} |r_{DA}-r_{BA}|}
\end{equation}
and
\begin{align}
I_\theta = \int_{-\infty}^0 d\theta_{AC} &\int_{\theta_0}^\infty d\theta_{DA} \int_{-\infty}^\infty d\theta_{BA} \nonumber \\
&\delta'(\theta_{AC} - \theta_{BA}) \delta(\theta_{DA}-\theta_{BA}),
\end{align}
where $\theta_0 <0$.   Note that in this case $I_\theta = 1$ and $I_r = \log^3 k_F L$ (where we cut off the logarithmic divergence in the same manner as Eq. \ref{ibcdlog}.   We conclude that $c^{BD}_{CABD} = 1$.

Applying this type of analysis to all cases, we find that the non-zero entries to table \ref{itable} are,
\begin{equation}
\begin{array}{ll}
c^{BD}_{ABDC} = 1, \quad &  c^{BD}_{CABD} = 1, \\
c^{BD}_{CADB} = 1/2, \quad & c^{BD}_{CBDA} = 1/3, \\
c^{BD}_{ACBD} = 2/3, \quad & c^{BD}_{ACDB} = 2/3,
\end{array}
\end{equation}
along with similar terms found by interchanging $A$ and $B$,
\begin{equation}
\begin{array}{ll}
c^{AD}_{BADC} = 1, \quad &  c^{AD}_{CBAD} = 1, \\
c^{AD}_{CBDA} = 1/2, \quad & c^{AD}_{CADB} = 1/3, \\
c^{AD}_{BCAD} = 2/3, \quad & c^{AD}_{BCDA} = 2/3.
\end{array}
\end{equation}
This completes the analysis of the contributions to $I$ from the double contact planes.

\section{\label{sec D} Optimizing the Fitting Analysis}
In Sec. \ref{sec IVB} we have presented a quadratic-fitting analysis for extracting the coefficient of $\log^2 L$ in the finite-size scaling of $\mathcal{I}^Q_3(L)$. For tight-binding models on both the triangular and square lattices, with one electron/hole-like Fermi surface, we have found $\chi_F \approx \pm 1$ as the best-fit value with roughly $5\%$ deviation from the true $\chi_F$. The uncertainty of fitting has been tamed to about $10\%$. Such results have been \textit{consistently} obtained when the size of  Fermi sea is varied (see Fig. \ref{fitting}), as well as when the interval of $\log L$ (in which the fitting is performed) is varied. To obtain this kind of numerical stability, we have adopted two approaches to optimize the data set. 

\begin{figure}[t!]
   \includegraphics[width=\columnwidth]{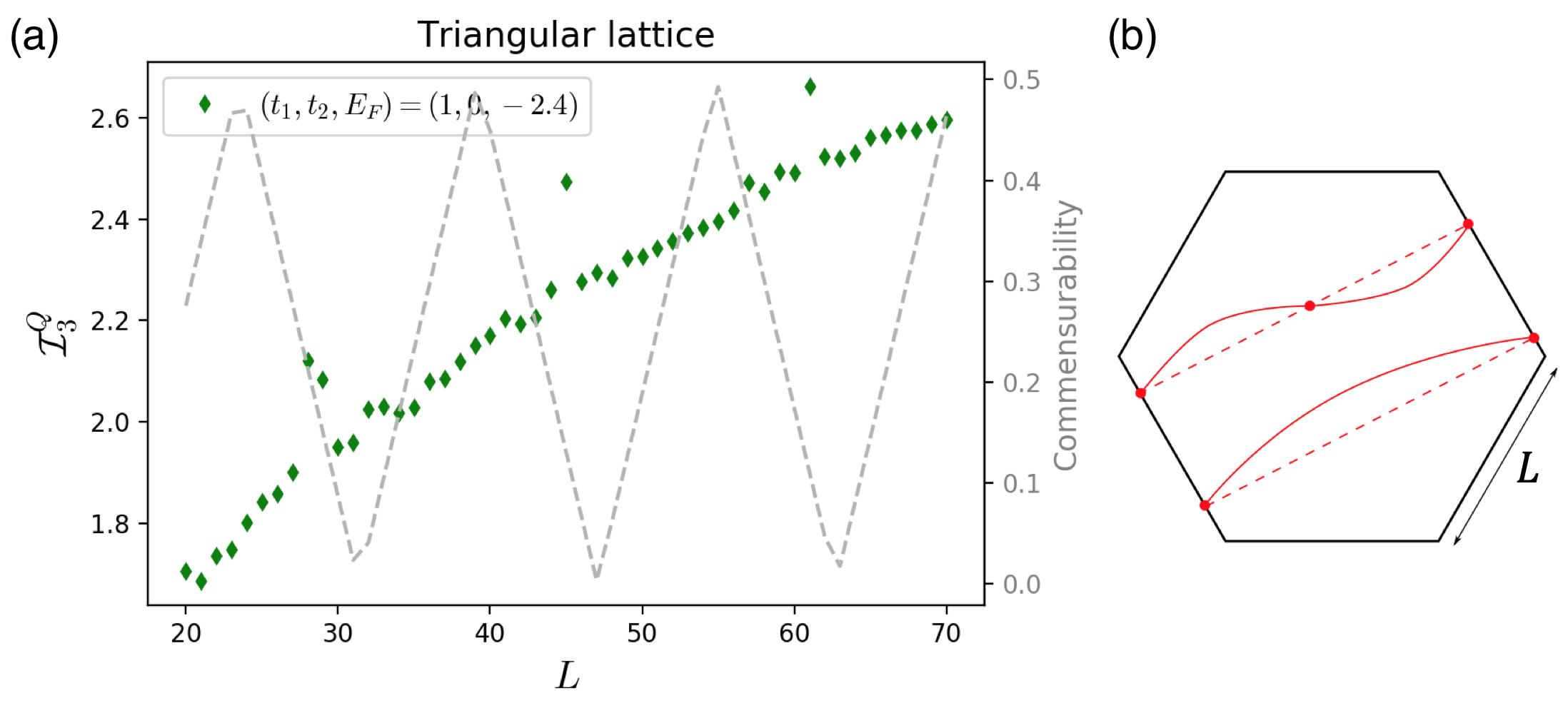}\centering
  \caption{\small{Outliers due to the hard-wall boundary condition. (a) shows an example of how $\mathcal{I}^Q_3$ (green diamond) increases with $L$ for the densest sampling ($\Delta L=1$). There is a smooth background, comparable to our scaling prediction, together with outliers appearing periodically. The outliers are related to electrons on the Fermi surface bouncing back-and-forth between the open boundary, as shown in (b), forming standing-wave when the condition in (\ref{outliercondition}) is satisfied (see the dashed line in (a), and discussion in the text). We conjecture that these non-local states enhance the entanglement and cause the jumps seen in (a).}}
  \label{commensurate}
\end{figure}

The first approach is to drop the outliers. When we do a very dense sampling of system size $L$, we observe that $\mathcal{I}^Q_3(L)$ contains drastic jumps above a smoothly-varying background, see Fig. \ref{commensurate} for an example on the triangular lattice. These jumps happen only at very specific $L$'s, and can be predicted from the following commensurability condition:
\begin{equation}\label{outliercondition}
\sqrt{3}Lk_F = \pi N, \quad (N\in \mathbb{Z})
\end{equation}
Semiclassically, this condition is satisfied when electrons on the Fermi surface (with wavelength $\lambda_F = 2\pi/k_F$) bounces between the open boundaries of our hexagonal system (separated by $\sqrt{3}L$) to form an extended standing-wave. The non-local nature of these states seems to enhance the entanglement and leads to the jumps observed in Fig. \ref{commensurate}(a). In Fig. \ref{commensurate}, we have also plotted the quantity $\abs{\sqrt{3}Lk_F/\pi - \floor{\sqrt{3}Lk_F/\pi+0.5}}$ as a function of $L$. As shown, around places where this quantity is minimized, i.e. the commensurability condition is approximately satisfied, outliers appear in $\mathcal{I}^Q_3(L)$. As $L$ increases towards the thermodynamic limit, we expect these jump of the outliers to become less significant, since any effects from the boundary should become insignificant as compared to the bulk in this limit. We thus decide to drop these outliers in the fitting analysis, and compare our theoretic prediction only to the smooth background obtained in the simulation. 

\begin{figure}[t!]
   \includegraphics[width=0.9\columnwidth]{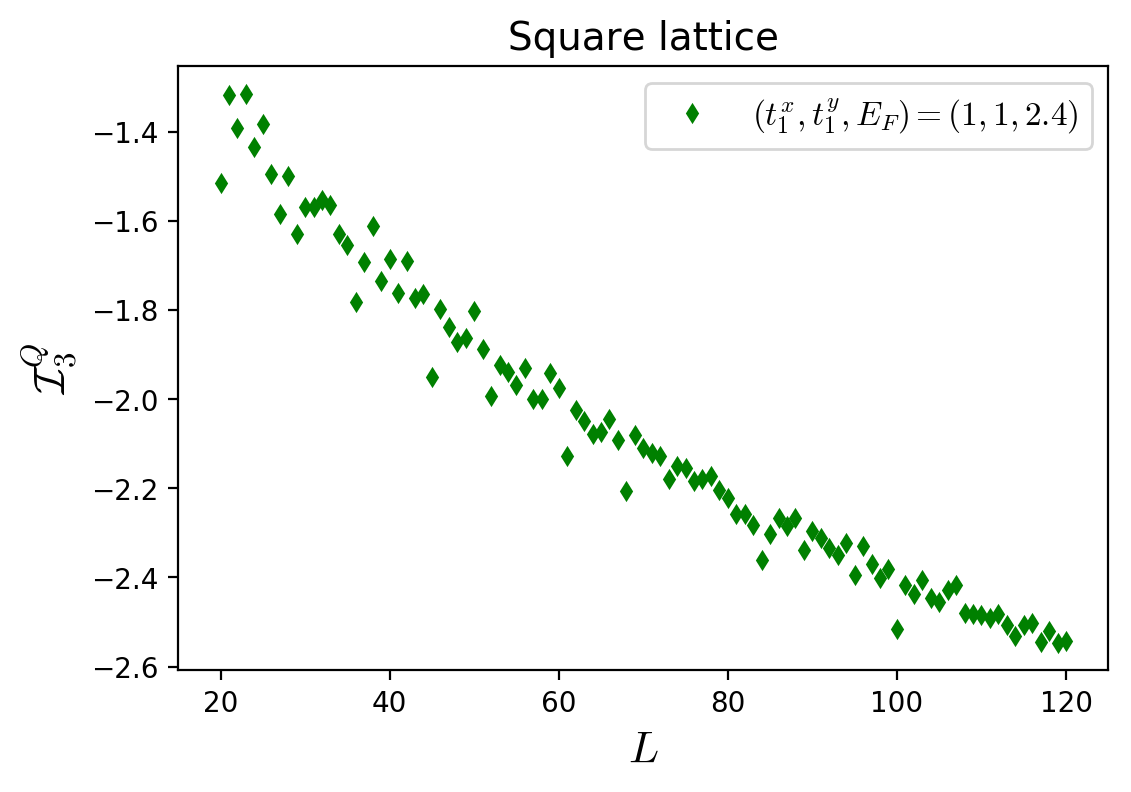}\centering
  \caption{\small{$\mathcal{I}^Q_3(L)$ in the small $L$ regime with the densest sampling ($\Delta L=1$), for the case of square lattice with one hole-like Fermi surface. Aside from the outliers caused by the boundary effect, there exists additional finite-size noise, presumably caused by the asymmetric tripartition. We try to reduce such noise in our fitting analysis by averaging data points among neighboring $L$'s.}}
  \label{sqnoise}
\end{figure}

Such finite-size noise also appears in the case of square lattice. However, the situation there is slightly worse, as illustrated in Fig. \ref{sqnoise}. Our triparition scheme in Fig. \ref{partition}(b) dictates that region $C$ is smaller in size relative to regions $A$ and $B$. Notice also that, for the case of square lattice, while the linear system size is $L$, the size of the triple contact is only $L/2$. The simulation for the square lattice thus suffers more finite-size noise. Again, we want to compare our theoretical prediction with the smooth background in the simulation, which is believed to continue into the thermodynamic limit. Therefore, we want to reduce the finite-size noise by an averaging procedure, namely taking
\begin{equation}
\widetilde{\mathcal{I}}^Q_3(L) = \frac{1}{3}(\mathcal{I}^Q_3(L-1)+\mathcal{I}^Q_3(L)+\mathcal{I}^Q_3(L+1))
\end{equation}
to be the fitting data. The fitting results (for square lattice) presented in Sec. \ref{sec IVB} are obtained after we have treated the raw data in this way. The best-fit value is then numerically stable when the fitting interval is varied, with fitting uncertainty restricted to about $10\%$, and the inferred $\chi_F$ matches reasonably well with the predicted quantized value.



\begin{thebibliography}{95}%
\makeatletter
\providecommand \@ifxundefined [1]{%
 \@ifx{#1\undefined}
}%
\providecommand \@ifnum [1]{%
 \ifnum #1\expandafter \@firstoftwo
 \else \expandafter \@secondoftwo
 \fi
}%
\providecommand \@ifx [1]{%
 \ifx #1\expandafter \@firstoftwo
 \else \expandafter \@secondoftwo
 \fi
}%
\providecommand \natexlab [1]{#1}%
\providecommand \enquote  [1]{``#1''}%
\providecommand \bibnamefont  [1]{#1}%
\providecommand \bibfnamefont [1]{#1}%
\providecommand \citenamefont [1]{#1}%
\providecommand \href@noop [0]{\@secondoftwo}%
\providecommand \href [0]{\begingroup \@sanitize@url \@href}%
\providecommand \@href[1]{\@@startlink{#1}\@@href}%
\providecommand \@@href[1]{\endgroup#1\@@endlink}%
\providecommand \@sanitize@url [0]{\catcode `\\12\catcode `\$12\catcode
  `\&12\catcode `\#12\catcode `\^12\catcode `\_12\catcode `\%12\relax}%
\providecommand \@@startlink[1]{}%
\providecommand \@@endlink[0]{}%
\providecommand \url  [0]{\begingroup\@sanitize@url \@url }%
\providecommand \@url [1]{\endgroup\@href {#1}{\urlprefix }}%
\providecommand \urlprefix  [0]{URL }%
\providecommand \Eprint [0]{\href }%
\providecommand \doibase [0]{http://dx.doi.org/}%
\providecommand \selectlanguage [0]{\@gobble}%
\providecommand \bibinfo  [0]{\@secondoftwo}%
\providecommand \bibfield  [0]{\@secondoftwo}%
\providecommand \translation [1]{[#1]}%
\providecommand \BibitemOpen [0]{}%
\providecommand \bibitemStop [0]{}%
\providecommand \bibitemNoStop [0]{.\EOS\space}%
\providecommand \EOS [0]{\spacefactor3000\relax}%
\providecommand \BibitemShut  [1]{\csname bibitem#1\endcsname}%
\let\auto@bib@innerbib\@empty
\bibitem [{\citenamefont {Kitaev}\ and\ \citenamefont
  {Preskill}(2006)}]{Kitaev2006}%
  \BibitemOpen
  \bibfield  {author} {\bibinfo {author} {\bibfnamefont {Alexei}\ \bibnamefont
  {Kitaev}}\ and\ \bibinfo {author} {\bibfnamefont {John}\ \bibnamefont
  {Preskill}},\ }\bibfield  {title} {\enquote {\bibinfo {title} {Topological
  entanglement entropy},}\ }\href {\doibase 10.1103/PhysRevLett.96.110404}
  {\bibfield  {journal} {\bibinfo  {journal} {Phys. Rev. Lett.}\ }\textbf
  {\bibinfo {volume} {96}},\ \bibinfo {pages} {110404} (\bibinfo {year}
  {2006})}\BibitemShut {NoStop}%
\bibitem [{\citenamefont {Levin}\ and\ \citenamefont {Wen}(2006)}]{Levin2006}%
  \BibitemOpen
  \bibfield  {author} {\bibinfo {author} {\bibfnamefont {Michael}\ \bibnamefont
  {Levin}}\ and\ \bibinfo {author} {\bibfnamefont {Xiao-Gang}\ \bibnamefont
  {Wen}},\ }\bibfield  {title} {\enquote {\bibinfo {title} {Detecting
  topological order in a ground state wave function},}\ }\href {\doibase
  10.1103/PhysRevLett.96.110405} {\bibfield  {journal} {\bibinfo  {journal}
  {Phys. Rev. Lett.}\ }\textbf {\bibinfo {volume} {96}},\ \bibinfo {pages}
  {110405} (\bibinfo {year} {2006})}\BibitemShut {NoStop}%
\bibitem [{\citenamefont {Castelnovo}\ and\ \citenamefont
  {Chamon}(2008)}]{Castelnovo2008}%
  \BibitemOpen
  \bibfield  {author} {\bibinfo {author} {\bibfnamefont {Claudio}\ \bibnamefont
  {Castelnovo}}\ and\ \bibinfo {author} {\bibfnamefont {Claudio}\ \bibnamefont
  {Chamon}},\ }\bibfield  {title} {\enquote {\bibinfo {title} {Topological
  order in a three-dimensional toric code at finite temperature},}\ }\href
  {\doibase 10.1103/PhysRevB.78.155120} {\bibfield  {journal} {\bibinfo
  {journal} {Phys. Rev. B}\ }\textbf {\bibinfo {volume} {78}},\ \bibinfo
  {pages} {155120} (\bibinfo {year} {2008})}\BibitemShut {NoStop}%
\bibitem [{\citenamefont {Grover}\ \emph {et~al.}(2011)\citenamefont {Grover},
  \citenamefont {Turner},\ and\ \citenamefont {Vishwanath}}]{Grover2011}%
  \BibitemOpen
  \bibfield  {author} {\bibinfo {author} {\bibfnamefont {Tarun}\ \bibnamefont
  {Grover}}, \bibinfo {author} {\bibfnamefont {Ari~M.}\ \bibnamefont {Turner}},
  \ and\ \bibinfo {author} {\bibfnamefont {Ashvin}\ \bibnamefont
  {Vishwanath}},\ }\bibfield  {title} {\enquote {\bibinfo {title} {Entanglement
  entropy of gapped phases and topological order in three dimensions},}\ }\href
  {\doibase 10.1103/PhysRevB.84.195120} {\bibfield  {journal} {\bibinfo
  {journal} {Phys. Rev. B}\ }\textbf {\bibinfo {volume} {84}},\ \bibinfo
  {pages} {195120} (\bibinfo {year} {2011})}\BibitemShut {NoStop}%
\bibitem [{\citenamefont {Holzhey}\ \emph {et~al.}(1994)\citenamefont
  {Holzhey}, \citenamefont {Larsen},\ and\ \citenamefont
  {Wilczek}}]{Holzhey1994}%
  \BibitemOpen
  \bibfield  {author} {\bibinfo {author} {\bibfnamefont {Christoph}\
  \bibnamefont {Holzhey}}, \bibinfo {author} {\bibfnamefont {Finn}\
  \bibnamefont {Larsen}}, \ and\ \bibinfo {author} {\bibfnamefont {Frank}\
  \bibnamefont {Wilczek}},\ }\bibfield  {title} {\enquote {\bibinfo {title}
  {Geometric and renormalized entropy in conformal field theory},}\ }\href
  {\doibase https://doi.org/10.1016/0550-3213(94)90402-2} {\bibfield  {journal}
  {\bibinfo  {journal} {Nuclear Physics B}\ }\textbf {\bibinfo {volume}
  {424}},\ \bibinfo {pages} {443--467} (\bibinfo {year} {1994})}\BibitemShut
  {NoStop}%
\bibitem [{\citenamefont {Vidal}\ \emph {et~al.}(2003)\citenamefont {Vidal},
  \citenamefont {Latorre}, \citenamefont {Rico},\ and\ \citenamefont
  {Kitaev}}]{Vidal2003}%
  \BibitemOpen
  \bibfield  {author} {\bibinfo {author} {\bibfnamefont {G.}~\bibnamefont
  {Vidal}}, \bibinfo {author} {\bibfnamefont {J.~I.}\ \bibnamefont {Latorre}},
  \bibinfo {author} {\bibfnamefont {E.}~\bibnamefont {Rico}}, \ and\ \bibinfo
  {author} {\bibfnamefont {A.}~\bibnamefont {Kitaev}},\ }\bibfield  {title}
  {\enquote {\bibinfo {title} {Entanglement in quantum critical phenomena},}\
  }\href {\doibase 10.1103/PhysRevLett.90.227902} {\bibfield  {journal}
  {\bibinfo  {journal} {Phys. Rev. Lett.}\ }\textbf {\bibinfo {volume} {90}},\
  \bibinfo {pages} {227902} (\bibinfo {year} {2003})}\BibitemShut {NoStop}%
\bibitem [{\citenamefont {Calabrese}\ and\ \citenamefont
  {Cardy}(2004)}]{Calabrese2004}%
  \BibitemOpen
  \bibfield  {author} {\bibinfo {author} {\bibfnamefont {Pasquale}\
  \bibnamefont {Calabrese}}\ and\ \bibinfo {author} {\bibfnamefont {John}\
  \bibnamefont {Cardy}},\ }\bibfield  {title} {\enquote {\bibinfo {title}
  {Entanglement entropy and quantum field theory},}\ }\href {\doibase
  10.1088/1742-5468/2004/06/p06002} {\bibfield  {journal} {\bibinfo  {journal}
  {Journal of Statistical Mechanics: Theory and Experiment}\ }\textbf {\bibinfo
  {volume} {2004}},\ \bibinfo {pages} {P06002} (\bibinfo {year}
  {2004})}\BibitemShut {NoStop}%
\bibitem [{\citenamefont {Calabrese}\ and\ \citenamefont
  {Cardy}(2009)}]{Calabrese2009}%
  \BibitemOpen
  \bibfield  {author} {\bibinfo {author} {\bibfnamefont {Pasquale}\
  \bibnamefont {Calabrese}}\ and\ \bibinfo {author} {\bibfnamefont {John}\
  \bibnamefont {Cardy}},\ }\bibfield  {title} {\enquote {\bibinfo {title}
  {Entanglement entropy and conformal field theory},}\ }\href {\doibase
  10.1088/1751-8113/42/50/504005} {\bibfield  {journal} {\bibinfo  {journal}
  {Journal of Physics A: Mathematical and Theoretical}\ }\textbf {\bibinfo
  {volume} {42}},\ \bibinfo {pages} {504005} (\bibinfo {year}
  {2009})}\BibitemShut {NoStop}%
\bibitem [{\citenamefont {Cardy}(1988)}]{Cardy1988ctheorem}%
  \BibitemOpen
  \bibfield  {author} {\bibinfo {author} {\bibfnamefont {John~L.}\ \bibnamefont
  {Cardy}},\ }\bibfield  {title} {\enquote {\bibinfo {title} {Is there a
  c-theorem in four dimensions?}}\ }\href {\doibase
  https://doi.org/10.1016/0370-2693(88)90054-8} {\bibfield  {journal} {\bibinfo
   {journal} {Physics Letters B}\ }\textbf {\bibinfo {volume} {215}},\ \bibinfo
  {pages} {749--752} (\bibinfo {year} {1988})}\BibitemShut {NoStop}%
\bibitem [{\citenamefont {Ryu}\ and\ \citenamefont
  {Takayanagi}(2006)}]{Ryu2006}%
  \BibitemOpen
  \bibfield  {author} {\bibinfo {author} {\bibfnamefont {Shinsei}\ \bibnamefont
  {Ryu}}\ and\ \bibinfo {author} {\bibfnamefont {Tadashi}\ \bibnamefont
  {Takayanagi}},\ }\bibfield  {title} {\enquote {\bibinfo {title} {Aspects of
  holographic entanglement entropy},}\ }\href {\doibase
  10.1088/1126-6708/2006/08/045} {\bibfield  {journal} {\bibinfo  {journal}
  {Journal of High Energy Physics}\ }\textbf {\bibinfo {volume} {2006}},\
  \bibinfo {pages} {045--045} (\bibinfo {year} {2006})}\BibitemShut {NoStop}%
\bibitem [{\citenamefont {Solodukhin}(2008)}]{Solodukhin2008}%
  \BibitemOpen
  \bibfield  {author} {\bibinfo {author} {\bibfnamefont {Sergey~N.}\
  \bibnamefont {Solodukhin}},\ }\bibfield  {title} {\enquote {\bibinfo {title}
  {Entanglement entropy, conformal invariance and extrinsic geometry},}\ }\href
  {\doibase https://doi.org/10.1016/j.physletb.2008.05.071} {\bibfield
  {journal} {\bibinfo  {journal} {Physics Letters B}\ }\textbf {\bibinfo
  {volume} {665}},\ \bibinfo {pages} {305--309} (\bibinfo {year}
  {2008})}\BibitemShut {NoStop}%
\bibitem [{\citenamefont {Casini}\ and\ \citenamefont
  {Huerta}(2009{\natexlab{a}})}]{Casini2009}%
  \BibitemOpen
  \bibfield  {author} {\bibinfo {author} {\bibfnamefont {H}~\bibnamefont
  {Casini}}\ and\ \bibinfo {author} {\bibfnamefont {M}~\bibnamefont {Huerta}},\
  }\bibfield  {title} {\enquote {\bibinfo {title} {Entanglement entropy in free
  quantum field theory},}\ }\href {\doibase 10.1088/1751-8113/42/50/504007}
  {\bibfield  {journal} {\bibinfo  {journal} {Journal of Physics A:
  Mathematical and Theoretical}\ }\textbf {\bibinfo {volume} {42}},\ \bibinfo
  {pages} {504007} (\bibinfo {year} {2009}{\natexlab{a}})}\BibitemShut
  {NoStop}%
\bibitem [{\citenamefont {Myers}\ and\ \citenamefont
  {Sinha}(2011)}]{Myers2011}%
  \BibitemOpen
  \bibfield  {author} {\bibinfo {author} {\bibfnamefont {Robert~C.}\
  \bibnamefont {Myers}}\ and\ \bibinfo {author} {\bibfnamefont {Aninda}\
  \bibnamefont {Sinha}},\ }\bibfield  {title} {\enquote {\bibinfo {title}
  {Holographic c-theorems in arbitrary dimensions},}\ }\href {\doibase
  10.1007/JHEP01(2011)125} {\bibfield  {journal} {\bibinfo  {journal} {Journal
  of High Energy Physics}\ }\textbf {\bibinfo {volume} {2011}},\ \bibinfo
  {pages} {125} (\bibinfo {year} {2011})}\BibitemShut {NoStop}%
\bibitem [{\citenamefont {Casini}\ \emph {et~al.}(2011)\citenamefont {Casini},
  \citenamefont {Huerta},\ and\ \citenamefont {Myers}}]{Casini2011}%
  \BibitemOpen
  \bibfield  {author} {\bibinfo {author} {\bibfnamefont {Horacio}\ \bibnamefont
  {Casini}}, \bibinfo {author} {\bibfnamefont {Marina}\ \bibnamefont {Huerta}},
  \ and\ \bibinfo {author} {\bibfnamefont {Robert~C.}\ \bibnamefont {Myers}},\
  }\bibfield  {title} {\enquote {\bibinfo {title} {Towards a derivation of
  holographic entanglement entropy},}\ }\href {\doibase
  10.1007/JHEP05(2011)036} {\bibfield  {journal} {\bibinfo  {journal} {Journal
  of High Energy Physics}\ }\textbf {\bibinfo {volume} {2011}},\ \bibinfo
  {pages} {36} (\bibinfo {year} {2011})}\BibitemShut {NoStop}%
\bibitem [{\citenamefont {Liu}\ and\ \citenamefont {Mezei}(2013)}]{Liu2013}%
  \BibitemOpen
  \bibfield  {author} {\bibinfo {author} {\bibfnamefont {Hong}\ \bibnamefont
  {Liu}}\ and\ \bibinfo {author} {\bibfnamefont {M{\'a}rk}\ \bibnamefont
  {Mezei}},\ }\bibfield  {title} {\enquote {\bibinfo {title} {A refinement of
  entanglement entropy and the number of degrees of freedom},}\ }\href
  {\doibase 10.1007/JHEP04(2013)162} {\bibfield  {journal} {\bibinfo  {journal}
  {Journal of High Energy Physics}\ }\textbf {\bibinfo {volume} {2013}},\
  \bibinfo {pages} {162} (\bibinfo {year} {2013})}\BibitemShut {NoStop}%
\bibitem [{\citenamefont {Casini}\ \emph {et~al.}(2015)\citenamefont {Casini},
  \citenamefont {Huerta}, \citenamefont {Myers},\ and\ \citenamefont
  {Yale}}]{Casini2015}%
  \BibitemOpen
  \bibfield  {author} {\bibinfo {author} {\bibfnamefont {Horacio}\ \bibnamefont
  {Casini}}, \bibinfo {author} {\bibfnamefont {Marina}\ \bibnamefont {Huerta}},
  \bibinfo {author} {\bibfnamefont {Robert~C.}\ \bibnamefont {Myers}}, \ and\
  \bibinfo {author} {\bibfnamefont {Alexandre}\ \bibnamefont {Yale}},\
  }\bibfield  {title} {\enquote {\bibinfo {title} {{Mutual information and the
  F-theorem}},}\ }\href {\doibase 10.1007/JHEP10(2015)003} {\bibfield
  {journal} {\bibinfo  {journal} {JHEP}\ }\textbf {\bibinfo {volume} {10}},\
  \bibinfo {pages} {003} (\bibinfo {year} {2015})},\ \Eprint
  {http://arxiv.org/abs/1506.06195} {arXiv:1506.06195 [hep-th]} \BibitemShut
  {NoStop}%
\bibitem [{\citenamefont {Wolf}(2006)}]{Wolf2006}%
  \BibitemOpen
  \bibfield  {author} {\bibinfo {author} {\bibfnamefont {Michael~M.}\
  \bibnamefont {Wolf}},\ }\bibfield  {title} {\enquote {\bibinfo {title}
  {Violation of the entropic area law for fermions},}\ }\href {\doibase
  10.1103/PhysRevLett.96.010404} {\bibfield  {journal} {\bibinfo  {journal}
  {Phys. Rev. Lett.}\ }\textbf {\bibinfo {volume} {96}},\ \bibinfo {pages}
  {010404} (\bibinfo {year} {2006})}\BibitemShut {NoStop}%
\bibitem [{\citenamefont {Gioev}\ and\ \citenamefont
  {Klich}(2006)}]{Klich2006}%
  \BibitemOpen
  \bibfield  {author} {\bibinfo {author} {\bibfnamefont {Dimitri}\ \bibnamefont
  {Gioev}}\ and\ \bibinfo {author} {\bibfnamefont {Israel}\ \bibnamefont
  {Klich}},\ }\bibfield  {title} {\enquote {\bibinfo {title} {Entanglement
  entropy of fermions in any dimension and the widom conjecture},}\ }\href
  {\doibase 10.1103/PhysRevLett.96.100503} {\bibfield  {journal} {\bibinfo
  {journal} {Phys. Rev. Lett.}\ }\textbf {\bibinfo {volume} {96}},\ \bibinfo
  {pages} {100503} (\bibinfo {year} {2006})}\BibitemShut {NoStop}%
\bibitem [{\citenamefont {Swingle}(2010{\natexlab{a}})}]{Swingle2010}%
  \BibitemOpen
  \bibfield  {author} {\bibinfo {author} {\bibfnamefont {Brian}\ \bibnamefont
  {Swingle}},\ }\bibfield  {title} {\enquote {\bibinfo {title} {Entanglement
  entropy and the fermi surface},}\ }\href {\doibase
  10.1103/PhysRevLett.105.050502} {\bibfield  {journal} {\bibinfo  {journal}
  {Phys. Rev. Lett.}\ }\textbf {\bibinfo {volume} {105}},\ \bibinfo {pages}
  {050502} (\bibinfo {year} {2010}{\natexlab{a}})}\BibitemShut {NoStop}%
\bibitem [{\citenamefont {Ding}\ \emph {et~al.}(2012)\citenamefont {Ding},
  \citenamefont {Seidel},\ and\ \citenamefont {Yang}}]{Ding2012}%
  \BibitemOpen
  \bibfield  {author} {\bibinfo {author} {\bibfnamefont {Wenxin}\ \bibnamefont
  {Ding}}, \bibinfo {author} {\bibfnamefont {Alexander}\ \bibnamefont
  {Seidel}}, \ and\ \bibinfo {author} {\bibfnamefont {Kun}\ \bibnamefont
  {Yang}},\ }\bibfield  {title} {\enquote {\bibinfo {title} {Entanglement
  entropy of fermi liquids via multidimensional bosonization},}\ }\href
  {\doibase 10.1103/PhysRevX.2.011012} {\bibfield  {journal} {\bibinfo
  {journal} {Phys. Rev. X}\ }\textbf {\bibinfo {volume} {2}},\ \bibinfo {pages}
  {011012} (\bibinfo {year} {2012})}\BibitemShut {NoStop}%
\bibitem [{\citenamefont {Calabrese}\ \emph {et~al.}(2012)\citenamefont
  {Calabrese}, \citenamefont {Mintchev},\ and\ \citenamefont
  {Vicari}}]{Calabrese2012a}%
  \BibitemOpen
  \bibfield  {author} {\bibinfo {author} {\bibfnamefont {P.}~\bibnamefont
  {Calabrese}}, \bibinfo {author} {\bibfnamefont {M.}~\bibnamefont {Mintchev}},
  \ and\ \bibinfo {author} {\bibfnamefont {E.}~\bibnamefont {Vicari}},\
  }\bibfield  {title} {\enquote {\bibinfo {title} {Entanglement entropies in
  free-fermion gases for arbitrary dimension},}\ }\href {\doibase
  10.1209/0295-5075/97/20009} {\bibfield  {journal} {\bibinfo  {journal} {{EPL}
  (Europhysics Letters)}\ }\textbf {\bibinfo {volume} {97}},\ \bibinfo {pages}
  {20009} (\bibinfo {year} {2012})}\BibitemShut {NoStop}%
\bibitem [{\citenamefont {Lai}\ and\ \citenamefont {Yang}(2016)}]{Yang2016}%
  \BibitemOpen
  \bibfield  {author} {\bibinfo {author} {\bibfnamefont {Hsin-Hua}\
  \bibnamefont {Lai}}\ and\ \bibinfo {author} {\bibfnamefont {Kun}\
  \bibnamefont {Yang}},\ }\bibfield  {title} {\enquote {\bibinfo {title}
  {Probing critical surfaces in momentum space using real-space entanglement
  entropy: Bose versus fermi},}\ }\href {\doibase 10.1103/PhysRevB.93.121109}
  {\bibfield  {journal} {\bibinfo  {journal} {Phys. Rev. B}\ }\textbf {\bibinfo
  {volume} {93}},\ \bibinfo {pages} {121109} (\bibinfo {year}
  {2016})}\BibitemShut {NoStop}%
\bibitem [{\citenamefont {Kane}(2022)}]{Kane2022}%
  \BibitemOpen
  \bibfield  {author} {\bibinfo {author} {\bibfnamefont {C.~L.}\ \bibnamefont
  {Kane}},\ }\bibfield  {title} {\enquote {\bibinfo {title} {Quantized
  nonlinear conductance in ballistic metals},}\ }\href {\doibase
  10.1103/PhysRevLett.128.076801} {\bibfield  {journal} {\bibinfo  {journal}
  {Phys. Rev. Lett.}\ }\textbf {\bibinfo {volume} {128}},\ \bibinfo {pages}
  {076801} (\bibinfo {year} {2022})}\BibitemShut {NoStop}%
\bibitem [{\citenamefont {Nakahara}(1990)}]{Nakahara1990}%
  \BibitemOpen
  \bibfield  {author} {\bibinfo {author} {\bibfnamefont {Mikio}\ \bibnamefont
  {Nakahara}},\ }\href {https://cds.cern.ch/record/206619} {\emph {\bibinfo
  {title} {{Geometry, topology and physics}}}},\ Graduate student series in
  physics\ (\bibinfo  {publisher} {Hilger},\ \bibinfo {address} {Bristol},\
  \bibinfo {year} {1990})\BibitemShut {NoStop}%
\bibitem [{\citenamefont {Milnor}(1963)}]{Milnor1963}%
  \BibitemOpen
  \bibfield  {author} {\bibinfo {author} {\bibfnamefont {J.W.}\ \bibnamefont
  {Milnor}},\ }\href {https://books.google.com/books?id=A9QZZ3S\_QxwC} {\emph
  {\bibinfo {title} {Morse Theory}}},\ Annals of Mathematics Studies\ (\bibinfo
   {publisher} {Princeton University Press},\ \bibinfo {year}
  {1963})\BibitemShut {NoStop}%
\bibitem [{\citenamefont {Nash}\ and\ \citenamefont {Sen}(1988)}]{Nash1988}%
  \BibitemOpen
  \bibfield  {author} {\bibinfo {author} {\bibfnamefont {C.}~\bibnamefont
  {Nash}}\ and\ \bibinfo {author} {\bibfnamefont {S.}~\bibnamefont {Sen}},\
  }\href {https://books.google.bi/books?id=nnnNCgAAQBAJ} {\emph {\bibinfo
  {title} {Topology and Geometry for Physicists}}}\ (\bibinfo  {publisher}
  {Elsevier Science},\ \bibinfo {year} {1988})\BibitemShut {NoStop}%
\bibitem [{\citenamefont {Lifshitz}\ \emph {et~al.}(1960)\citenamefont
  {Lifshitz} \emph {et~al.}}]{Lifshitz1960}%
  \BibitemOpen
  \bibfield  {author} {\bibinfo {author} {\bibfnamefont {IM}~\bibnamefont
  {Lifshitz}} \emph {et~al.},\ }\bibfield  {title} {\enquote {\bibinfo {title}
  {Anomalies of electron characteristics of a metal in the high pressure
  region},}\ }\href@noop {} {\bibfield  {journal} {\bibinfo  {journal} {Sov.
  Phys. JETP}\ }\textbf {\bibinfo {volume} {11}},\ \bibinfo {pages}
  {1130--1135} (\bibinfo {year} {1960})}\BibitemShut {NoStop}%
\bibitem [{\citenamefont {Volovik}(2017)}]{Volovik2017}%
  \BibitemOpen
  \bibfield  {author} {\bibinfo {author} {\bibfnamefont {GE}~\bibnamefont
  {Volovik}},\ }\bibfield  {title} {\enquote {\bibinfo {title} {Topological
  lifshitz transitions},}\ }\href@noop {} {\bibfield  {journal} {\bibinfo
  {journal} {Low Temperature Physics}\ }\textbf {\bibinfo {volume} {43}},\
  \bibinfo {pages} {47--55} (\bibinfo {year} {2017})}\BibitemShut {NoStop}%
\bibitem [{\citenamefont {Walter}\ \emph {et~al.}(2016)\citenamefont {Walter},
  \citenamefont {Gross},\ and\ \citenamefont {Eisert}}]{Walter2016}%
  \BibitemOpen
  \bibfield  {author} {\bibinfo {author} {\bibfnamefont {Michael}\ \bibnamefont
  {Walter}}, \bibinfo {author} {\bibfnamefont {David}\ \bibnamefont {Gross}}, \
  and\ \bibinfo {author} {\bibfnamefont {Jens}\ \bibnamefont {Eisert}},\
  }\enquote {\bibinfo {title} {Multipartite entanglement},}\ in\ \href
  {\doibase https://doi.org/10.1002/9783527805785.ch14} {\emph {\bibinfo
  {booktitle} {Quantum Information}}}\ (\bibinfo  {publisher} {John Wiley \&
  Sons, Ltd},\ \bibinfo {year} {2016})\ Chap.~\bibinfo {chapter} {14}, pp.\
  \bibinfo {pages} {293--330}\BibitemShut {NoStop}%
\bibitem [{\citenamefont {Rota}(2016)}]{Rota2016}%
  \BibitemOpen
  \bibfield  {author} {\bibinfo {author} {\bibfnamefont {Massimiliano}\
  \bibnamefont {Rota}},\ }\bibfield  {title} {\enquote {\bibinfo {title}
  {Tripartite information of highly entangled states},}\ }\href {\doibase
  10.1007/JHEP04(2016)075} {\bibfield  {journal} {\bibinfo  {journal} {Journal
  of High Energy Physics}\ }\textbf {\bibinfo {volume} {2016}},\ \bibinfo
  {pages} {75} (\bibinfo {year} {2016})}\BibitemShut {NoStop}%
\bibitem [{\citenamefont {Bayat}(2017)}]{Bayat2017}%
  \BibitemOpen
  \bibfield  {author} {\bibinfo {author} {\bibfnamefont {Abolfazl}\
  \bibnamefont {Bayat}},\ }\bibfield  {title} {\enquote {\bibinfo {title}
  {Scaling of tripartite entanglement at impurity quantum phase transitions},}\
  }\href {\doibase 10.1103/PhysRevLett.118.036102} {\bibfield  {journal}
  {\bibinfo  {journal} {Phys. Rev. Lett.}\ }\textbf {\bibinfo {volume} {118}},\
  \bibinfo {pages} {036102} (\bibinfo {year} {2017})}\BibitemShut {NoStop}%
\bibitem [{\citenamefont {Pezz\`e}\ \emph {et~al.}(2017)\citenamefont
  {Pezz\`e}, \citenamefont {Gabbrielli}, \citenamefont {Lepori},\ and\
  \citenamefont {Smerzi}}]{Pezze2017}%
  \BibitemOpen
  \bibfield  {author} {\bibinfo {author} {\bibfnamefont {Luca}\ \bibnamefont
  {Pezz\`e}}, \bibinfo {author} {\bibfnamefont {Marco}\ \bibnamefont
  {Gabbrielli}}, \bibinfo {author} {\bibfnamefont {Luca}\ \bibnamefont
  {Lepori}}, \ and\ \bibinfo {author} {\bibfnamefont {Augusto}\ \bibnamefont
  {Smerzi}},\ }\bibfield  {title} {\enquote {\bibinfo {title} {Multipartite
  entanglement in topological quantum phases},}\ }\href {\doibase
  10.1103/PhysRevLett.119.250401} {\bibfield  {journal} {\bibinfo  {journal}
  {Phys. Rev. Lett.}\ }\textbf {\bibinfo {volume} {119}},\ \bibinfo {pages}
  {250401} (\bibinfo {year} {2017})}\BibitemShut {NoStop}%
\bibitem [{\citenamefont {Wang}\ \emph {et~al.}(2018)\citenamefont {Wang},
  \citenamefont {Wang},\ and\ \citenamefont {Wang}}]{Wang2018}%
  \BibitemOpen
  \bibfield  {author} {\bibinfo {author} {\bibfnamefont {Qiang}\ \bibnamefont
  {Wang}}, \bibinfo {author} {\bibfnamefont {Da}~\bibnamefont {Wang}}, \ and\
  \bibinfo {author} {\bibfnamefont {Qiang-Hua}\ \bibnamefont {Wang}},\
  }\bibfield  {title} {\enquote {\bibinfo {title} {Entanglement in a
  second-order topological insulator on a square lattice},}\ }\href {\doibase
  10.1209/0295-5075/124/50005} {\bibfield  {journal} {\bibinfo  {journal}
  {{EPL} (Europhysics Letters)}\ }\textbf {\bibinfo {volume} {124}},\ \bibinfo
  {pages} {50005} (\bibinfo {year} {2018})}\BibitemShut {NoStop}%
\bibitem [{\citenamefont {Shirley}\ \emph {et~al.}(2019)\citenamefont
  {Shirley}, \citenamefont {Slagle},\ and\ \citenamefont {Chen}}]{Shirley2019}%
  \BibitemOpen
  \bibfield  {author} {\bibinfo {author} {\bibfnamefont {Wilbur}\ \bibnamefont
  {Shirley}}, \bibinfo {author} {\bibfnamefont {Kevin}\ \bibnamefont {Slagle}},
  \ and\ \bibinfo {author} {\bibfnamefont {Xie}\ \bibnamefont {Chen}},\
  }\bibfield  {title} {\enquote {\bibinfo {title} {{Universal entanglement
  signatures of foliated fracton phases}},}\ }\href {\doibase
  10.21468/SciPostPhys.6.1.015} {\bibfield  {journal} {\bibinfo  {journal}
  {SciPost Phys.}\ }\textbf {\bibinfo {volume} {6}},\ \bibinfo {pages} {15}
  (\bibinfo {year} {2019})}\BibitemShut {NoStop}%
\bibitem [{\citenamefont {Zou}\ \emph {et~al.}(2021)\citenamefont {Zou},
  \citenamefont {Siva}, \citenamefont {Soejima}, \citenamefont {Mong},\ and\
  \citenamefont {Zaletel}}]{Zou2021}%
  \BibitemOpen
  \bibfield  {author} {\bibinfo {author} {\bibfnamefont {Yijian}\ \bibnamefont
  {Zou}}, \bibinfo {author} {\bibfnamefont {Karthik}\ \bibnamefont {Siva}},
  \bibinfo {author} {\bibfnamefont {Tomohiro}\ \bibnamefont {Soejima}},
  \bibinfo {author} {\bibfnamefont {Roger S.~K.}\ \bibnamefont {Mong}}, \ and\
  \bibinfo {author} {\bibfnamefont {Michael~P.}\ \bibnamefont {Zaletel}},\
  }\bibfield  {title} {\enquote {\bibinfo {title} {Universal tripartite
  entanglement in one-dimensional many-body systems},}\ }\href {\doibase
  10.1103/PhysRevLett.126.120501} {\bibfield  {journal} {\bibinfo  {journal}
  {Phys. Rev. Lett.}\ }\textbf {\bibinfo {volume} {126}},\ \bibinfo {pages}
  {120501} (\bibinfo {year} {2021})}\BibitemShut {NoStop}%
\bibitem [{\citenamefont {Liu}\ \emph {et~al.}(2022)\citenamefont {Liu},
  \citenamefont {Sohal}, \citenamefont {Kudler-Flam},\ and\ \citenamefont
  {Ryu}}]{Liu2022}%
  \BibitemOpen
  \bibfield  {author} {\bibinfo {author} {\bibfnamefont {Yuhan}\ \bibnamefont
  {Liu}}, \bibinfo {author} {\bibfnamefont {Ramanjit}\ \bibnamefont {Sohal}},
  \bibinfo {author} {\bibfnamefont {Jonah}\ \bibnamefont {Kudler-Flam}}, \ and\
  \bibinfo {author} {\bibfnamefont {Shinsei}\ \bibnamefont {Ryu}},\ }\bibfield
  {title} {\enquote {\bibinfo {title} {Multipartitioning topological phases by
  vertex states and quantum entanglement},}\ }\href {\doibase
  10.1103/PhysRevB.105.115107} {\bibfield  {journal} {\bibinfo  {journal}
  {Phys. Rev. B}\ }\textbf {\bibinfo {volume} {105}},\ \bibinfo {pages}
  {115107} (\bibinfo {year} {2022})}\BibitemShut {NoStop}%
\bibitem [{\citenamefont {Casini}\ and\ \citenamefont
  {Huerta}(2009{\natexlab{b}})}]{Casini2009_MI}%
  \BibitemOpen
  \bibfield  {author} {\bibinfo {author} {\bibfnamefont {H}~\bibnamefont
  {Casini}}\ and\ \bibinfo {author} {\bibfnamefont {M}~\bibnamefont {Huerta}},\
  }\bibfield  {title} {\enquote {\bibinfo {title} {Remarks on the entanglement
  entropy for disconnected regions},}\ }\href {\doibase
  10.1088/1126-6708/2009/03/048} {\bibfield  {journal} {\bibinfo  {journal}
  {Journal of High Energy Physics}\ }\textbf {\bibinfo {volume} {2009}},\
  \bibinfo {pages} {048--048} (\bibinfo {year}
  {2009}{\natexlab{b}})}\BibitemShut {NoStop}%
\bibitem [{\citenamefont {Swingle}(2010{\natexlab{b}})}]{Swingle2010MI}%
  \BibitemOpen
  \bibfield  {author} {\bibinfo {author} {\bibfnamefont {Brian}\ \bibnamefont
  {Swingle}},\ }\bibfield  {title} {\enquote {\bibinfo {title} {Mutual
  information and the structure of entanglement in quantum field theory},}\
  }\href {https://arxiv.org/abs/1010.4038} {\bibfield  {journal} {\bibinfo
  {journal} {arXiv: Quantum Physics}\ } (\bibinfo {year}
  {2010}{\natexlab{b}})}\BibitemShut {NoStop}%
\bibitem [{\citenamefont {Swingle}(2012)}]{Swingle2012MI}%
  \BibitemOpen
  \bibfield  {author} {\bibinfo {author} {\bibfnamefont {Brian}\ \bibnamefont
  {Swingle}},\ }\bibfield  {title} {\enquote {\bibinfo {title} {R\'enyi
  entropy, mutual information, and fluctuation properties of fermi liquids},}\
  }\href {\doibase 10.1103/PhysRevB.86.045109} {\bibfield  {journal} {\bibinfo
  {journal} {Phys. Rev. B}\ }\textbf {\bibinfo {volume} {86}},\ \bibinfo
  {pages} {045109} (\bibinfo {year} {2012})}\BibitemShut {NoStop}%
\bibitem [{\citenamefont {Klich}\ and\ \citenamefont
  {Levitov}(2009)}]{Klich2009}%
  \BibitemOpen
  \bibfield  {author} {\bibinfo {author} {\bibfnamefont {Israel}\ \bibnamefont
  {Klich}}\ and\ \bibinfo {author} {\bibfnamefont {Leonid}\ \bibnamefont
  {Levitov}},\ }\bibfield  {title} {\enquote {\bibinfo {title} {Quantum noise
  as an entanglement meter},}\ }\href {\doibase 10.1103/PhysRevLett.102.100502}
  {\bibfield  {journal} {\bibinfo  {journal} {Phys. Rev. Lett.}\ }\textbf
  {\bibinfo {volume} {102}},\ \bibinfo {pages} {100502} (\bibinfo {year}
  {2009})}\BibitemShut {NoStop}%
\bibitem [{\citenamefont {Song}\ \emph {et~al.}(2011)\citenamefont {Song},
  \citenamefont {Flindt}, \citenamefont {Rachel}, \citenamefont {Klich},\ and\
  \citenamefont {Le~Hur}}]{Song2011}%
  \BibitemOpen
  \bibfield  {author} {\bibinfo {author} {\bibfnamefont {H.~Francis}\
  \bibnamefont {Song}}, \bibinfo {author} {\bibfnamefont {Christian}\
  \bibnamefont {Flindt}}, \bibinfo {author} {\bibfnamefont {Stephan}\
  \bibnamefont {Rachel}}, \bibinfo {author} {\bibfnamefont {Israel}\
  \bibnamefont {Klich}}, \ and\ \bibinfo {author} {\bibfnamefont {Karyn}\
  \bibnamefont {Le~Hur}},\ }\bibfield  {title} {\enquote {\bibinfo {title}
  {Entanglement entropy from charge statistics: Exact relations for
  noninteracting many-body systems},}\ }\href {\doibase
  10.1103/PhysRevB.83.161408} {\bibfield  {journal} {\bibinfo  {journal} {Phys.
  Rev. B}\ }\textbf {\bibinfo {volume} {83}},\ \bibinfo {pages} {161408}
  (\bibinfo {year} {2011})}\BibitemShut {NoStop}%
\bibitem [{\citenamefont {Song}\ \emph {et~al.}(2012)\citenamefont {Song},
  \citenamefont {Rachel}, \citenamefont {Flindt}, \citenamefont {Klich},
  \citenamefont {Laflorencie},\ and\ \citenamefont {Le~Hur}}]{Song2012}%
  \BibitemOpen
  \bibfield  {author} {\bibinfo {author} {\bibfnamefont {H.~Francis}\
  \bibnamefont {Song}}, \bibinfo {author} {\bibfnamefont {Stephan}\
  \bibnamefont {Rachel}}, \bibinfo {author} {\bibfnamefont {Christian}\
  \bibnamefont {Flindt}}, \bibinfo {author} {\bibfnamefont {Israel}\
  \bibnamefont {Klich}}, \bibinfo {author} {\bibfnamefont {Nicolas}\
  \bibnamefont {Laflorencie}}, \ and\ \bibinfo {author} {\bibfnamefont {Karyn}\
  \bibnamefont {Le~Hur}},\ }\bibfield  {title} {\enquote {\bibinfo {title}
  {Bipartite fluctuations as a probe of many-body entanglement},}\ }\href
  {\doibase 10.1103/PhysRevB.85.035409} {\bibfield  {journal} {\bibinfo
  {journal} {Phys. Rev. B}\ }\textbf {\bibinfo {volume} {85}},\ \bibinfo
  {pages} {035409} (\bibinfo {year} {2012})}\BibitemShut {NoStop}%
\bibitem [{\citenamefont {Chung}\ and\ \citenamefont
  {Peschel}(2001)}]{Peschel2001}%
  \BibitemOpen
  \bibfield  {author} {\bibinfo {author} {\bibfnamefont {Ming-Chiang}\
  \bibnamefont {Chung}}\ and\ \bibinfo {author} {\bibfnamefont {Ingo}\
  \bibnamefont {Peschel}},\ }\bibfield  {title} {\enquote {\bibinfo {title}
  {Density-matrix spectra of solvable fermionic systems},}\ }\href {\doibase
  10.1103/PhysRevB.64.064412} {\bibfield  {journal} {\bibinfo  {journal} {Phys.
  Rev. B}\ }\textbf {\bibinfo {volume} {64}},\ \bibinfo {pages} {064412}
  (\bibinfo {year} {2001})}\BibitemShut {NoStop}%
\bibitem [{\citenamefont {Peschel}(2003)}]{Peschel2003}%
  \BibitemOpen
  \bibfield  {author} {\bibinfo {author} {\bibfnamefont {Ingo}\ \bibnamefont
  {Peschel}},\ }\bibfield  {title} {\enquote {\bibinfo {title} {Calculation of
  reduced density matrices from correlation functions},}\ }\href {\doibase
  10.1088/0305-4470/36/14/101} {\bibfield  {journal} {\bibinfo  {journal}
  {Journal of Physics A: Mathematical and General}\ }\textbf {\bibinfo {volume}
  {36}},\ \bibinfo {pages} {L205--L208} (\bibinfo {year} {2003})}\BibitemShut
  {NoStop}%
\bibitem [{\citenamefont {Cheong}\ and\ \citenamefont
  {Henley}(2004)}]{Henley2004}%
  \BibitemOpen
  \bibfield  {author} {\bibinfo {author} {\bibfnamefont {Siew-Ann}\
  \bibnamefont {Cheong}}\ and\ \bibinfo {author} {\bibfnamefont
  {Christopher~L.}\ \bibnamefont {Henley}},\ }\bibfield  {title} {\enquote
  {\bibinfo {title} {Many-body density matrices for free fermions},}\ }\href
  {\doibase 10.1103/PhysRevB.69.075111} {\bibfield  {journal} {\bibinfo
  {journal} {Phys. Rev. B}\ }\textbf {\bibinfo {volume} {69}},\ \bibinfo
  {pages} {075111} (\bibinfo {year} {2004})}\BibitemShut {NoStop}%
\bibitem [{\citenamefont {Peschel}\ and\ \citenamefont
  {Eisler}(2009)}]{Peschel2009}%
  \BibitemOpen
  \bibfield  {author} {\bibinfo {author} {\bibfnamefont {Ingo}\ \bibnamefont
  {Peschel}}\ and\ \bibinfo {author} {\bibfnamefont {Viktor}\ \bibnamefont
  {Eisler}},\ }\bibfield  {title} {\enquote {\bibinfo {title} {Reduced density
  matrices and entanglement entropy in free lattice models},}\ }\href {\doibase
  10.1088/1751-8113/42/50/504003} {\bibfield  {journal} {\bibinfo  {journal}
  {Journal of Physics A: Mathematical and Theoretical}\ }\textbf {\bibinfo
  {volume} {42}},\ \bibinfo {pages} {504003} (\bibinfo {year}
  {2009})}\BibitemShut {NoStop}%
\bibitem [{\citenamefont {Giamarchi}\ and\ \citenamefont
  {Press}(2004)}]{Giamarchi2004}%
  \BibitemOpen
  \bibfield  {author} {\bibinfo {author} {\bibfnamefont {T.}~\bibnamefont
  {Giamarchi}}\ and\ \bibinfo {author} {\bibfnamefont {Oxford~University}\
  \bibnamefont {Press}},\ }\href
  {https://books.google.com/books?id=1MwTDAAAQBAJ} {\emph {\bibinfo {title}
  {Quantum Physics in One Dimension}}},\ International Series of Monographs on
  Physics\ (\bibinfo  {publisher} {Clarendon Press},\ \bibinfo {year}
  {2004})\BibitemShut {NoStop}%
\bibitem [{\citenamefont {Casini}\ \emph {et~al.}(2005)\citenamefont {Casini},
  \citenamefont {Fosco},\ and\ \citenamefont {Huerta}}]{Casini2005}%
  \BibitemOpen
  \bibfield  {author} {\bibinfo {author} {\bibfnamefont {H.}~\bibnamefont
  {Casini}}, \bibinfo {author} {\bibfnamefont {C.~D.}\ \bibnamefont {Fosco}}, \
  and\ \bibinfo {author} {\bibfnamefont {M.}~\bibnamefont {Huerta}},\
  }\bibfield  {title} {\enquote {\bibinfo {title} {Entanglement and alpha
  entropies for a massive dirac field in two dimensions},}\ }\href {\doibase
  10.1088/1742-5468/2005/07/p07007} {\bibfield  {journal} {\bibinfo  {journal}
  {Journal of Statistical Mechanics: Theory and Experiment}\ }\textbf {\bibinfo
  {volume} {2005}},\ \bibinfo {pages} {P07007--P07007} (\bibinfo {year}
  {2005})}\BibitemShut {NoStop}%
\bibitem [{\citenamefont {Larsen}\ and\ \citenamefont
  {Wilczek}(1995)}]{Larsen1995}%
  \BibitemOpen
  \bibfield  {author} {\bibinfo {author} {\bibfnamefont {F.}~\bibnamefont
  {Larsen}}\ and\ \bibinfo {author} {\bibfnamefont {F.}~\bibnamefont
  {Wilczek}},\ }\bibfield  {title} {\enquote {\bibinfo {title} {Geometric
  entropy, wave functionals, and fermions},}\ }\href {\doibase
  https://doi.org/10.1006/aphy.1995.1100} {\bibfield  {journal} {\bibinfo
  {journal} {Annals of Physics}\ }\textbf {\bibinfo {volume} {243}},\ \bibinfo
  {pages} {280--298} (\bibinfo {year} {1995})}\BibitemShut {NoStop}%
\bibitem [{\citenamefont {Delacretaz}\ \emph {et~al.}(2022)\citenamefont
  {Delacretaz}, \citenamefont {Du}, \citenamefont {Mehta},\ and\ \citenamefont
  {Son}}]{Son2022}%
  \BibitemOpen
  \bibfield  {author} {\bibinfo {author} {\bibfnamefont {Luca~V}\ \bibnamefont
  {Delacretaz}}, \bibinfo {author} {\bibfnamefont {Yi-Hsien}\ \bibnamefont
  {Du}}, \bibinfo {author} {\bibfnamefont {Umang}\ \bibnamefont {Mehta}}, \
  and\ \bibinfo {author} {\bibfnamefont {Dam~Thanh}\ \bibnamefont {Son}},\
  }\bibfield  {title} {\enquote {\bibinfo {title} {Nonlinear bosonization of
  fermi surfaces: The method of coadjoint orbits},}\ }\href
  {https://arxiv.org/abs/2203.05004} {\bibfield  {journal} {\bibinfo  {journal}
  {arXiv preprint arXiv:2203.05004}\ } (\bibinfo {year} {2022})}\BibitemShut
  {NoStop}%
\bibitem [{\citenamefont {Dieck}(2008)}]{Dieck2008}%
  \BibitemOpen
  \bibfield  {author} {\bibinfo {author} {\bibfnamefont {T.}~\bibnamefont
  {Dieck}},\ }\href {https://books.google.com/books?id=ruSqmB7LWOcC} {\emph
  {\bibinfo {title} {Algebraic Topology}}},\ EMS textbooks in mathematics\
  (\bibinfo  {publisher} {European Mathematical Society},\ \bibinfo {year}
  {2008})\BibitemShut {NoStop}%
\bibitem [{\citenamefont {Barthel}\ \emph {et~al.}(2006)\citenamefont
  {Barthel}, \citenamefont {Chung},\ and\ \citenamefont
  {Schollw\"ock}}]{Barthel2006}%
  \BibitemOpen
  \bibfield  {author} {\bibinfo {author} {\bibfnamefont {T.}~\bibnamefont
  {Barthel}}, \bibinfo {author} {\bibfnamefont {M.-C.}\ \bibnamefont {Chung}},
  \ and\ \bibinfo {author} {\bibfnamefont {U.}~\bibnamefont {Schollw\"ock}},\
  }\bibfield  {title} {\enquote {\bibinfo {title} {Entanglement scaling in
  critical two-dimensional fermionic and bosonic systems},}\ }\href {\doibase
  10.1103/PhysRevA.74.022329} {\bibfield  {journal} {\bibinfo  {journal} {Phys.
  Rev. A}\ }\textbf {\bibinfo {volume} {74}},\ \bibinfo {pages} {022329}
  (\bibinfo {year} {2006})}\BibitemShut {NoStop}%
\bibitem [{\citenamefont {Li}\ \emph {et~al.}(2006)\citenamefont {Li},
  \citenamefont {Ding}, \citenamefont {Yu}, \citenamefont {Roscilde},\ and\
  \citenamefont {Haas}}]{Li2006}%
  \BibitemOpen
  \bibfield  {author} {\bibinfo {author} {\bibfnamefont {Weifei}\ \bibnamefont
  {Li}}, \bibinfo {author} {\bibfnamefont {Letian}\ \bibnamefont {Ding}},
  \bibinfo {author} {\bibfnamefont {Rong}\ \bibnamefont {Yu}}, \bibinfo
  {author} {\bibfnamefont {Tommaso}\ \bibnamefont {Roscilde}}, \ and\ \bibinfo
  {author} {\bibfnamefont {Stephan}\ \bibnamefont {Haas}},\ }\bibfield  {title}
  {\enquote {\bibinfo {title} {Scaling behavior of entanglement in two- and
  three-dimensional free-fermion systems},}\ }\href {\doibase
  10.1103/PhysRevB.74.073103} {\bibfield  {journal} {\bibinfo  {journal} {Phys.
  Rev. B}\ }\textbf {\bibinfo {volume} {74}},\ \bibinfo {pages} {073103}
  (\bibinfo {year} {2006})}\BibitemShut {NoStop}%
\bibitem [{\citenamefont {Polchinski}(1992)}]{Polchinski1992}%
  \BibitemOpen
  \bibfield  {author} {\bibinfo {author} {\bibfnamefont {Joseph}\ \bibnamefont
  {Polchinski}},\ }\bibfield  {title} {\enquote {\bibinfo {title} {{Effective
  field theory and the Fermi surface}},}\ }in\ \href@noop {} {\emph {\bibinfo
  {booktitle} {{Theoretical Advanced Study Institute (TASI 92): From Black
  Holes and Strings to Particles}}}}\ (\bibinfo {year} {1992})\ pp.\ \bibinfo
  {pages} {235--276},\ \Eprint {http://arxiv.org/abs/hep-th/9210046}
  {arXiv:hep-th/9210046} \BibitemShut {NoStop}%
\bibitem [{\citenamefont {Shankar}(1994)}]{Shankar1994}%
  \BibitemOpen
  \bibfield  {author} {\bibinfo {author} {\bibfnamefont {R.}~\bibnamefont
  {Shankar}},\ }\bibfield  {title} {\enquote {\bibinfo {title}
  {Renormalization-group approach to interacting fermions},}\ }\href {\doibase
  10.1103/RevModPhys.66.129} {\bibfield  {journal} {\bibinfo  {journal} {Rev.
  Mod. Phys.}\ }\textbf {\bibinfo {volume} {66}},\ \bibinfo {pages} {129--192}
  (\bibinfo {year} {1994})}\BibitemShut {NoStop}%
\bibitem [{\citenamefont {Hastings}\ \emph {et~al.}(2010)\citenamefont
  {Hastings}, \citenamefont {Gonz\'alez}, \citenamefont {Kallin},\ and\
  \citenamefont {Melko}}]{Hastings2010}%
  \BibitemOpen
  \bibfield  {author} {\bibinfo {author} {\bibfnamefont {Matthew~B.}\
  \bibnamefont {Hastings}}, \bibinfo {author} {\bibfnamefont {Iv\'an}\
  \bibnamefont {Gonz\'alez}}, \bibinfo {author} {\bibfnamefont {Ann~B.}\
  \bibnamefont {Kallin}}, \ and\ \bibinfo {author} {\bibfnamefont {Roger~G.}\
  \bibnamefont {Melko}},\ }\bibfield  {title} {\enquote {\bibinfo {title}
  {Measuring renyi entanglement entropy in quantum monte carlo simulations},}\
  }\href {\doibase 10.1103/PhysRevLett.104.157201} {\bibfield  {journal}
  {\bibinfo  {journal} {Phys. Rev. Lett.}\ }\textbf {\bibinfo {volume} {104}},\
  \bibinfo {pages} {157201} (\bibinfo {year} {2010})}\BibitemShut {NoStop}%
\bibitem [{\citenamefont {Humeniuk}\ and\ \citenamefont
  {Roscilde}(2012)}]{Humeniuk2012}%
  \BibitemOpen
  \bibfield  {author} {\bibinfo {author} {\bibfnamefont {Stephan}\ \bibnamefont
  {Humeniuk}}\ and\ \bibinfo {author} {\bibfnamefont {Tommaso}\ \bibnamefont
  {Roscilde}},\ }\bibfield  {title} {\enquote {\bibinfo {title} {Quantum monte
  carlo calculation of entanglement r\'enyi entropies for generic quantum
  systems},}\ }\href {\doibase 10.1103/PhysRevB.86.235116} {\bibfield
  {journal} {\bibinfo  {journal} {Phys. Rev. B}\ }\textbf {\bibinfo {volume}
  {86}},\ \bibinfo {pages} {235116} (\bibinfo {year} {2012})}\BibitemShut
  {NoStop}%
\bibitem [{\citenamefont {Grover}(2013)}]{Grover2013}%
  \BibitemOpen
  \bibfield  {author} {\bibinfo {author} {\bibfnamefont {Tarun}\ \bibnamefont
  {Grover}},\ }\bibfield  {title} {\enquote {\bibinfo {title} {Entanglement of
  interacting fermions in quantum monte carlo calculations},}\ }\href {\doibase
  10.1103/PhysRevLett.111.130402} {\bibfield  {journal} {\bibinfo  {journal}
  {Phys. Rev. Lett.}\ }\textbf {\bibinfo {volume} {111}},\ \bibinfo {pages}
  {130402} (\bibinfo {year} {2013})}\BibitemShut {NoStop}%
\bibitem [{\citenamefont {Broecker}\ and\ \citenamefont
  {Trebst}(2014)}]{Broecker2014}%
  \BibitemOpen
  \bibfield  {author} {\bibinfo {author} {\bibfnamefont {Peter}\ \bibnamefont
  {Broecker}}\ and\ \bibinfo {author} {\bibfnamefont {Simon}\ \bibnamefont
  {Trebst}},\ }\bibfield  {title} {\enquote {\bibinfo {title} {R{\'{e}}nyi
  entropies of interacting fermions from determinantal quantum monte carlo
  simulations},}\ }\href {\doibase 10.1088/1742-5468/2014/08/p08015} {\bibfield
   {journal} {\bibinfo  {journal} {Journal of Statistical Mechanics: Theory and
  Experiment}\ }\textbf {\bibinfo {volume} {2014}},\ \bibinfo {pages} {P08015}
  (\bibinfo {year} {2014})}\BibitemShut {NoStop}%
\bibitem [{\citenamefont {Wang}\ and\ \citenamefont {Troyer}(2014)}]{Wang2014}%
  \BibitemOpen
  \bibfield  {author} {\bibinfo {author} {\bibfnamefont {Lei}\ \bibnamefont
  {Wang}}\ and\ \bibinfo {author} {\bibfnamefont {Matthias}\ \bibnamefont
  {Troyer}},\ }\bibfield  {title} {\enquote {\bibinfo {title} {Renyi
  entanglement entropy of interacting fermions calculated using the
  continuous-time quantum monte carlo method},}\ }\href {\doibase
  10.1103/PhysRevLett.113.110401} {\bibfield  {journal} {\bibinfo  {journal}
  {Phys. Rev. Lett.}\ }\textbf {\bibinfo {volume} {113}},\ \bibinfo {pages}
  {110401} (\bibinfo {year} {2014})}\BibitemShut {NoStop}%
\bibitem [{\citenamefont {Pichler}\ \emph {et~al.}(2013)\citenamefont
  {Pichler}, \citenamefont {Bonnes}, \citenamefont {Daley}, \citenamefont
  {Läuchli},\ and\ \citenamefont {Zoller}}]{Pichler2013}%
  \BibitemOpen
  \bibfield  {author} {\bibinfo {author} {\bibfnamefont {Hannes}\ \bibnamefont
  {Pichler}}, \bibinfo {author} {\bibfnamefont {Lars}\ \bibnamefont {Bonnes}},
  \bibinfo {author} {\bibfnamefont {Andrew~J}\ \bibnamefont {Daley}}, \bibinfo
  {author} {\bibfnamefont {Andreas~M}\ \bibnamefont {Läuchli}}, \ and\
  \bibinfo {author} {\bibfnamefont {Peter}\ \bibnamefont {Zoller}},\ }\bibfield
   {title} {\enquote {\bibinfo {title} {Thermal versus entanglement entropy: a
  measurement protocol for fermionic atoms with a quantum gas microscope},}\
  }\href {\doibase 10.1088/1367-2630/15/6/063003} {\bibfield  {journal}
  {\bibinfo  {journal} {New Journal of Physics}\ }\textbf {\bibinfo {volume}
  {15}},\ \bibinfo {pages} {063003} (\bibinfo {year} {2013})}\BibitemShut
  {NoStop}%
\bibitem [{\citenamefont {Islam}\ \emph {et~al.}(2015)\citenamefont {Islam},
  \citenamefont {Ma}, \citenamefont {Preiss}, \citenamefont {Eric~Tai},
  \citenamefont {Lukin}, \citenamefont {Rispoli},\ and\ \citenamefont
  {Greiner}}]{Islam2015}%
  \BibitemOpen
  \bibfield  {author} {\bibinfo {author} {\bibfnamefont {Rajibul}\ \bibnamefont
  {Islam}}, \bibinfo {author} {\bibfnamefont {Ruichao}\ \bibnamefont {Ma}},
  \bibinfo {author} {\bibfnamefont {Philipp~M.}\ \bibnamefont {Preiss}},
  \bibinfo {author} {\bibfnamefont {M.}~\bibnamefont {Eric~Tai}}, \bibinfo
  {author} {\bibfnamefont {Alexander}\ \bibnamefont {Lukin}}, \bibinfo {author}
  {\bibfnamefont {Matthew}\ \bibnamefont {Rispoli}}, \ and\ \bibinfo {author}
  {\bibfnamefont {Markus}\ \bibnamefont {Greiner}},\ }\bibfield  {title}
  {\enquote {\bibinfo {title} {Measuring entanglement entropy in a quantum
  many-body system},}\ }\href {\doibase 10.1038/nature15750} {\bibfield
  {journal} {\bibinfo  {journal} {Nature}\ }\textbf {\bibinfo {volume} {528}},\
  \bibinfo {pages} {77--83} (\bibinfo {year} {2015})}\BibitemShut {NoStop}%
\bibitem [{\citenamefont {Brydges}\ \emph {et~al.}(2019)\citenamefont
  {Brydges}, \citenamefont {Elben}, \citenamefont {Jurcevic}, \citenamefont
  {Vermersch}, \citenamefont {Maier}, \citenamefont {Lanyon}, \citenamefont
  {Zoller}, \citenamefont {Blatt},\ and\ \citenamefont {Roos}}]{Brydges2019}%
  \BibitemOpen
  \bibfield  {author} {\bibinfo {author} {\bibfnamefont {Tiff}\ \bibnamefont
  {Brydges}}, \bibinfo {author} {\bibfnamefont {Andreas}\ \bibnamefont
  {Elben}}, \bibinfo {author} {\bibfnamefont {Petar}\ \bibnamefont {Jurcevic}},
  \bibinfo {author} {\bibfnamefont {Benoît}\ \bibnamefont {Vermersch}},
  \bibinfo {author} {\bibfnamefont {Christine}\ \bibnamefont {Maier}}, \bibinfo
  {author} {\bibfnamefont {Ben~P.}\ \bibnamefont {Lanyon}}, \bibinfo {author}
  {\bibfnamefont {Peter}\ \bibnamefont {Zoller}}, \bibinfo {author}
  {\bibfnamefont {Rainer}\ \bibnamefont {Blatt}}, \ and\ \bibinfo {author}
  {\bibfnamefont {Christian~F.}\ \bibnamefont {Roos}},\ }\bibfield  {title}
  {\enquote {\bibinfo {title} {Probing renyi entanglement entropy via
  randomized measurements},}\ }\href {\doibase 10.1126/science.aau4963}
  {\bibfield  {journal} {\bibinfo  {journal} {Science}\ }\textbf {\bibinfo
  {volume} {364}},\ \bibinfo {pages} {260--263} (\bibinfo {year}
  {2019})}\BibitemShut {NoStop}%
\bibitem [{\citenamefont {Cornfeld}\ \emph {et~al.}(2019)\citenamefont
  {Cornfeld}, \citenamefont {Sela},\ and\ \citenamefont
  {Goldstein}}]{Cornfeld2019}%
  \BibitemOpen
  \bibfield  {author} {\bibinfo {author} {\bibfnamefont {Eyal}\ \bibnamefont
  {Cornfeld}}, \bibinfo {author} {\bibfnamefont {Eran}\ \bibnamefont {Sela}}, \
  and\ \bibinfo {author} {\bibfnamefont {Moshe}\ \bibnamefont {Goldstein}},\
  }\bibfield  {title} {\enquote {\bibinfo {title} {Measuring fermionic
  entanglement: Entropy, negativity, and spin structure},}\ }\href {\doibase
  10.1103/PhysRevA.99.062309} {\bibfield  {journal} {\bibinfo  {journal} {Phys.
  Rev. A}\ }\textbf {\bibinfo {volume} {99}},\ \bibinfo {pages} {062309}
  (\bibinfo {year} {2019})}\BibitemShut {NoStop}%
\bibitem [{\citenamefont {Hasan}\ and\ \citenamefont {Kane}(2010)}]{Hasan2010}%
  \BibitemOpen
  \bibfield  {author} {\bibinfo {author} {\bibfnamefont {M.~Z.}\ \bibnamefont
  {Hasan}}\ and\ \bibinfo {author} {\bibfnamefont {C.~L.}\ \bibnamefont
  {Kane}},\ }\bibfield  {title} {\enquote {\bibinfo {title} {Colloquium:
  Topological insulators},}\ }\href {\doibase 10.1103/RevModPhys.82.3045}
  {\bibfield  {journal} {\bibinfo  {journal} {Rev. Mod. Phys.}\ }\textbf
  {\bibinfo {volume} {82}},\ \bibinfo {pages} {3045--3067} (\bibinfo {year}
  {2010})}\BibitemShut {NoStop}%
\bibitem [{\citenamefont {Qi}\ and\ \citenamefont {Zhang}(2011)}]{Qi2011}%
  \BibitemOpen
  \bibfield  {author} {\bibinfo {author} {\bibfnamefont {Xiao-Liang}\
  \bibnamefont {Qi}}\ and\ \bibinfo {author} {\bibfnamefont {Shou-Cheng}\
  \bibnamefont {Zhang}},\ }\bibfield  {title} {\enquote {\bibinfo {title}
  {Topological insulators and superconductors},}\ }\href {\doibase
  10.1103/RevModPhys.83.1057} {\bibfield  {journal} {\bibinfo  {journal} {Rev.
  Mod. Phys.}\ }\textbf {\bibinfo {volume} {83}},\ \bibinfo {pages}
  {1057--1110} (\bibinfo {year} {2011})}\BibitemShut {NoStop}%
\bibitem [{\citenamefont {Castro~Neto}\ and\ \citenamefont
  {Fradkin}(1994)}]{CastroNeto1994}%
  \BibitemOpen
  \bibfield  {author} {\bibinfo {author} {\bibfnamefont {A.~H.}\ \bibnamefont
  {Castro~Neto}}\ and\ \bibinfo {author} {\bibfnamefont {Eduardo}\ \bibnamefont
  {Fradkin}},\ }\bibfield  {title} {\enquote {\bibinfo {title} {Bosonization of
  fermi liquids},}\ }\href {\doibase 10.1103/PhysRevB.49.10877} {\bibfield
  {journal} {\bibinfo  {journal} {Phys. Rev. B}\ }\textbf {\bibinfo {volume}
  {49}},\ \bibinfo {pages} {10877--10892} (\bibinfo {year} {1994})}\BibitemShut
  {NoStop}%
\bibitem [{\citenamefont {Houghton}\ \emph {et~al.}(2000)\citenamefont
  {Houghton}, \citenamefont {Kwon},\ and\ \citenamefont
  {Marston}}]{Houghton2000}%
  \BibitemOpen
  \bibfield  {author} {\bibinfo {author} {\bibfnamefont {A.}~\bibnamefont
  {Houghton}}, \bibinfo {author} {\bibfnamefont {H.-J.}\ \bibnamefont {Kwon}},
  \ and\ \bibinfo {author} {\bibfnamefont {J.~B.}\ \bibnamefont {Marston}},\
  }\bibfield  {title} {\enquote {\bibinfo {title} {Multidimensional
  bosonization},}\ }\href {\doibase 10.1080/000187300243363} {\bibfield
  {journal} {\bibinfo  {journal} {Advances in Physics}\ }\textbf {\bibinfo
  {volume} {49}},\ \bibinfo {pages} {141--228} (\bibinfo {year}
  {2000})}\BibitemShut {NoStop}%
\bibitem [{\citenamefont {Myers}\ and\ \citenamefont
  {Singh}(2012)}]{Myers2012}%
  \BibitemOpen
  \bibfield  {author} {\bibinfo {author} {\bibfnamefont {Robert~C.}\
  \bibnamefont {Myers}}\ and\ \bibinfo {author} {\bibfnamefont {Ajay}\
  \bibnamefont {Singh}},\ }\bibfield  {title} {\enquote {\bibinfo {title}
  {Entanglement entropy for singular surfaces},}\ }\href {\doibase
  10.1007/JHEP09(2012)013} {\bibfield  {journal} {\bibinfo  {journal} {Journal
  of High Energy Physics}\ }\textbf {\bibinfo {volume} {2012}},\ \bibinfo
  {pages} {13} (\bibinfo {year} {2012})}\BibitemShut {NoStop}%
\bibitem [{\citenamefont {Bednik}\ \emph {et~al.}(2019)\citenamefont {Bednik},
  \citenamefont {Hayward~Sierens}, \citenamefont {Guo}, \citenamefont {Myers},\
  and\ \citenamefont {Melko}}]{Bednik2019}%
  \BibitemOpen
  \bibfield  {author} {\bibinfo {author} {\bibfnamefont {Grigory}\ \bibnamefont
  {Bednik}}, \bibinfo {author} {\bibfnamefont {Lauren~E.}\ \bibnamefont
  {Hayward~Sierens}}, \bibinfo {author} {\bibfnamefont {Minyong}\ \bibnamefont
  {Guo}}, \bibinfo {author} {\bibfnamefont {Robert~C.}\ \bibnamefont {Myers}},
  \ and\ \bibinfo {author} {\bibfnamefont {Roger~G.}\ \bibnamefont {Melko}},\
  }\bibfield  {title} {\enquote {\bibinfo {title} {Probing trihedral corner
  entanglement for dirac fermions},}\ }\href {\doibase
  10.1103/PhysRevB.99.155153} {\bibfield  {journal} {\bibinfo  {journal} {Phys.
  Rev. B}\ }\textbf {\bibinfo {volume} {99}},\ \bibinfo {pages} {155153}
  (\bibinfo {year} {2019})}\BibitemShut {NoStop}%
\bibitem [{\citenamefont {Bueno}\ \emph {et~al.}(2019)\citenamefont {Bueno},
  \citenamefont {Casini},\ and\ \citenamefont {Witczak-Krempa}}]{Bueno2019}%
  \BibitemOpen
  \bibfield  {author} {\bibinfo {author} {\bibfnamefont {Pablo}\ \bibnamefont
  {Bueno}}, \bibinfo {author} {\bibfnamefont {Horacio}\ \bibnamefont {Casini}},
  \ and\ \bibinfo {author} {\bibfnamefont {William}\ \bibnamefont
  {Witczak-Krempa}},\ }\bibfield  {title} {\enquote {\bibinfo {title}
  {Generalizing the entanglement entropy of singular regions in conformal field
  theories},}\ }\href {\doibase 10.1007/JHEP08(2019)069} {\bibfield  {journal}
  {\bibinfo  {journal} {Journal of High Energy Physics}\ }\textbf {\bibinfo
  {volume} {2019}},\ \bibinfo {pages} {69} (\bibinfo {year}
  {2019})}\BibitemShut {NoStop}%
\bibitem [{\citenamefont {Kraus}\ \emph {et~al.}(2010)\citenamefont {Kraus},
  \citenamefont {Schuch}, \citenamefont {Verstraete},\ and\ \citenamefont
  {Cirac}}]{Kraus2010}%
  \BibitemOpen
  \bibfield  {author} {\bibinfo {author} {\bibfnamefont {Christina~V.}\
  \bibnamefont {Kraus}}, \bibinfo {author} {\bibfnamefont {Norbert}\
  \bibnamefont {Schuch}}, \bibinfo {author} {\bibfnamefont {Frank}\
  \bibnamefont {Verstraete}}, \ and\ \bibinfo {author} {\bibfnamefont
  {J.~Ignacio}\ \bibnamefont {Cirac}},\ }\bibfield  {title} {\enquote {\bibinfo
  {title} {Fermionic projected entangled pair states},}\ }\href {\doibase
  10.1103/PhysRevA.81.052338} {\bibfield  {journal} {\bibinfo  {journal} {Phys.
  Rev. A}\ }\textbf {\bibinfo {volume} {81}},\ \bibinfo {pages} {052338}
  (\bibinfo {year} {2010})}\BibitemShut {NoStop}%
\bibitem [{\citenamefont {Mortier}\ \emph {et~al.}(2020)\citenamefont
  {Mortier}, \citenamefont {Schuch}, \citenamefont {Verstraete},\ and\
  \citenamefont {Haegeman}}]{Mortier2020}%
  \BibitemOpen
  \bibfield  {author} {\bibinfo {author} {\bibfnamefont {Quinten}\ \bibnamefont
  {Mortier}}, \bibinfo {author} {\bibfnamefont {Norbert}\ \bibnamefont
  {Schuch}}, \bibinfo {author} {\bibfnamefont {Frank}\ \bibnamefont
  {Verstraete}}, \ and\ \bibinfo {author} {\bibfnamefont {Jutho}\ \bibnamefont
  {Haegeman}},\ }\bibfield  {title} {\enquote {\bibinfo {title} {Tensor
  networks can resolve fermi surfaces},}\ }\href
  {https://arxiv.org/abs/2008.11176} {\bibfield  {journal} {\bibinfo  {journal}
  {arXiv preprint arXiv:2008.11176}\ } (\bibinfo {year} {2020})}\BibitemShut
  {NoStop}%
\bibitem [{\citenamefont {Corboz}\ and\ \citenamefont
  {Vidal}(2009)}]{Corboz2009}%
  \BibitemOpen
  \bibfield  {author} {\bibinfo {author} {\bibfnamefont {Philippe}\
  \bibnamefont {Corboz}}\ and\ \bibinfo {author} {\bibfnamefont {Guifr\'e}\
  \bibnamefont {Vidal}},\ }\bibfield  {title} {\enquote {\bibinfo {title}
  {Fermionic multiscale entanglement renormalization ansatz},}\ }\href
  {\doibase 10.1103/PhysRevB.80.165129} {\bibfield  {journal} {\bibinfo
  {journal} {Phys. Rev. B}\ }\textbf {\bibinfo {volume} {80}},\ \bibinfo
  {pages} {165129} (\bibinfo {year} {2009})}\BibitemShut {NoStop}%
\bibitem [{\citenamefont {Corboz}\ \emph {et~al.}(2010)\citenamefont {Corboz},
  \citenamefont {Evenbly}, \citenamefont {Verstraete},\ and\ \citenamefont
  {Vidal}}]{Corboz2010}%
  \BibitemOpen
  \bibfield  {author} {\bibinfo {author} {\bibfnamefont {Philippe}\
  \bibnamefont {Corboz}}, \bibinfo {author} {\bibfnamefont {Glen}\ \bibnamefont
  {Evenbly}}, \bibinfo {author} {\bibfnamefont {Frank}\ \bibnamefont
  {Verstraete}}, \ and\ \bibinfo {author} {\bibfnamefont {Guifr\'e}\
  \bibnamefont {Vidal}},\ }\bibfield  {title} {\enquote {\bibinfo {title}
  {Simulation of interacting fermions with entanglement renormalization},}\
  }\href {\doibase 10.1103/PhysRevA.81.010303} {\bibfield  {journal} {\bibinfo
  {journal} {Phys. Rev. A}\ }\textbf {\bibinfo {volume} {81}},\ \bibinfo
  {pages} {010303} (\bibinfo {year} {2010})}\BibitemShut {NoStop}%
\bibitem [{\citenamefont {Barthel}\ \emph {et~al.}(2009)\citenamefont
  {Barthel}, \citenamefont {Pineda},\ and\ \citenamefont
  {Eisert}}]{Barthel2009}%
  \BibitemOpen
  \bibfield  {author} {\bibinfo {author} {\bibfnamefont {Thomas}\ \bibnamefont
  {Barthel}}, \bibinfo {author} {\bibfnamefont {Carlos}\ \bibnamefont
  {Pineda}}, \ and\ \bibinfo {author} {\bibfnamefont {Jens}\ \bibnamefont
  {Eisert}},\ }\bibfield  {title} {\enquote {\bibinfo {title} {Contraction of
  fermionic operator circuits and the simulation of strongly correlated
  fermions},}\ }\href {\doibase 10.1103/PhysRevA.80.042333} {\bibfield
  {journal} {\bibinfo  {journal} {Phys. Rev. A}\ }\textbf {\bibinfo {volume}
  {80}},\ \bibinfo {pages} {042333} (\bibinfo {year} {2009})}\BibitemShut
  {NoStop}%
\bibitem [{\citenamefont {Pineda}\ \emph {et~al.}(2010)\citenamefont {Pineda},
  \citenamefont {Barthel},\ and\ \citenamefont {Eisert}}]{Pineda2010}%
  \BibitemOpen
  \bibfield  {author} {\bibinfo {author} {\bibfnamefont {Carlos}\ \bibnamefont
  {Pineda}}, \bibinfo {author} {\bibfnamefont {Thomas}\ \bibnamefont
  {Barthel}}, \ and\ \bibinfo {author} {\bibfnamefont {Jens}\ \bibnamefont
  {Eisert}},\ }\bibfield  {title} {\enquote {\bibinfo {title} {Unitary circuits
  for strongly correlated fermions},}\ }\href {\doibase
  10.1103/PhysRevA.81.050303} {\bibfield  {journal} {\bibinfo  {journal} {Phys.
  Rev. A}\ }\textbf {\bibinfo {volume} {81}},\ \bibinfo {pages} {050303}
  (\bibinfo {year} {2010})}\BibitemShut {NoStop}%
\bibitem [{\citenamefont {Barthel}\ \emph {et~al.}(2010)\citenamefont
  {Barthel}, \citenamefont {Kliesch},\ and\ \citenamefont
  {Eisert}}]{Barthel2010}%
  \BibitemOpen
  \bibfield  {author} {\bibinfo {author} {\bibfnamefont {Thomas}\ \bibnamefont
  {Barthel}}, \bibinfo {author} {\bibfnamefont {Martin}\ \bibnamefont
  {Kliesch}}, \ and\ \bibinfo {author} {\bibfnamefont {Jens}\ \bibnamefont
  {Eisert}},\ }\bibfield  {title} {\enquote {\bibinfo {title} {Real-space
  renormalization yields finite correlations},}\ }\href {\doibase
  10.1103/PhysRevLett.105.010502} {\bibfield  {journal} {\bibinfo  {journal}
  {Phys. Rev. Lett.}\ }\textbf {\bibinfo {volume} {105}},\ \bibinfo {pages}
  {010502} (\bibinfo {year} {2010})}\BibitemShut {NoStop}%
\bibitem [{\citenamefont {Evenbly}\ and\ \citenamefont
  {Vidal}(2014{\natexlab{a}})}]{Evenbly2014}%
  \BibitemOpen
  \bibfield  {author} {\bibinfo {author} {\bibfnamefont {G.}~\bibnamefont
  {Evenbly}}\ and\ \bibinfo {author} {\bibfnamefont {G.}~\bibnamefont
  {Vidal}},\ }\bibfield  {title} {\enquote {\bibinfo {title} {Class of highly
  entangled many-body states that can be efficiently simulated},}\ }\href
  {\doibase 10.1103/PhysRevLett.112.240502} {\bibfield  {journal} {\bibinfo
  {journal} {Phys. Rev. Lett.}\ }\textbf {\bibinfo {volume} {112}},\ \bibinfo
  {pages} {240502} (\bibinfo {year} {2014}{\natexlab{a}})}\BibitemShut
  {NoStop}%
\bibitem [{\citenamefont {Evenbly}\ and\ \citenamefont
  {Vidal}(2014{\natexlab{b}})}]{Evenbly2014B}%
  \BibitemOpen
  \bibfield  {author} {\bibinfo {author} {\bibfnamefont {G.}~\bibnamefont
  {Evenbly}}\ and\ \bibinfo {author} {\bibfnamefont {G.}~\bibnamefont
  {Vidal}},\ }\bibfield  {title} {\enquote {\bibinfo {title} {Scaling of
  entanglement entropy in the (branching) multiscale entanglement
  renormalization ansatz},}\ }\href {\doibase 10.1103/PhysRevB.89.235113}
  {\bibfield  {journal} {\bibinfo  {journal} {Phys. Rev. B}\ }\textbf {\bibinfo
  {volume} {89}},\ \bibinfo {pages} {235113} (\bibinfo {year}
  {2014}{\natexlab{b}})}\BibitemShut {NoStop}%
\bibitem [{\citenamefont {Haegeman}\ \emph {et~al.}(2018)\citenamefont
  {Haegeman}, \citenamefont {Swingle}, \citenamefont {Walter}, \citenamefont
  {Cotler}, \citenamefont {Evenbly},\ and\ \citenamefont
  {Scholz}}]{Haegeman2018}%
  \BibitemOpen
  \bibfield  {author} {\bibinfo {author} {\bibfnamefont {Jutho}\ \bibnamefont
  {Haegeman}}, \bibinfo {author} {\bibfnamefont {Brian}\ \bibnamefont
  {Swingle}}, \bibinfo {author} {\bibfnamefont {Michael}\ \bibnamefont
  {Walter}}, \bibinfo {author} {\bibfnamefont {Jordan}\ \bibnamefont {Cotler}},
  \bibinfo {author} {\bibfnamefont {Glen}\ \bibnamefont {Evenbly}}, \ and\
  \bibinfo {author} {\bibfnamefont {Volkher~B.}\ \bibnamefont {Scholz}},\
  }\bibfield  {title} {\enquote {\bibinfo {title} {Rigorous free-fermion
  entanglement renormalization from wavelet theory},}\ }\href {\doibase
  10.1103/PhysRevX.8.011003} {\bibfield  {journal} {\bibinfo  {journal} {Phys.
  Rev. X}\ }\textbf {\bibinfo {volume} {8}},\ \bibinfo {pages} {011003}
  (\bibinfo {year} {2018})}\BibitemShut {NoStop}%
\bibitem [{\citenamefont {Reulet}\ \emph {et~al.}(2003)\citenamefont {Reulet},
  \citenamefont {Senzier},\ and\ \citenamefont {Prober}}]{Reulet2003}%
  \BibitemOpen
  \bibfield  {author} {\bibinfo {author} {\bibfnamefont {B.}~\bibnamefont
  {Reulet}}, \bibinfo {author} {\bibfnamefont {J.}~\bibnamefont {Senzier}}, \
  and\ \bibinfo {author} {\bibfnamefont {D.~E.}\ \bibnamefont {Prober}},\
  }\bibfield  {title} {\enquote {\bibinfo {title} {Environmental effects in the
  third moment of voltage fluctuations in a tunnel junction},}\ }\href
  {\doibase 10.1103/PhysRevLett.91.196601} {\bibfield  {journal} {\bibinfo
  {journal} {Phys. Rev. Lett.}\ }\textbf {\bibinfo {volume} {91}},\ \bibinfo
  {pages} {196601} (\bibinfo {year} {2003})}\BibitemShut {NoStop}%
\bibitem [{\citenamefont {Sukhorukov}\ \emph {et~al.}(2007)\citenamefont
  {Sukhorukov}, \citenamefont {Jordan}, \citenamefont {Gustavsson},
  \citenamefont {Leturcq}, \citenamefont {Ihn},\ and\ \citenamefont
  {Ensslin}}]{Sukhorukov2007}%
  \BibitemOpen
  \bibfield  {author} {\bibinfo {author} {\bibfnamefont {Eugene~V.}\
  \bibnamefont {Sukhorukov}}, \bibinfo {author} {\bibfnamefont {Andrew~N.}\
  \bibnamefont {Jordan}}, \bibinfo {author} {\bibfnamefont {Simon}\
  \bibnamefont {Gustavsson}}, \bibinfo {author} {\bibfnamefont {Renaud}\
  \bibnamefont {Leturcq}}, \bibinfo {author} {\bibfnamefont {Thomas}\
  \bibnamefont {Ihn}}, \ and\ \bibinfo {author} {\bibfnamefont {Klaus}\
  \bibnamefont {Ensslin}},\ }\bibfield  {title} {\enquote {\bibinfo {title}
  {Conditional statistics of electron transport in interacting nanoscale
  conductors},}\ }\href {\doibase 10.1038/nphys564} {\bibfield  {journal}
  {\bibinfo  {journal} {Nature Physics}\ }\textbf {\bibinfo {volume} {3}},\
  \bibinfo {pages} {243--247} (\bibinfo {year} {2007})}\BibitemShut {NoStop}%
\bibitem [{\citenamefont {Gershon}\ \emph {et~al.}(2008)\citenamefont
  {Gershon}, \citenamefont {Bomze}, \citenamefont {Sukhorukov},\ and\
  \citenamefont {Reznikov}}]{Gershon2008}%
  \BibitemOpen
  \bibfield  {author} {\bibinfo {author} {\bibfnamefont {G.}~\bibnamefont
  {Gershon}}, \bibinfo {author} {\bibfnamefont {Yu.}\ \bibnamefont {Bomze}},
  \bibinfo {author} {\bibfnamefont {E.~V.}\ \bibnamefont {Sukhorukov}}, \ and\
  \bibinfo {author} {\bibfnamefont {M.}~\bibnamefont {Reznikov}},\ }\bibfield
  {title} {\enquote {\bibinfo {title} {Detection of non-gaussian fluctuations
  in a quantum point contact},}\ }\href {\doibase
  10.1103/PhysRevLett.101.016803} {\bibfield  {journal} {\bibinfo  {journal}
  {Phys. Rev. Lett.}\ }\textbf {\bibinfo {volume} {101}},\ \bibinfo {pages}
  {016803} (\bibinfo {year} {2008})}\BibitemShut {NoStop}%
\bibitem [{\citenamefont {Flindt}\ \emph {et~al.}(2009)\citenamefont {Flindt},
  \citenamefont {Fricke}, \citenamefont {Hohls}, \citenamefont {Novotný},
  \citenamefont {Netočný}, \citenamefont {Brandes},\ and\ \citenamefont
  {Haug}}]{Flindt2009}%
  \BibitemOpen
  \bibfield  {author} {\bibinfo {author} {\bibfnamefont {C.}~\bibnamefont
  {Flindt}}, \bibinfo {author} {\bibfnamefont {C.}~\bibnamefont {Fricke}},
  \bibinfo {author} {\bibfnamefont {F.}~\bibnamefont {Hohls}}, \bibinfo
  {author} {\bibfnamefont {T.}~\bibnamefont {Novotný}}, \bibinfo {author}
  {\bibfnamefont {K.}~\bibnamefont {Netočný}}, \bibinfo {author}
  {\bibfnamefont {T.}~\bibnamefont {Brandes}}, \ and\ \bibinfo {author}
  {\bibfnamefont {R.~J.}\ \bibnamefont {Haug}},\ }\bibfield  {title} {\enquote
  {\bibinfo {title} {Universal oscillations in counting statistics},}\ }\href
  {\doibase 10.1073/pnas.0901002106} {\bibfield  {journal} {\bibinfo  {journal}
  {Proceedings of the National Academy of Sciences}\ }\textbf {\bibinfo
  {volume} {106}},\ \bibinfo {pages} {10116--10119} (\bibinfo {year}
  {2009})}\BibitemShut {NoStop}%
\bibitem [{\citenamefont {Bl\"ote}\ \emph {et~al.}(1986)\citenamefont
  {Bl\"ote}, \citenamefont {Cardy},\ and\ \citenamefont
  {Nightingale}}]{Nightingale1986}%
  \BibitemOpen
  \bibfield  {author} {\bibinfo {author} {\bibfnamefont {H.~W.~J.}\
  \bibnamefont {Bl\"ote}}, \bibinfo {author} {\bibfnamefont {John~L.}\
  \bibnamefont {Cardy}}, \ and\ \bibinfo {author} {\bibfnamefont {M.~P.}\
  \bibnamefont {Nightingale}},\ }\bibfield  {title} {\enquote {\bibinfo {title}
  {Conformal invariance, the central charge, and universal finite-size
  amplitudes at criticality},}\ }\href {\doibase 10.1103/PhysRevLett.56.742}
  {\bibfield  {journal} {\bibinfo  {journal} {Phys. Rev. Lett.}\ }\textbf
  {\bibinfo {volume} {56}},\ \bibinfo {pages} {742--745} (\bibinfo {year}
  {1986})}\BibitemShut {NoStop}%
\bibitem [{\citenamefont {Kane}\ and\ \citenamefont {Fisher}(1997)}]{Kane1997}%
  \BibitemOpen
  \bibfield  {author} {\bibinfo {author} {\bibfnamefont {C.~L.}\ \bibnamefont
  {Kane}}\ and\ \bibinfo {author} {\bibfnamefont {Matthew P.~A.}\ \bibnamefont
  {Fisher}},\ }\bibfield  {title} {\enquote {\bibinfo {title} {Quantized
  thermal transport in the fractional quantum hall effect},}\ }\href {\doibase
  10.1103/PhysRevB.55.15832} {\bibfield  {journal} {\bibinfo  {journal} {Phys.
  Rev. B}\ }\textbf {\bibinfo {volume} {55}},\ \bibinfo {pages} {15832--15837}
  (\bibinfo {year} {1997})}\BibitemShut {NoStop}%
\bibitem [{\citenamefont {Read}\ and\ \citenamefont {Green}(2000)}]{Read2000}%
  \BibitemOpen
  \bibfield  {author} {\bibinfo {author} {\bibfnamefont {N.}~\bibnamefont
  {Read}}\ and\ \bibinfo {author} {\bibfnamefont {Dmitry}\ \bibnamefont
  {Green}},\ }\bibfield  {title} {\enquote {\bibinfo {title} {Paired states of
  fermions in two dimensions with breaking of parity and time-reversal
  symmetries and the fractional quantum hall effect},}\ }\href {\doibase
  10.1103/PhysRevB.61.10267} {\bibfield  {journal} {\bibinfo  {journal} {Phys.
  Rev. B}\ }\textbf {\bibinfo {volume} {61}},\ \bibinfo {pages} {10267--10297}
  (\bibinfo {year} {2000})}\BibitemShut {NoStop}%
\bibitem [{\citenamefont {Cappelli}\ \emph {et~al.}(2002)\citenamefont
  {Cappelli}, \citenamefont {Huerta},\ and\ \citenamefont
  {Zemba}}]{Cappelli2002}%
  \BibitemOpen
  \bibfield  {author} {\bibinfo {author} {\bibfnamefont {Andrea}\ \bibnamefont
  {Cappelli}}, \bibinfo {author} {\bibfnamefont {Marina}\ \bibnamefont
  {Huerta}}, \ and\ \bibinfo {author} {\bibfnamefont {Guillermo~R.}\
  \bibnamefont {Zemba}},\ }\bibfield  {title} {\enquote {\bibinfo {title}
  {Thermal transport in chiral conformal theories and hierarchical quantum hall
  states},}\ }\href {\doibase https://doi.org/10.1016/S0550-3213(02)00340-1}
  {\bibfield  {journal} {\bibinfo  {journal} {Nuclear Physics B}\ }\textbf
  {\bibinfo {volume} {636}},\ \bibinfo {pages} {568--582} (\bibinfo {year}
  {2002})}\BibitemShut {NoStop}%
\bibitem [{\citenamefont {Luttinger}(1964)}]{Luttinger1964}%
  \BibitemOpen
  \bibfield  {author} {\bibinfo {author} {\bibfnamefont {J.~M.}\ \bibnamefont
  {Luttinger}},\ }\bibfield  {title} {\enquote {\bibinfo {title} {Theory of
  thermal transport coefficients},}\ }\href {\doibase
  10.1103/PhysRev.135.A1505} {\bibfield  {journal} {\bibinfo  {journal} {Phys.
  Rev.}\ }\textbf {\bibinfo {volume} {135}},\ \bibinfo {pages} {A1505--A1514}
  (\bibinfo {year} {1964})}\BibitemShut {NoStop}%
\bibitem [{\citenamefont {Ryu}\ \emph {et~al.}(2012)\citenamefont {Ryu},
  \citenamefont {Moore},\ and\ \citenamefont {Ludwig}}]{Ryu2012_anomaly}%
  \BibitemOpen
  \bibfield  {author} {\bibinfo {author} {\bibfnamefont {Shinsei}\ \bibnamefont
  {Ryu}}, \bibinfo {author} {\bibfnamefont {Joel~E.}\ \bibnamefont {Moore}}, \
  and\ \bibinfo {author} {\bibfnamefont {Andreas W.~W.}\ \bibnamefont
  {Ludwig}},\ }\bibfield  {title} {\enquote {\bibinfo {title} {Electromagnetic
  and gravitational responses and anomalies in topological insulators and
  superconductors},}\ }\href {\doibase 10.1103/PhysRevB.85.045104} {\bibfield
  {journal} {\bibinfo  {journal} {Phys. Rev. B}\ }\textbf {\bibinfo {volume}
  {85}},\ \bibinfo {pages} {045104} (\bibinfo {year} {2012})}\BibitemShut
  {NoStop}%
\bibitem [{\citenamefont {Stone}(2012)}]{Stone2012_anomaly}%
  \BibitemOpen
  \bibfield  {author} {\bibinfo {author} {\bibfnamefont {Michael}\ \bibnamefont
  {Stone}},\ }\bibfield  {title} {\enquote {\bibinfo {title} {Gravitational
  anomalies and thermal hall effect in topological insulators},}\ }\href
  {\doibase 10.1103/PhysRevB.85.184503} {\bibfield  {journal} {\bibinfo
  {journal} {Phys. Rev. B}\ }\textbf {\bibinfo {volume} {85}},\ \bibinfo
  {pages} {184503} (\bibinfo {year} {2012})}\BibitemShut {NoStop}%
\bibitem [{\citenamefont {Ginsparg}(1988)}]{Ginsparg1988}%
  \BibitemOpen
  \bibfield  {author} {\bibinfo {author} {\bibfnamefont {Paul~H.}\ \bibnamefont
  {Ginsparg}},\ }\bibfield  {title} {\enquote {\bibinfo {title} {{APPLIED
  CONFORMAL FIELD THEORY}},}\ }in\ \href@noop {} {\emph {\bibinfo {booktitle}
  {{Les Houches Summer School in Theoretical Physics: Fields, Strings, Critical
  Phenomena Les Houches, France, June 28-August 5, 1988}}}}\ (\bibinfo {year}
  {1988})\ pp.\ \bibinfo {pages} {1--168},\ \Eprint
  {http://arxiv.org/abs/hep-th/9108028} {arXiv:hep-th/9108028 [hep-th]}
  \BibitemShut {NoStop}%
\bibitem [{\citenamefont {Di~Francesco}\ \emph {et~al.}(1997)\citenamefont
  {Di~Francesco}, \citenamefont {Mathieu},\ and\ \citenamefont
  {Senechal}}]{BYB}%
  \BibitemOpen
  \bibfield  {author} {\bibinfo {author} {\bibfnamefont {P.}~\bibnamefont
  {Di~Francesco}}, \bibinfo {author} {\bibfnamefont {P.}~\bibnamefont
  {Mathieu}}, \ and\ \bibinfo {author} {\bibfnamefont {D.}~\bibnamefont
  {Senechal}},\ }\href {\doibase 10.1007/978-1-4612-2256-9} {\emph {\bibinfo
  {title} {{Conformal Field Theory}}}},\ Graduate Texts in Contemporary
  Physics\ (\bibinfo  {publisher} {Springer-Verlag},\ \bibinfo {address} {New
  York},\ \bibinfo {year} {1997})\BibitemShut {NoStop}%
\bibitem [{\citenamefont {Else}\ \emph {et~al.}(2021)\citenamefont {Else},
  \citenamefont {Thorngren},\ and\ \citenamefont {Senthil}}]{Else2021}%
  \BibitemOpen
  \bibfield  {author} {\bibinfo {author} {\bibfnamefont {Dominic~V.}\
  \bibnamefont {Else}}, \bibinfo {author} {\bibfnamefont {Ryan}\ \bibnamefont
  {Thorngren}}, \ and\ \bibinfo {author} {\bibfnamefont {T.}~\bibnamefont
  {Senthil}},\ }\bibfield  {title} {\enquote {\bibinfo {title} {Non-fermi
  liquids as ersatz fermi liquids: General constraints on compressible
  metals},}\ }\href {\doibase 10.1103/PhysRevX.11.021005} {\bibfield  {journal}
  {\bibinfo  {journal} {Phys. Rev. X}\ }\textbf {\bibinfo {volume} {11}},\
  \bibinfo {pages} {021005} (\bibinfo {year} {2021})}\BibitemShut {NoStop}%
\end{thebibliography}

%

\end{document}